\documentclass[aps,prb,twocolumn,superscriptaddress,floatfix]{revtex4-2}
\usepackage{placeins}
\usepackage{graphicx}
\usepackage{multirow}
\usepackage{color}
\usepackage{amsmath}
\usepackage{amsfonts}
\usepackage{amssymb}
\usepackage{graphicx}
\graphicspath{{Figures/}}
\usepackage{epstopdf}
\usepackage{empheq}
\usepackage{float}
\usepackage{cases} 
\usepackage{mathtools}
\usepackage{dcolumn} 
\usepackage{bbold}
\usepackage{xfrac}
\usepackage{bm}
\usepackage{siunitx}

\usepackage[dvipsnames]{xcolor}
\usepackage[colorlinks]{hyperref}
\hypersetup{
colorlinks=True,
linkbordercolor=White,
linkcolor=BrickRed,
citecolor=Blue,	 
}

\DeclarePairedDelimiter\ket{\lvert}{\rangle}

\renewcommand{\vec}[1]{\mathbf{#1}}

\newcommand{\addcomm}[1]{{#1}}
\newcommand\YMN[1]{{#1}}

\begin{document}

\title{Strain engineering in Ge/GeSi spin qubits heterostructures}

\author{Lorenzo Mauro}
\affiliation{Univ. Grenoble Alpes, CEA, IRIG-MEM-L\_Sim, Grenoble, France.}%
\author{Esteban A. Rodr\'iguez-Mena}
\affiliation{Univ. Grenoble Alpes, CEA, IRIG-MEM-L\_Sim, Grenoble, France.}%
\author{Biel Martinez}
\affiliation{Univ. Grenoble Alpes, CEA, LETI, F-38000, Grenoble, France}%
\author{Yann-Michel Niquet}
\email{yniquet@cea.fr}
\affiliation{Univ. Grenoble Alpes, CEA, IRIG-MEM-L\_Sim, Grenoble, France.}%

\date{\today}

\begin{abstract}
The heavy-holes in Ge/GeSi heterostructures show highly anisotropic gyromagnetic response with in-plane $g$-factors $g_{x,y}^*\lesssim 0.3$ and out-of-plane $g$-factor $g_z^*\gtrsim 10$. As a consequence, Rabi hot spots and dephasing sweet lines are extremely sharp and call for a careful alignment of the magnetic field in Ge spin qubit devices. We investigate how the $g$-factors can be engineered by strains. We show that uniaxial strains can raise in-plane $g$-factors above unity while leaving $g_z^*$ essentially constant. We discuss how the etching of an elongated mesa in a strained buffer can actually induce uniaxial (but inhomogeneous) strains in the heterostructure. This broadens the operational magnetic field range and enables spin manipulation by shuttling holes between neighboring dots with different $g$-factors.
\end{abstract}

\maketitle

\section{Introduction}

Hole spin qubits \cite{Loss98,Burkard2023Review,Fang2023Review} in germanium and silicon/germanium heterostructures \cite{Sammak19,Scappucci20} have shown outstanding progress in the past few years \cite{Ares13,Watzinger18,Hendrickx20b,Hendrickx20,Froning21,Hendrickx21,wang2022ultrafast,Borsoi22,Wang2023,Jirovec23,Zhang23,Hendrickx2024,Wang2024}. High fidelity single and two-qubit gates have been demonstrated \cite{Hendrickx20b,Hendrickx20}, as well as the operation of a four-qubit processor \cite{Hendrickx21} and simulator \cite{Wang2023}, the control of charge in a sixteen-dot array \cite{Borsoi22}, and the coherent shuttling in a ten-qubit array \cite{Wang2024}. Planar germanium heterostructures actually benefit from many assets: epitaxial materials are usually of much better quality than semiconductor/oxide interfaces, which reduces disorder near the qubits \cite{Martinez2022,Varley23,Massai23,Martinez24}; moreover, the effective mass of holes is smaller in germanium than in silicon \cite{Sammak19,Scappucci20}, which allows for larger quantum dots (at given orbital splitting) and eases constraints on fabrication. The intrinsic spin-orbit coupling (SOC) in the valence band of semiconductor materials \cite{Winkler03,Marcellina17,Terrazos21,Wang21,Bosco21b,martinez2022hole,Sarkar23,Abadillo2023,Rodriguez2023} enables full electrical control of hole spin qubits \cite{Ares13,Watzinger18,Hendrickx20b,Froning21,wang2022ultrafast}; yet the strength of the SOC is tempered (and made manageable) in germanium heterostructures by the built-in strains that open the heavy-hole (HH)/light-hole (LH) bandgap \cite{Sammak19,Scappucci20}. 

Nonetheless, germanium spin qubits in planar heterostructures may suffer from the strong anisotropy of the gyromagnetic $g$-factors of heavy-holes. Indeed, the out-of-plane $g$-factor can be up to 100 times larger than the in-plane $g$-factors \cite{Scappucci20,Hendrickx2024,martinez2022hole}. While such an anisotropy is characteristic of heavy-holes, it is particularly prominent in germanium. As a consequence, the qubit properties vary rapidly when the magnetic field crosses the heterostructure plane (where the devices are usually operated). The width of the dephasing ``sweet lines'' or Rabi ``hot spots'' can actually be as small as a few degrees on the unit sphere describing the magnetic field orientation \cite{martinez2022hole,Hendrickx2024,Mauro24}. This calls for a careful alignment of the magnetic field, and is a source of variability that may hinder the optimal operation of large arrays of spin qubits. Planar germanium heterostructures would, therefore, benefit from a controllable enhancement of the in-plane $g$-factors.

It has been emphasized recently that strains play an ubiquitous role in spin physics \cite{Venitucci18,Morton18,Pla18,Liles20,Adelsberger23,Woods2024}, especially in silicon/germanium heterostructures \cite{Abadillo2023,Corley2023,Mauro24,Wang2024}. Beyond their impact on the HH/LH bandgap, they are directly responsible for various SOC mechanisms owing to their interplay with the kinetic and Zeeman hamiltonians of the holes. In particular, uniaxial and shear strains directly modulate the in-plane $g$-factors; moreover, the gradients of strains give rise to specific Rashba- and Dresselhaus-like spin-orbit interactions.

In this work, we explore the prospects for strain engineering in planar silicon/germanium heterostructures. We show in section \ref{sec:uniax} that uniaxial strains may significantly increase the in-plane $g$-factors. We support this conclusion with analytical models as well as numerical simulations on realistic setups. We next discuss in section \ref{sec:mesa} how to achieve large uniaxial strains in silicon/germanium heterostructures. We show, in particular, that the relaxation of a rectangular mesa etched in a strained buffer is essentially uniaxial. Non-intentional, inhomogeneous uniaxial strains due to mesa relaxation or ohmic contacts may in fact already play a role in some existing devices. We discuss how such inhomogeneous uniaxial strains can improve the control over the qubits and soften the strong dependence on the magnetic field orientation.

\section{Effects of uniaxial strains on the $g$-factors}
\label{sec:uniax}

The effects of strains on the $g$-factors of a hole can be understood within the $g$-matrix formalism \cite{Abragam1970,Crippa18,Venitucci18}. The effective Zeeman Hamiltonian of a HH confined in a $[001]$-oriented Ge well is
\begin{equation}
H=\frac{1}{2}\mu_B\boldsymbol{\sigma}\cdot g\vec{B}    
\end{equation}
where $\mu_B$ is Bohr's magneton, $\boldsymbol{\sigma}$ is the vector of Pauli matrices, $\vec{B}$ is the magnetic field and $g$ is the gyromagnetic $g$-matrix. In the $\ket{\pm 3/2}$ basis set for the HH states and in the $x=[110]$, $y=[\bar{1}10]$, $z=[001]$ axes for the magnetic field, the diagonal elements of $g$ are \cite{Michal21,martinez2022hole,Abadillo2023}: 
\begin{subequations}
\label{eq:gHH}
\begin{align}
g_{xx}&\approx -3q+\delta g_{xx}^{(c)}+\delta g_{xx}^{(\varepsilon)} \\
g_{yy}&\approx +3q+\delta g_{yy}^{(c)}+\delta g_{yy}^{(\varepsilon)} \\
g_{zz}&\approx6\kappa+\frac{27}{2}q-2\gamma_h\,,
\end{align}
\end{subequations}
where $\kappa=3.41$ and $q=0.06$ are the isotropic and cubic Zeeman parameters of Ge, $\delta g_i^{(c)}$ describe the effects of in-plane confinement on the Zeeman splitting, and $\delta g_i^{(\varepsilon)}$ the effects of strains. We point out that the present in-plane axes $x=[110]$ and $y=[\bar{1}10]$ are different from those ($x=[100]$, $y=[010]$) of Refs.~\cite{martinez2022hole,Abadillo2023} (see later discussion). The in-plane corrections $\delta g_{xx}^{(c)}$ and $\delta g_{yy}^{(c)}$ read \cite{Michal21}
\begin{subequations}
\label{eq:dgc}
\begin{align}
\delta g_{xx}^{(c)}&=+\frac{6}{m_0\Delta_\mathrm{LH}}\left(\lambda\langle p_x^2\rangle-\lambda^\prime\langle p_y^2\rangle\right) \\
\delta g_{yy}^{(c)}&=-\frac{6}{m_0\Delta_\mathrm{LH}}\left(\lambda\langle p_y^2\rangle-\lambda^\prime\langle p_x^2\rangle\right)\,,
\end{align}
\end{subequations}
where $m_0$ is the free electron mass, $\Delta_\mathrm{LH}$ is the HH/LH bandgap, $\lambda=\kappa\gamma_3-2\eta_h\gamma_2\gamma_3\approx -0.38$, and $\lambda^\prime=\kappa\gamma_3-2\eta_h\gamma_3^2\approx -7.15$, with $\gamma_2=4.24$ and $\gamma_3=5.69$ the Luttinger parameters of bulk Ge \addcomm{\cite{Abadillo2023,Winkler03}}. The squared momentum operators $\langle p_x^2\rangle$ and $\langle p_y^2\rangle$ are averaged over the ground-state HH envelope \footnote{We assume here that $\langle p_xp_y\rangle=0$; if not the case there are additional $\delta g_{xy}^{(c)}$ and $\delta g_{yx}^{(c)}$ corrections given in Ref.~\cite{Michal21}. We have also discarded small $\propto\langle p_x^2\rangle/\Delta_\mathrm{LH}$ and $\propto\langle p_y^2\rangle/\Delta_\mathrm{LH}$ corrections to $g_{zz}$.}. The factors $\gamma_h\approx 2.62$ and $\eta_h\approx 0.41$ depend on vertical confinement and account for the action of the magnetic vector potential on the orbital motion of the holes \cite{Michal21}. The contributions from strains are \cite{Piot22}
\begin{equation}
\delta g_{xx}^{(\varepsilon)}=\delta g_{yy}^{(\varepsilon)}=\frac{2\sqrt{3}d_v\kappa}{\Delta_\mathrm{LH}}\left(\langle\varepsilon_{yy}\rangle-\langle\varepsilon_{xx}\rangle\right)\,,
\label{eq:dgs}
\end{equation}
where $d_v=-6.06$\,eV is the shear deformation potential of the valence band and $\langle\varepsilon_{xx}\rangle$, $\langle\varepsilon_{yy}\rangle$ are the expectation values of the in-plane strains \addcomm{\cite{Fischetti96}}. There may also be off-diagonal $g$-matrix corrections $\delta g_{\alpha\beta}^{(\varepsilon)}$ ($\alpha\ne\beta$) proportional to the shear strains $\varepsilon_{\alpha\beta}$ that tilt the principal magnetic axes of the system \cite{Abadillo2023,Mauro24}. Of particular importance in the following is \footnote{For completeness, 
\begin{equation}
\delta g_{zy}^{(\varepsilon)}=-\frac{4\sqrt{3}\kappa d_v}{\Delta_\mathrm{LH}}\langle\varepsilon_{yz}\rangle\,,
\end{equation}
and:
\begin{equation}
\delta g_{xy}^{(\varepsilon)}=-\delta g_{yx}^{(\varepsilon)}=\frac{12b_v\kappa}{\Delta_\mathrm{LH}}\langle\varepsilon_{xy}\rangle\,.
\end{equation}}
\begin{equation}
\delta g_{zx}^{(\varepsilon)}=-\frac{4\sqrt{3}\kappa d_v}{\Delta_\mathrm{LH}}\langle\varepsilon_{xz}\rangle\,.
\label{eq:gzx}
\end{equation}
Eqs.~\eqref{eq:dgc} and \eqref{eq:dgs} account for mixing between HHs and LHs by the kinetic, strain and Zeeman Hamiltonians of the holes to first order in $1/\Delta_\mathrm{LH}$. The HH/LH bandgap can be approximated as
\begin{equation}
\Delta_\mathrm{LH}\approx\frac{2\pi^2\hbar^2\gamma_2}{m_0L_\mathrm{W}^2}+b_v(\langle\varepsilon_{xx}\rangle+\langle\varepsilon_{yy}\rangle-2\langle\varepsilon_{zz}\rangle)\,,
\label{eq:delta}
\end{equation}
where the first term accounts for vertical confinement, and the second for strains in the Ge well (with $L_\mathrm{W}$ the thickness of the well and $b_v=-2.16$\,eV the uniaxial deformation potential of the valence band \addcomm{\cite{Fischetti96}}).

\begin{figure}[t]
\centering
\includegraphics[width=0.75\linewidth]{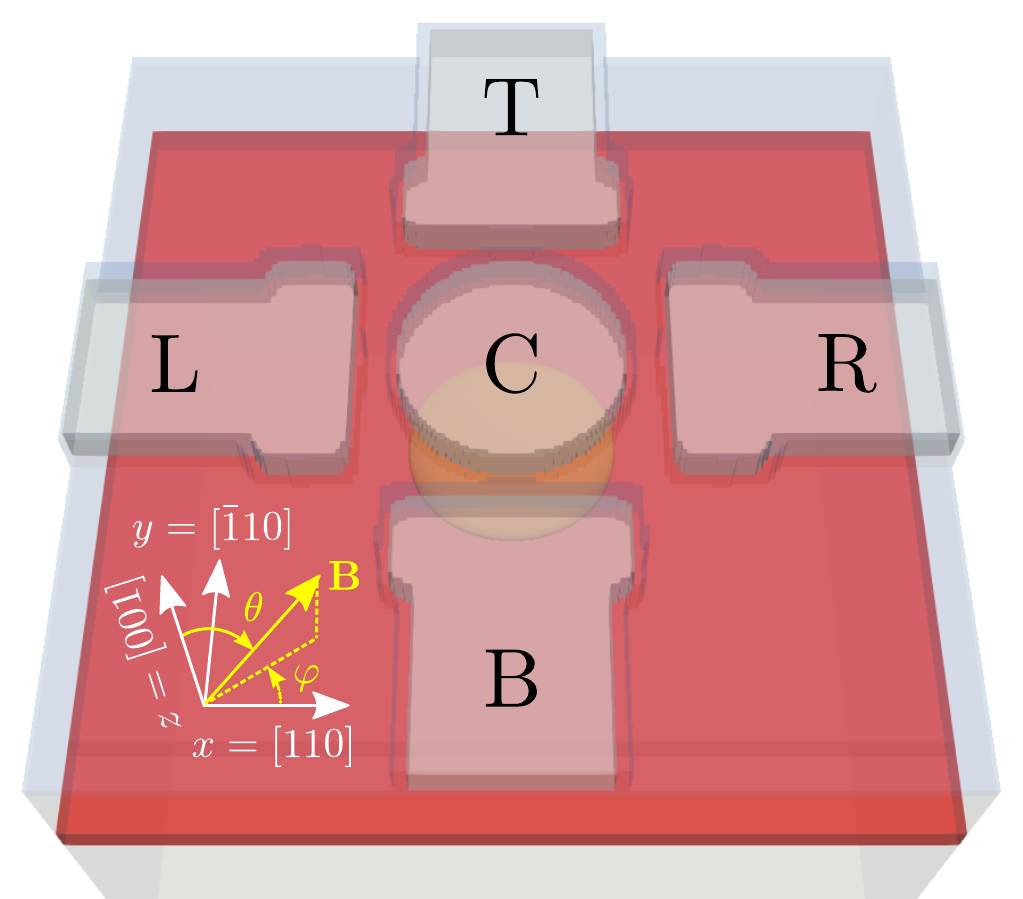}
\caption{The test device is made of a 16-nm-thick Ge quantum well (red) grown on a thick Ge$_{0.8}$Si$_{0.2}$ buffer capped with a 50-nm-thick Ge$_{0.8}$Si$_{0.2}$ barrier (blue). The dot is shaped by five Al gates (gray) embedded in 5 nm of Al$_2$O$_3$. The diameter of the central gate is $d=100$\,nm. The yellow contour is the isodensity surface that encloses $90\%$ of the ground-state hole charge at bias $V_\mathrm{C}=-40$\,mV (side gates grounded). The orientation of the magnetic field $\vec{B}$ is characterized by the angles $\theta$ and $\varphi$ in the crystallographic axes set $x=[110]$, $y=[\bar{1}10]$ and $z=[001]$.}
\label{fig:device}
\end{figure}

The $g$-factors of a ``pure'' heavy hole $|g_{xx}|=|g_{yy}|=3q\approx 0.18$ and $g_{zz}=6\kappa=20.46$ are highly anisotropic. The net anisotropy is much stronger in Ge than in Si quantum dots due to the large $\kappa$ and the weak HH/LH mixings. The HH/LH bandgap $\Delta_\mathrm{LH}$ is, indeed, usually opened by the lattice mismatch in the heterostructure. We emphasize that in a disk-shaped quantum dot in biaxial strains ($\langle p_x^2\rangle=\langle p_y^2\rangle$, $\varepsilon_{xx}=\varepsilon_{yy}$), the in-plane $g$-matrix elements $g_{yy}=-g_{xx}$ are opposite and smaller in magnitude than $3q$. In general, squeezing the dot along $x$ or $y$ ($\langle p_x^2\rangle\ne\langle p_y^2\rangle$) or applying uniaxial strains ($\varepsilon_{xx}\ne\varepsilon_{yy}$) increases or decreases both in-plane $g$-matrix elements. If $\varepsilon_{xx}>\varepsilon_{yy}$ for example, $g_{xx}$ and $g_{yy}$ do increase; $g_{xx}$ thus ultimately changes sign as $g_{xx}-g_{yy}$ is independent on strains. \addcomm{The same can in principle be achieved by squeezing the dot along $y$ \cite{Bosco21b}; however it is hardly possible to achieve large $|g_{xx}|$ and $|g_{yy}|$ without compromising the electrostatic stability of the dot} \footnote{\addcomm{In the geometry of Fig.~\ref{fig:device}, the dot can be squeezed along $y$ by applying a positive voltage on the B and T gates. However, the squeezed hole tends to be pulled out of the well by the B and T gates, and ultimately localizes at the top GeSi/Al$_2$O$_3$ interface where the action of the side gates is screened by the C gate. In the present device, we can at best reach $\langle p_y^2\rangle\approx 9\langle p_x^2\rangle$ and $|g_{xx}|\approx 0.3$, $|g_{yy}|\approx 0$ before losing control of the squeezed dot \cite{martinez2022hole,Mauro24}.}}.

We further illustrate how efficiently $g$-factors respond to uniaxial strains on the device of Fig.~\ref{fig:device}, similar to the one studied in Refs.~\cite{martinez2022hole,Abadillo2023,Rodriguez2023,Mauro24}. This device comprises a 16-nm-thick Ge quantum well grown on a Ge$_{0.8}$Si$_{0.2}$ buffer and capped with a 50-nm-thick Ge$_{0.8}$Si$_{0.2}$ barrier. The quantum dot is shaped by a negative bias on the central plunger gate C with all side L/R/T/B gates grounded. We assume, as a reference, that there are residual tensile strains $\varepsilon_{xx}=\varepsilon_{yy}=\varepsilon_\mathrm{buf}=0.26\%$ in the Ge$_{0.8}$Si$_{0.2}$ buffer \cite{Sammak19}. The compressive biaxial strains in the Ge well are, therefore, $\varepsilon_{xx}=\varepsilon_{yy}=\varepsilon_\parallel=-0.61\%$ and $\varepsilon_{zz}=\varepsilon_\perp=+0.45\%$. We solve Poisson's equation for the potential landscape with a finite volumes method and the Luttinger-Kohn (LK) model \cite{Luttinger56,Winkler03,KP09} for the hole wave functions on a finite-differences grid \cite{Venitucci18}. We next compute the $g$-matrices of a single hole following Ref.~\cite{Venitucci18}.

\begin{figure}[t]
\centering
\includegraphics[width=1.0\linewidth]{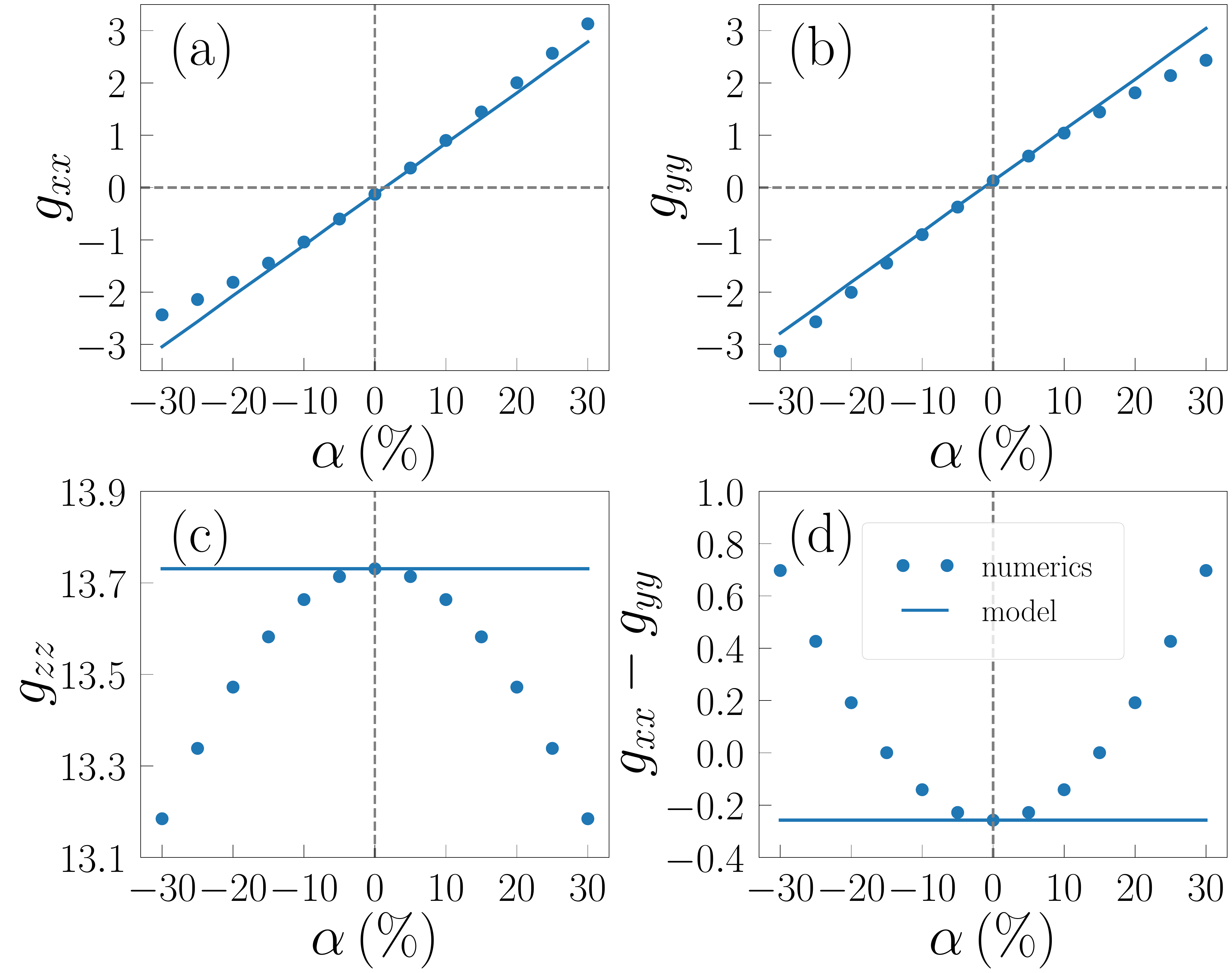}
\caption{The diagonal $g$-matrix elements $g_{ii}(\alpha)$ as a function of the uniaxial strength $\alpha$, at $V_\mathrm{C}=-40$\,mV. The results from the full numerical calculations (dots) are compared to the analytical model (solid lines) of Eqs.~\eqref{eq:gHH}, \eqref{eq:dgc} and \eqref{eq:dgs}.} 
\label{fig:gdelta}
\end{figure}

Starting from this reference device, we apply a uniaxial imbalance between the in-plane strains of all layers. Namely, we substitute $\varepsilon_{xx}\to\varepsilon_\parallel(1-\alpha)$ and $\varepsilon_{yy}\to\varepsilon_\parallel(1+\alpha)$ in the Ge well, and adjust the strains in Ge$_{0.8}$Si$_{0.2}$ accordingly. The uniaxial strength $\alpha$ can therefore be defined as:
\begin{equation}
\alpha=\frac{\varepsilon_{yy}-\varepsilon_{xx}}{\varepsilon_{yy}+\varepsilon_{xx}}\,.
\label{eq:alpha}
\end{equation}
Such an imbalance does not change $\varepsilon_{zz}$ nor $\Delta_\mathrm{LH}$ to first order in $\alpha$ \footnote{Indeed, $\varepsilon_{zz}=-(c_{12}/c_{11})(\varepsilon_{xx}+\varepsilon_{yy})$ with $c_{11}$ and $c_{12}$ the elastic constants of the material, so that $\Delta_\mathrm{LH}$ as given by Eq.~\eqref{eq:deltalh} is independent on $\alpha$ as long as $\varepsilon_{xx}+\varepsilon_{yy}$ is constant.}. We plot in Fig.~\ref{fig:gdelta} the diagonal $g$-matrix elements $g_{ii}(\alpha)=g_{ii}(0)+\delta g_{ii}^{(\varepsilon)}(\alpha)$ obtained from Eq. \eqref{eq:dgc} and from the full numerical calculations. The reference $g$-matrix elements in biaxial strains are $g_{xx}(0)=-0.13$, $g_{yy}(0)=0.13$ and $g_{zz}(0)=13.73$. The HH/LH bandgap $\Delta_\mathrm{LH}=89.4$\,meV  used in Eq.~\eqref{eq:dgc} is fitted to the slope $g_{xx}^\prime(0)=g_{yy}^\prime(0)=4\sqrt{3} d_v\kappa\varepsilon_\parallel/\Delta_\mathrm{LH}=9.72$ of the numerical data \footnote{The fitted $\Delta_\mathrm{LH}$ is actually larger than the $\approx 70$\,meV splitting between the uncoupled, ground HH and LH states expected from Eq.~\eqref{eq:delta} because Eqs.~\eqref{eq:dgc} and \eqref{eq:dgs} are approximations to sums over LH states with different excitation energies.}. As expected from Eq.~\eqref{eq:dgs}, $g_{xx}$ and $g_{yy}$ show a strong (and almost linear) response to uniaxial strains, while $g_{zz}$ experiences smaller (and quadratic) variations. The difference $g_{xx}-g_{yy}$, expected constant from Eq.~\eqref{eq:dgs}, also shows a quadratic dependence on $\alpha$. These non-linearities result from higher order corrections to Eqs.~\eqref{eq:gHH}. $g_{xx}$ changes sign for small $\alpha\approx 1.3\%$ and exceeds 3 once $\alpha\gtrsim 30\%$ (while $g_{yy}$ exceeds 2.4). The situation is symmetric [$g_{xx}(-\alpha)=-g_{yy}(\alpha)$ and $g_{yy}(-\alpha)=-g_{xx}(\alpha)$] for negative $\alpha$'s. 

If the uniaxial strain imbalance $\varepsilon_{xx}-\varepsilon_{yy}$ is applied along $x=[100]$ and $y=[010]$ (instead of $x=[110]$ and $y=[\bar{1}10]$), the in-plane $g$-matrix corrections become \cite{Abadillo2023}:
\begin{equation}
\delta g_{xx}^{(\varepsilon)}=\delta g_{yy}^{(\varepsilon)}=\frac{6b_v\kappa}{\Delta_\mathrm{LH}}\left(\langle\varepsilon_{yy}\rangle-\langle\varepsilon_{xx}\rangle\right)\,.
\label{eq:dgs100}
\end{equation}
The effects of strains are, therefore, qualitatively the same as in the $x=[110]$ and $y=[\bar{1}10]$ axes, but their strength is smaller, since $|d_v|>\sqrt{3}|b_v|$. In particular, the in-plane $g$-factors only reach $\approx 2$ at $\alpha=30\%$ (see Appendix \ref{appendix:100}).

\section{Strain engineering in Ge/GeSi heterostructures}
\label{sec:mesa}

Fig.~\ref{fig:gdelta} opens the way for strain engineering in Ge/GeSi spin qubits. In this section, we discuss how to demonstrate experimentally, and possibly exploit the strong dependence of the in-plane $g$-factors on strains.

Strain engineering has become customary in advanced micro-electronic technologies \cite{Thompson04,Tsutsui19}. Uniaxial strains can be induced by stressors grown or deposited on the devices \cite{Ito00,Shimizu01}, \YMN{or by pulling on a micro-bridge \cite{Pilon2019,Woods2024}}. For demonstration purposes, stress can also be applied with various techniques such as wafer bending \cite{Richter08,Chen11} (hardly suitable for mK fridges though). In the following, we discuss how the relaxation of a strained mesa can induce sizable uniaxial strains in the heterostructure. Although hardly scalable, mesa engineering can be used to test the predictions of this work, and mesa relaxation may (unintentionally) play a role in existing devices. 

\subsection{Example: strain relaxation in etched mesas}

When a mesa is etched in a strained Ge/GeSi heterostructure, the free lateral flanks can deform to relax part of the strains built up in the materials. This is particularly relevant when the GeSi buffer itself is undergoing residual strains \cite{Sammak19}: upon etching, the bulky buffer indeed drives the relaxation of the whole mesa, possibly enhancing the strains in the Ge well itself. If the shape of the mesa is highly anisotropic, so is the resulting strain distribution. Namely, if the width $W$ of the mesa along $x$ is much smaller than its length $L$ along $y$, then the average $\varepsilon_{xx}$ in the buffer decreases much faster than the average $\varepsilon_{yy}$ when increasing the etch depth $t$. 

\begin{figure}[t]
\centering
\includegraphics[width=1.0\linewidth]{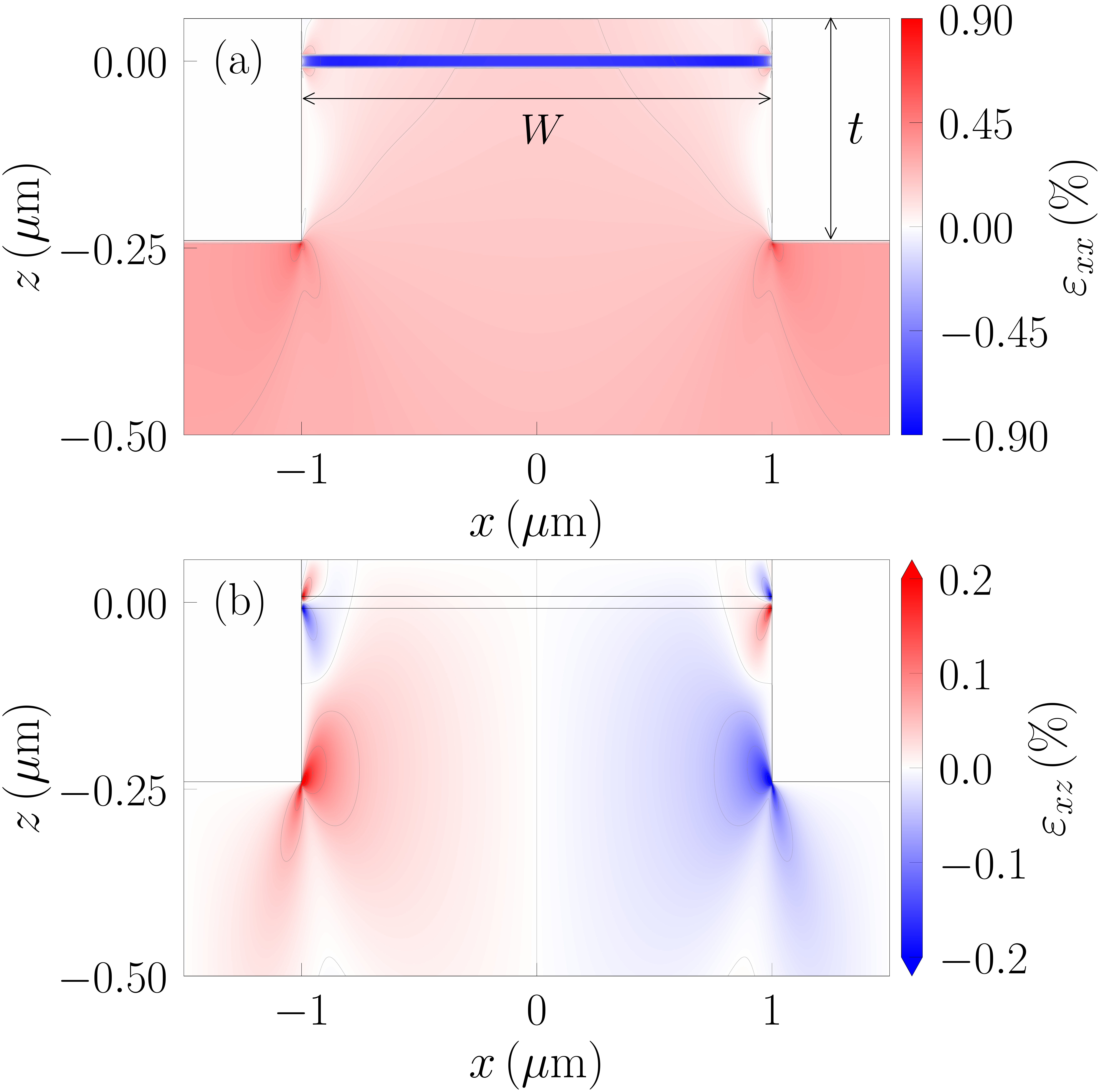}
\caption{(a) Map of the strain $\varepsilon_{xx}$ in the cross section of a mesa with width $W=2\,\mu$m, depth $t=0.3\,\mu$m and length $L\to\infty$. (b) Map of the shear strain $\varepsilon_{xz}$ in the same mesa. The residual strain in the GeSi buffer (before etching) is $\varepsilon_{xx}=\varepsilon_{yy}=\varepsilon_\mathrm{buf}=0.26\%$.} 
\label{fig:mesa}
\end{figure}

In order to compute the resulting, inhomogeneous strain distribution, it is best to consider the limit $L\to\infty$ where the elastic problem becomes bi-dimensional (2D). We thus consider the prototypical mesa structure of Fig.~\ref{fig:mesa}a with fixed width $W=2\,\mu$m, varying depth $t$, and residual in-plane strain $\varepsilon_\mathrm{buf}=0.26\%$ in the buffer (before etching) \cite{Sammak19}. This figure also displays the map of $\varepsilon_{xx}$ computed for depth $t=0.3\,\mu$m with a finite element discretization of the continuum elasticity equations \cite{Abadillo2023}. The strain $\varepsilon_{yy}$ can not relax and remains $\varepsilon_{yy}=0.26\%$ in GeSi and $\varepsilon_{yy}=-0.61\%$ in Ge. The strain $\varepsilon_{xx}$ is very inhomogeneous, being relaxed in GeSi near the flanks but the stronger along the $z$ axis the deeper in the mesa. As a consequence, the strain $\varepsilon_{zz}$ (not shown) is also inhomogeneous and the mesa gets significantly sheared near the edges ($\varepsilon_{xz}\ne 0$, see Fig.~\ref{fig:mesa}b). The Ge well itself gets further stressed by the relaxation of the thicker buffer, with in-plane strain reaching up to $\varepsilon_{xx}=-0.79\%$. \addcomm{The relaxation is the stronger the larger the $t/W$ ratio; the width $W=2\,\mu$m was actually chosen to achieve significant relaxation at small $t<0.5\,\mu$m while leaving room for the qubits.} \YMN{Although similar to the micro-bridge design of Ref.~\cite{Woods2024}, the present mesa structure addresses a different limit $L\gg W$ where there is no pull along $y$ by the bulky anchors at both ends of the bridge}.

\begin{figure}[t]
\centering
\includegraphics[width=0.8\linewidth]{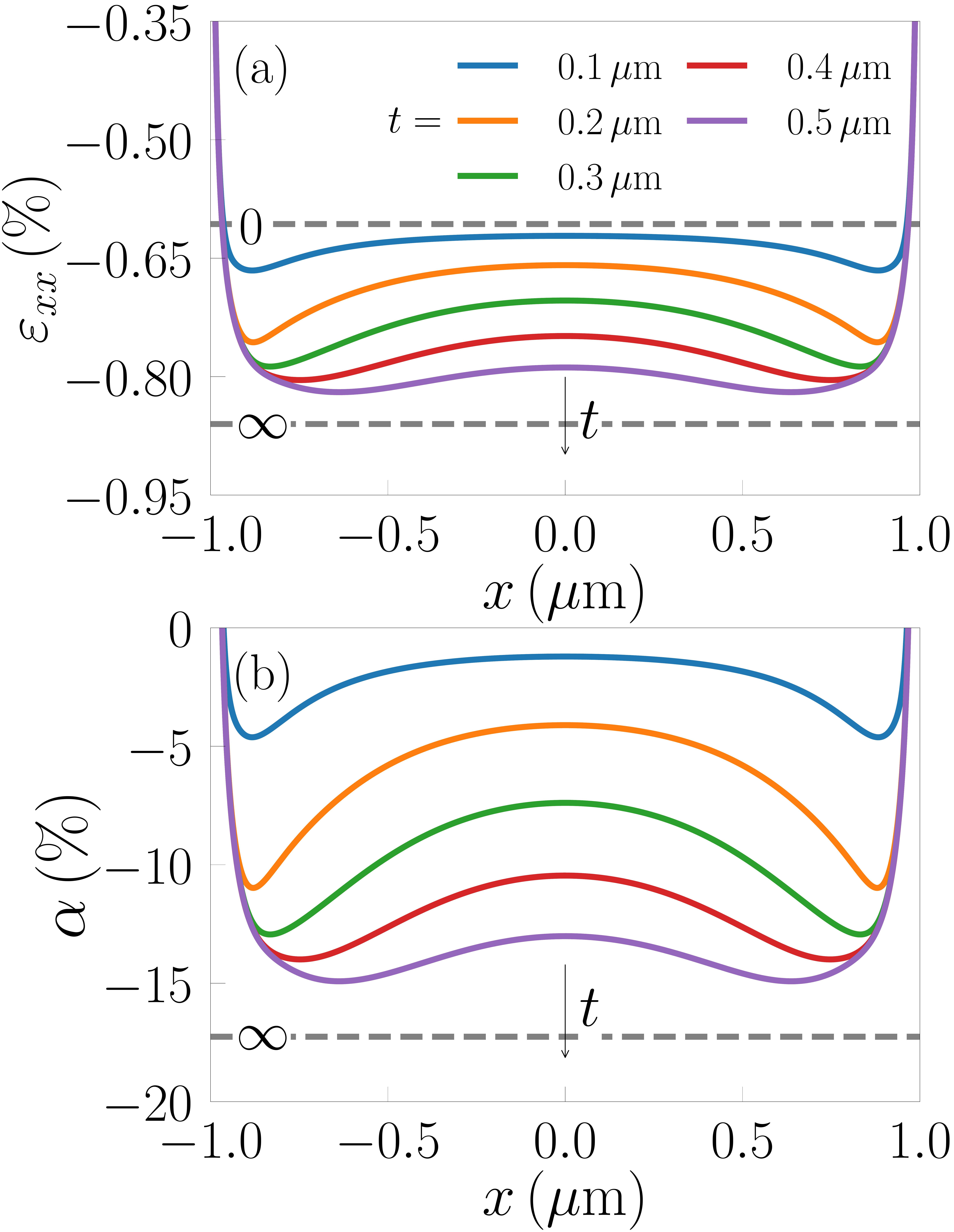}
\caption{(a) Strain $\varepsilon_{xx}(x)$ as a function of the lateral position $x$ along the middle of the Ge well $(z=0)$ for mesas with width $W=2\,\mu$m and depth $t$ varying from $0.1$ to $0.5\,\mu$m. (b) The corresponding uniaxial strengths $\alpha(x)$. In both panels, the horizontal dashed line are the limits $t\to 0$ and $t\to\infty$.}
\label{fig:depth}
\end{figure}

The strain profile $\varepsilon_{xx}(x)$ along the middle of the Ge well ($z=0$) is plotted for various etch depths $t$ in Fig.~\ref{fig:depth}a. The corresponding uniaxial strength $\alpha(x)$ defined by Eq.~\eqref{eq:alpha} is plotted in Fig.~\ref{fig:depth}b. The limits $t\to 0$ ($\varepsilon_{xx}=-0.26\%$ in the buffer and $\varepsilon_{xx}=-0.61\%$ in the Ge well) and $t\to\infty$ (defined here as $\varepsilon_{xx}=0$ in the buffer and $\varepsilon_{xx}=-0.86\%$ in the Ge well \footnote{This definition is introduced for convenience. When $t\to\infty$, the strain $\varepsilon_{xx}$ in the buffer is not expected to tend to zero (but to a negative value, as the buffer remains strained along $y$) and remains inhomogeneous along $x$ owing, notably, to the competition between the relaxation of the buffer and Ge well.}) are reported on these figures as dashed lines. As expected, the larger the $t$, the deeper the relaxation and the more uniaxial the resulting strain profile in the Ge well. The uniaxial strength $\alpha(0)$ at the center of the well is significant even for moderate depths $t$. 

\begin{figure}[t]
\centering
\includegraphics[width=1.0\linewidth]{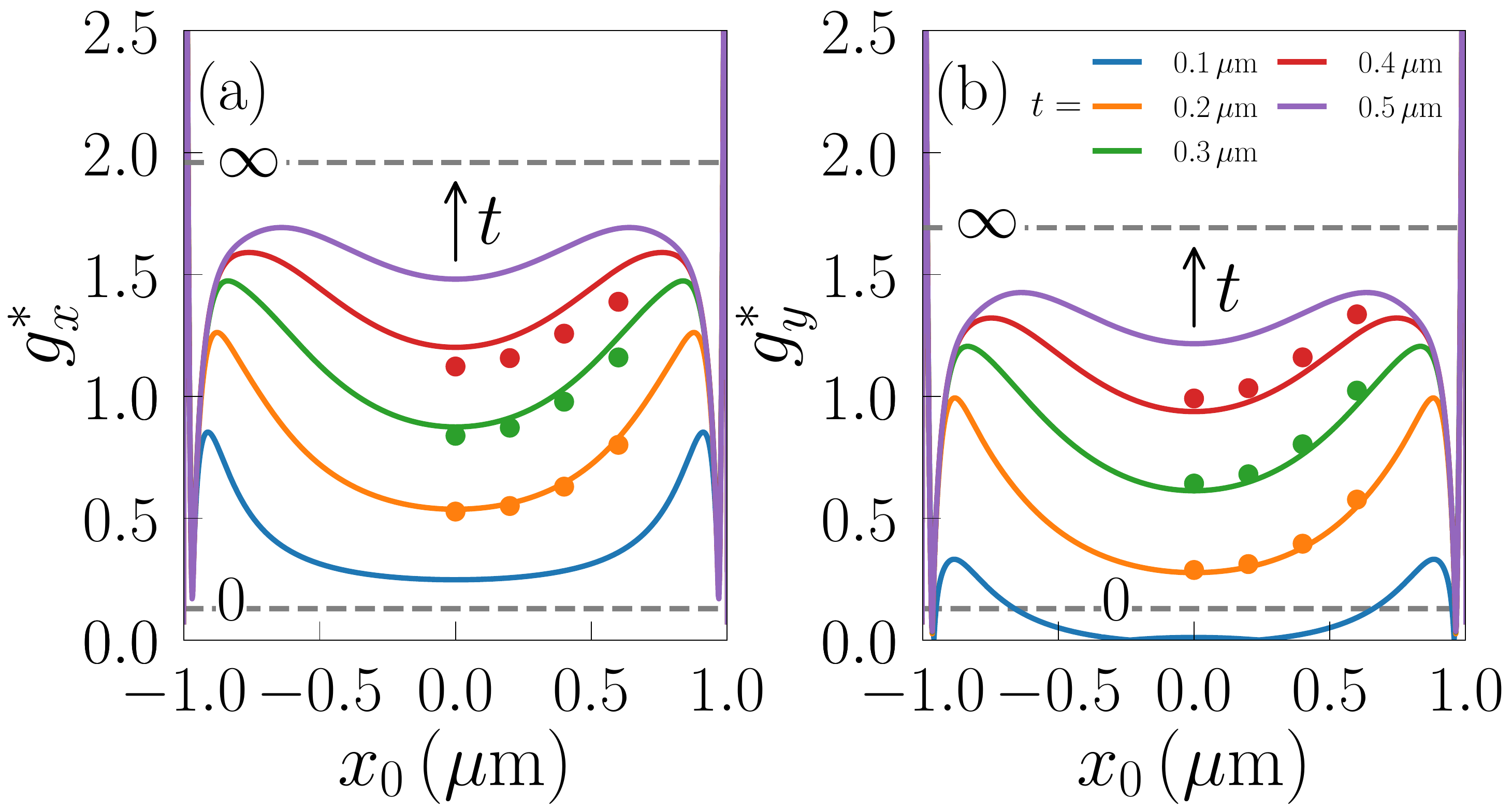}
\caption{The effective in-plane $g$-factors $g_x^*=\sqrt{g_{xx}^2+g_{zx}^2}$ and $g_y^*=|g_{yy}|$ as a function of the lateral position $x_0$ of the dot, for mesas with width $W=2\,\mu$m and depth $t$ varying from $0.1$ to $0.5\,\mu$m. They are either computed from Eqs.~\eqref{eq:gHH} and~\eqref{eq:deltalh} (lines), or from a full solution of the LK equations in the calculated inhomogeneous strains (dots). In both panels, the horizontal dashed lines are the limits $t\to 0$ and $t\to\infty$.}
\label{fig:gmesa}
\end{figure}

We may reconstruct the $g$-matrix of a dot from the calculated strain distributions. For that purpose, we can input the calculated inhomogeneous strain distribution in the LK solver assuming that the plunger gate of Fig.~\ref{fig:device} is centered at some lateral position $x=x_0$. Alternatively, we can use Eqs.~\eqref{eq:gHH} and \eqref{eq:gzx} for the $g$-matrix assuming that the strains $\varepsilon_{ij}(x,z)\approx\varepsilon_{ij}(x_0,z=0)$ are quasi-homogeneous at the scale ($\sim 100$\,nm) of the dot. For better accuracy, we account in the model for the spatial variations of the HH/LH bandgap
\begin{equation}
\Delta_\mathrm{LH}=\Delta_\mathrm{LH}^\mathrm{biax}+b_v(\delta\varepsilon_{xx}+\delta\varepsilon_{yy}-2\delta\varepsilon_{zz})\,,
\label{eq:deltalh}
\end{equation}
where $\Delta_\mathrm{LH}^\mathrm{biax}=89.4$\,meV is the biaxial bandgap and $\delta\varepsilon_{xx}=\varepsilon_{xx}-\varepsilon_\parallel$, $\delta\varepsilon_{yy}=\varepsilon_{yy}-\varepsilon_\parallel$ and $\delta\varepsilon_{zz}=\varepsilon_{zz}-\varepsilon_\perp$ are the deviations of the strains with respect to the biaxial reference.

\begin{figure}[t]
\centering
\includegraphics[width=0.8\linewidth]{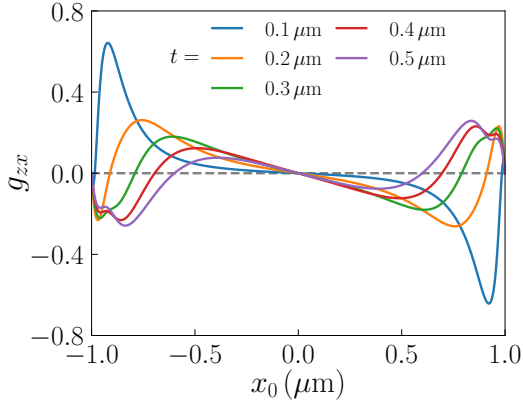}
\caption{The $g$-matrix element $g_{zx}$ as a function of the lateral position $x_0$ of the dot, for mesas with width $W=2\,\mu$m and depth $t$ varying from $0.1$ to $0.5\,\mu$m, computed from Eqs.~\eqref{eq:dgc}.}
\label{fig:gzxmesa}
\end{figure}

We plot in Fig.~\ref{fig:gmesa} the effective in-plane $g$-factors $g_x^*=|g\vec{x}|=\sqrt{g_{xx}^2+g_{zx}^2}$ and $g_y^*=|g\vec{y}|=|g_{yy}|$ computed with both methods as a function of the dot position $x_0$, for various etch depths $t$. The numerical calculations are in good agreement with the expectations of the analytical model. The effective $g$-factors $g_x^*$ and $g_y^*$ exhibit significant variations across the mesa consistent with those of the uniaxial strength $\alpha$. Note that $g_y^*$ shows zeroes when $t=0.1\,\mu$m due to the change of sign of $g_{yy}$ for small uniaxial strengths $\alpha<0$. The off-diagonal $g_{zx}$ resulting from the shear strains $\varepsilon_{xz}$ can be significant near the edges of the mesa (see Fig.~\ref{fig:gzxmesa}). It gives rise to a slight rotation of the principal magnetic axes of the $g$-matrix around $y$, by an angle $\theta\approx g_{zx}/g_{zz}$ ranging from 0 (in the center) to $\pm 0.67^\circ$ (at $x=\pm 0.6\,\mu$m) for $t=0.3\,\mu$m.

The effective in-plane $g$-factors $g_x^*$, and even more so, $g_y^*$ thus get considerably enhanced by the mesa relaxation. The Larmor frequency becomes, moreover, dependent on the position of the dot across the mesa. This dependence is, however, systematic and can in principle be accounted for in the quantum manipulation protocols and algorithms. It may also be leveraged to detune the qubits one from the other and drive them selectively at different Larmor frequencies. \addcomm{We discuss the dependence of $g_x^*$ and $g_y^*$ on $W$ and $t$ in more details in Appendix \ref{appendix:tW}. We also show in Appendix \ref{appendix:FiniteSlope} that a smaller, yet still considerable enhancement can be achieved when the mesas are tapered and the side walls make a finite angle with the vertical.} In the following, we address the consequences for spin manipulation in single and double quantum dots in more detail. 

\subsection{Spin manipulation in single dots}
\label{sec:singledots}

\begin{figure}[t]
\begin{center}
\includegraphics[width=0.8\linewidth]{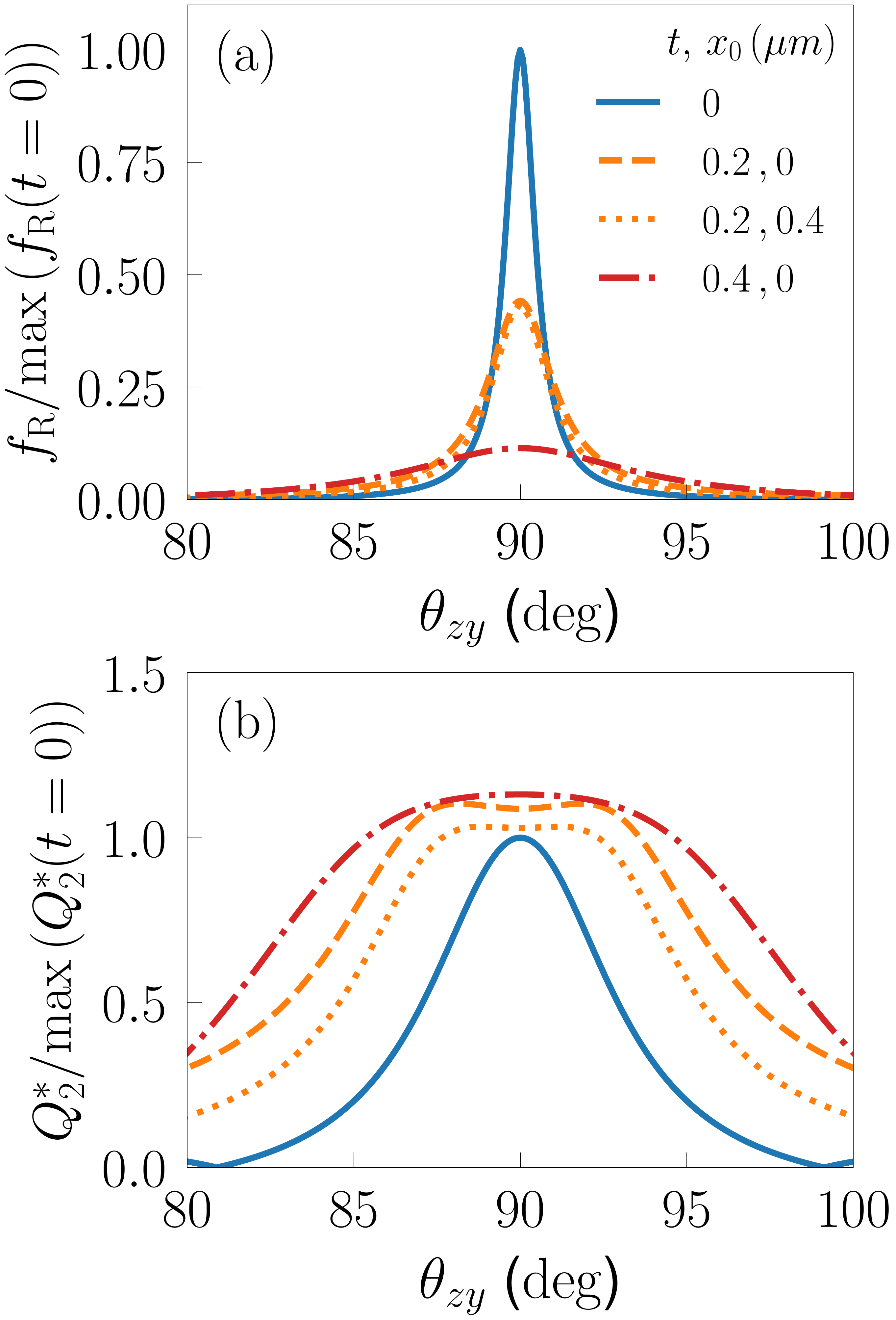}
\caption{(a) Normalized Rabi frequency $f_\mathrm{R}/\mathrm{max}\left(f_\mathrm{R}(t=0)\right)$ for BT drive as a function of the polar angle $\theta_{zy}$ of the magnetic field in the $yz$ plane. (b) Normalized quality factor $Q_2^*/\mathrm{max}\left(Q_2^*(t=0)\right)$ for BT drive as a function of $\theta_{zy}$. The data are plotted for two etch depths $t=0.2\,\mu$m and $t=0.4\,\mu$m ($W=2\,\mu$m), and for two dot positions $x_0=0$ and $x_0=0.4\,\mu$m.}
\label{fig:RabiQ}
\end{center}
\end{figure}

The hole spin can be manipulated by a radiofrequency electric field resonant with the Larmor frequency (electric dipole spin resonance or EDSR \cite{Rashba03,Kato03,Golovach06,Rashba08,Crippa18}). We characterize the resulting spin rotations by the Rabi frequency $f_\mathrm{R}$ and by the quality factor  
\begin{equation}
Q_2^*=2f_\mathrm{R}T_2^*\,,
\end{equation}
which is the number of $\pi$ rotations that can be achieved within one pure dephasing time $T_2^*$. The latter is computed as in Ref.~\cite{Mauro24}, lumping all electrical fluctuations in effective gate voltage noises:
\begin{equation}
\frac{1}{T_2^*}=\Gamma_2^*=\sqrt{2}\pi\sqrt{\sum_\mathrm{G}\left(\delta V_\mathrm{G}^\mathrm{rms}\frac{\partial f_\mathrm{L}}{\partial V_\mathrm{G}}\right)^2}\,.
\label{eq:gammatot}
\end{equation}
Here the sum runs over the gates $\mathrm{G}\in\{\mathrm{C},\mathrm{L},\mathrm{R},\mathrm{T},\mathrm{B}\}$, $f_\mathrm{L}$ is the Larmor frequency whose response to gate G is characterized by the longitudinal spin electric susceptibility (LSES) $\partial f_\mathrm{L}/\partial V_\mathrm{G}$ \cite{Piot22,Mauro24}, and $\delta V_\mathrm{G}^\mathrm{rms}$ is the rms amplitude of the noise on gate G. 

Rabi frequencies and quality factors are plotted on Fig.~\ref{fig:RabiQ} as a function of the polar angle $\theta$ of the magnetic field. They are computed with the $g$-matrix formalism \cite{Venitucci18,Mauro24} in the device of Fig.~\ref{fig:device} centered at $x=x_0$, using the inhomogeneous strains in the etched mesa as input. The additional strains brought by the differential thermal contraction of materials upon cool down are not accounted for in this study (see later discussion) \cite{Abadillo2023}. The hole is driven with opposite gate voltage modulations on the B and T gates (BT drive $\delta V_\mathrm{B}=-\delta V_\mathrm{T}=\tfrac{1}{2}V_\mathrm{ac}\sin 2\pi f_\mathrm{L}t$).

The BT drive is optimal for magnetic fields in the $yz$ plane. For each polar angle $\theta_{zy}$ in this plane, and for each etch depth $t$, the magnetic field strength $|\vec{B}(\theta_{zy},t)|$ is chosen so that the Larmor frequency of the device centered at $x_0=0$ is $f_\mathrm{L}=5$\,GHz. The Larmor frequencies of devices centered at $x_0\ne 0$ are, therefore, different as the effective $g$-factors depend on $x_0$ (but the magnetic field is the same for all devices of a given mesa). We normalize the Rabi frequencies and quality factors of Fig.~\ref{fig:RabiQ} with respect to their maximum values in the unetched device ($t\to0$ irrespective of $x_0$). These normalized figures are indeed independent on the choice of $f_\mathrm{L}$, $V_\mathrm{ac}$ and $\delta V_\mathrm{G}^\mathrm{rms}$'s (assumed the same for all gates), enabling a straightforward comparison of the line shapes.

The Rabi frequencies show a distinct peak near the equatorial plane $\theta_{zy}=90^\circ$. In the present case, the Rabi oscillations essentially result from the coupling between the in-plane and out-of-plane motions of the hole in the non-separable confinement potential of the heterostructure \cite{martinez2022hole}. This gives rise to a modulation of $g_{zy}$ that tilts the principal magnetic axes of the $g$-matrix. Namely, the principal axes $y\to y^\prime$ and $z\to z^\prime$ slightly rock around $x$ when the hole is driven along $y$, so that the magnetic field makes small excursions around the effective equatorial plane $xy^\prime$. These are converted into much larger swings of the hole pseudo-spin precession axis $\vec{\Omega}\propto g\vec{B}$ by the large $g_{zz}\gg |g_{yy}|$ \cite{Mauro24}. 

The motion of the hole in the inhomogeneous shear strains \cite{Abadillo2023} resulting from the mesa relaxation may also give rise to a similar $g$-tensor modulation resonance ($g$-TMR) \cite{Kato03}. However, $\delta g_{zy}^{(\varepsilon)}\propto\langle\varepsilon_{yz}\rangle\approx 0$ in the mesa, whose relaxation does not, therefore, stimulate Rabi oscillations for BT drive. For LR drive $\delta V_\mathrm{L}=-\delta V_\mathrm{R}=\tfrac{1}{2}V_\mathrm{ac}\sin 2\pi f_\mathrm{L}t$, the displacement of the hole in the inhomogeneous $\varepsilon_{xz}$ modulates $g_{zx}$ [see Eq.~\eqref{eq:gzx} and Fig.~\ref{fig:gzxmesa}], as does the coupling between in- and out-of-plane motions. However, the gradient of shear strains due to the mesa relaxation ($\partial\varepsilon_{xz}/\partial x\approx 2\times 10^{-7}$\,nm$^{-1}$ at $t=0.3\,\mu$m) is small and would typically be exceeded by the inhomogeneous strains locally induced by the thermal contraction of the metal gates \cite{Abadillo2023}. We focus, therefore, on BT drive here and discuss how to take best advantage of the position-dependent $g_{zx}$ by shuttling the holes in section \ref{sec:doubledots}. Although cool-down strains may strongly enhance the Rabi frequencies, they would hardly change the conclusions drawn from Fig.~\ref{fig:RabiQ}.

The maximal Rabi frequencies decrease markedly in the mesas. Indeed, $f_\mathrm{R}$ is, to lowest order, proportional to $B_y$ in the equatorial plane \cite{martinez2022hole,Abadillo2023}. The increase of the effective $g$-factor $g_y^*$ in the mesas thus results in a decrease of the magnetic field (at given $f_\mathrm{L}=5$\,GHz at $x_0=0$) and in a slow-down of the Rabi oscillations. The width of the Rabi peak is, on the other hand, $\delta\theta_{zy}\propto |g_{yy}|/g_{zz}$. When $g_{zz}\gg |g_{yy}|$, the magnetic field strength, thus the Rabi frequencies indeed collapse once the magnetic field goes out-of-plane. Moreover, the angular momentum of the HH gets effectively locked along the $z$-axis, and the driving mechanisms discussed above become inefficient \cite{martinez2022hole,Abadillo2023,Mauro24}. The Rabi peak is therefore considerably broadened in the mesas. This softens the constraints on the alignment of the magnetic field and extends the operational range of the qubits. 

This extension is even more visible on the quality factors of Fig.~\ref{fig:RabiQ} b. First of all, the maximum $Q_2^*$ does not vary significantly in the mesas (in contrast to the Rabi frequencies) because both $f_\mathrm{R}$ and $\Gamma_2^*=1/T_2^*$ are proportional to $|\vec{B}|$ (hence their ratio is independent on $|\vec{B}|$). It also peaks in or near the equatorial plane \cite{Mauro24}. The full width at half maximum (FWHM) of the peak is markedly enhanced in the mesas and weakly dependent on $x_0$ (at least when $|x_0|<0.5\,\mu$m). It ranges from $6^\circ$ at $t=0$ to $16.4^\circ$ at $t=0.4\,\mu$m ($x_0=0$). The quality factor remains almost unity at $\theta_{zy}=90\pm3^\circ$ when $t=0.2-0.4\,\mu$m, whereas it is halved in the biaxial limit $t\to 0$. We emphasize that the peaks are significantly broader for BT drive than for the L/R/B/T only drives investigated in Ref.~\onlinecite{Mauro24} (because the Rabi maps are different). A softer angular dependence shall particularly benefit to multiple qubit systems, as it facilitates the search for a global field orientation that fits in the operational range of all individual qubits despite disorder.

\subsection{Spin manipulation in double dots}
\label{sec:doubledots}

\begin{figure}[t]
\centering
\includegraphics[width=1.0\linewidth]{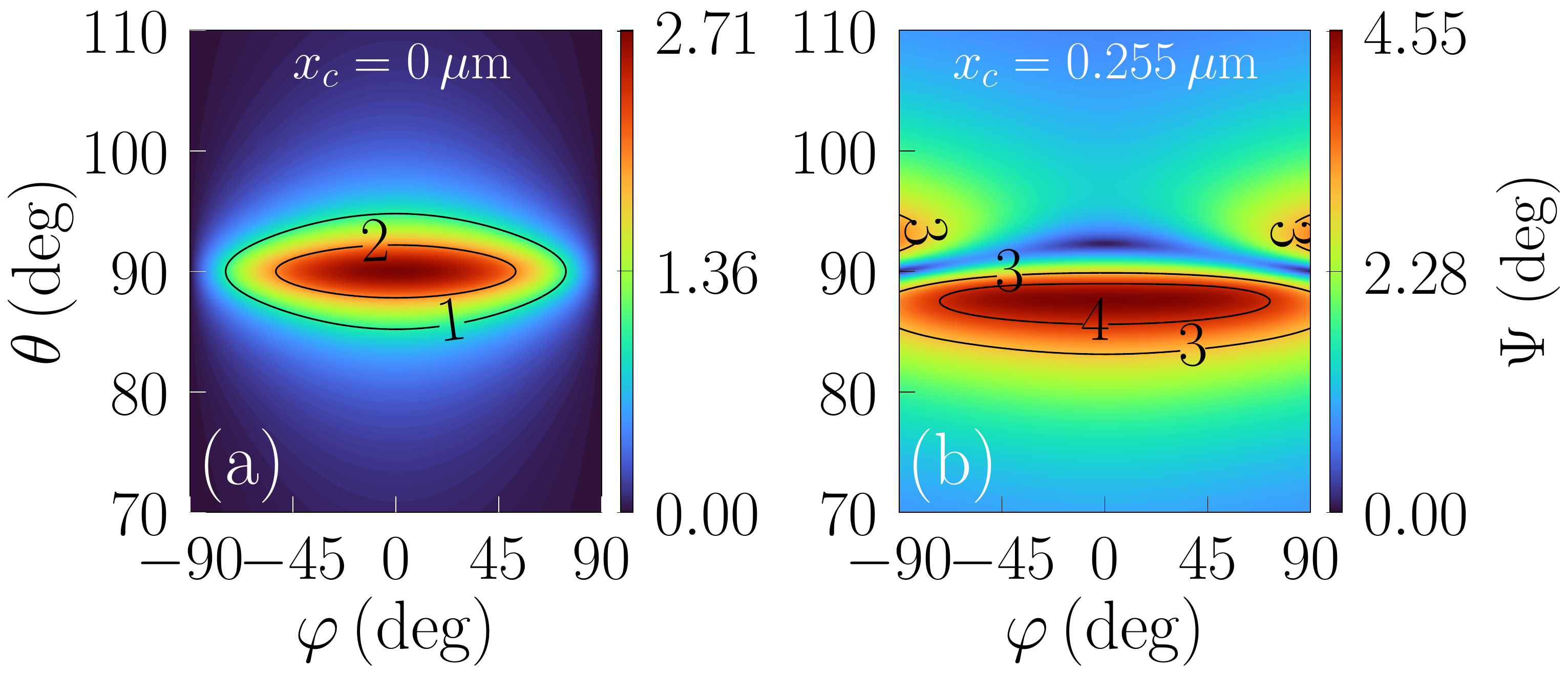}
\caption{Rotation angle $\Psi$ of the pseudo-spin precession axis when shuttling spins, as a function of the orientation of the magnetic field for a mesa with width $W=2\,\mu$m and depth $t=0.3\,\mu$m. (a) The left dot is at $x_1=-0.085\,\mu$m and the right dot at $x_2=0.085\,\mu$m. (b) Same for a left dot at $x_1=0.17\,\mu$m and a right dot at $x_2=0.34\,\mu$m.}
\label{fig:shuttle1}
\end{figure}

Besides the conventional EDSR discussed above, the spins can also be manipulated by shuttling the holes between dots with different $g$-matrices $g_1$ and $g_2$. The protocol for such manipulations is described in Ref.~\cite{Wang2024}: the hole in dot 1, whose pseudo-spin precesses around Larmor vector $\hbar\vec{\Omega}_1=\mu_B g_1\vec{B}$ at Larmor frequency $f_{\mathrm{L}1}=|\vec{\Omega}_1|/(2\pi)$, is shuttled to dot 2. The pseudo-spin, if preserved by shuttling, then starts to precess around Larmor vector $\hbar\vec{\Omega}_2=\mu_B g_2\vec{B}$ at Larmor frequency $f_{\mathrm{L}2}=|\vec{\Omega}_2|/(2\pi)$. After half a precession over time $\Delta t_2=1/(2f_{\mathrm{L}2})$, the hole is shuttled back to dot 1. This back and forth shuttling cycle can be repeated after an other half precession in dot 1 over time $\Delta t_1=1/(2f_{\mathrm{L}1})$. At the end of each cycle, the pseudo-spin in dot 1 has completed a rotation by an angle $2\Psi$ such that \cite{Wang2024}:
\begin{equation}
\cos\Psi=\frac{(g_1\vec{B})\cdot(g_2\vec{B})}{|g_1\vec{B}||g_2\vec{B}|}\,.
\end{equation}
Namely, $\Psi$ is the angle between the precession axes in dots 1 and 2 (measured in the Bloch spheres of the spin-orbit basis sets that are preserved by shuttling). This manipulation scheme calls for a precise timing of the shuttling events and is, therefore, more adapted to ``slow'' Larmor frequencies $f_\mathrm{L}<1$\,GHz. The average Rabi frequency is:
\begin{equation}
\bar{f}_\mathrm{R}=\frac{\Psi}{\pi}\bar{f}_\mathrm{L}
\end{equation}
with
\begin{equation}
2\bar{f}_\mathrm{L}^{-1}=f_{\mathrm{L}1}^{-1}+f_{\mathrm{L}2}^{-1}\,.
\end{equation}

\YMN{The relaxation of the mesa as well as the anisotropies of dots with different shapes may contribute to the mismatch between the $g$-matrices $g_1$ and $g_2$. We focus, however, on the effects of strains alone, and show that they are sufficient to manipulate the hole spins}. We consider two circular dots similar to Fig.~\ref{fig:device}, centered at positions $x_1$ and $x_2$ such that $x_2-x_1=0.17\,\mu$m. We compute the corrections $\delta g_{xx}^{(\varepsilon)}$, $\delta g_{yy}^{(\varepsilon)}$ and $\delta g_{zx}^{(\varepsilon)}$ for each dot using Eqs.~\eqref{eq:dgs} and \eqref{eq:gzx}, assuming again that the strains $\varepsilon_{ij}(x,z)\approx\varepsilon_{ij}(x_n,z=0)$ are quasi-homogeneous at the scale of dot $n=1,\,2$. We discard in a first approximation Rashba spin-orbit interactions \cite{Winkler03,Marcellina17,Terrazos21,Bosco21b,Abadillo2023}, and assume, therefore, that inter-dot tunneling is diagonal in the $\ket{\pm 3/2}$ basis sets of the $g$-matrices.
\begin{figure}[t]
\centering
\includegraphics[width=0.8\linewidth]{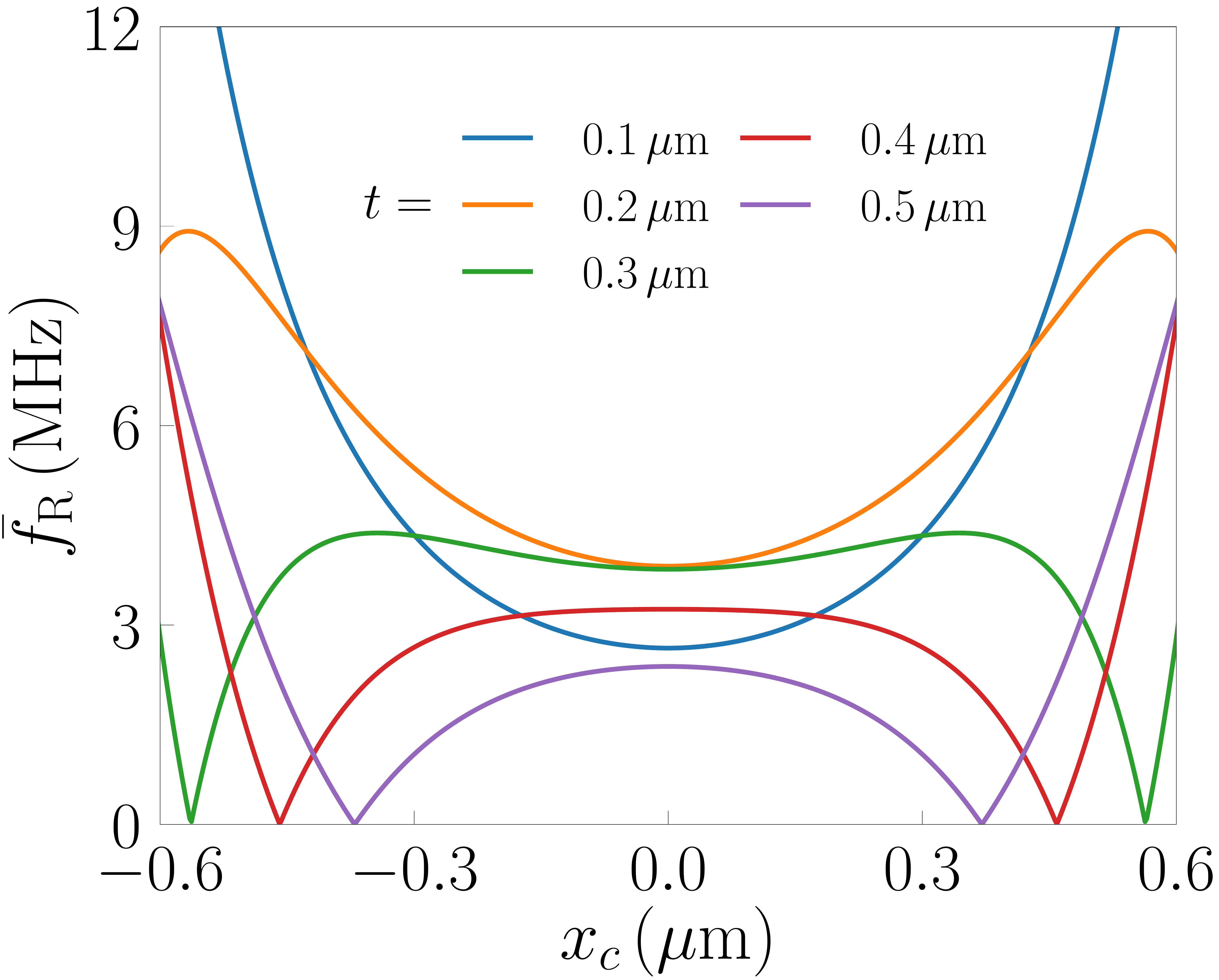}
\caption{Shuttling Rabi frequency $\bar{f}_\mathrm{R}$ as a function of the average position of the dots $x_c=(x_1+x_2)/2$ for mesas with width $W=2\,\mu$m and different depths $t$. The magnetic field is oriented along $x$ and the Larmor frequency of the dots at $x_c=0$ is $f_\mathrm{L}=250$\,MHz.}
\label{fig:shuttle2}
\end{figure}

We plot in Fig.~\ref{fig:shuttle1} the angle $\Psi$ as a function of the magnetic field orientation for two pairs of dots with different average positions $x_c=(x_1+x_2)/2$ in a $t=0.3\,\mu$m deep mesa. This angle peaks in the $xz$ plane ($\varphi=0$), at polar angle $\theta<90^\circ$ when $x_c>0$, $\theta>90^\circ$ when $x_c<0$, and $\theta=90^\circ$ when $x_c=0$. It consistently reaches $\ge 2.5^\circ$ in the equatorial plane as long as $-0.42<x_c<0.42\,\mu$m. $\Psi$ picks contributions from the differences $\Delta g_{xx}=g_{xx}(x_2)-g_{xx}(x_1)$ and $\Delta g_{zx}=g_{zx}(x_2)-g_{zx}(x_1)$ between the $g$-matrix elements of two dots, but is actually dominated by the latter. $\Delta g_{zx}$ results from the gradient of shear strains $\varepsilon_{xz}$ across the mesa [Eq.~\eqref{eq:gzx} and Fig.~\ref{fig:mesa}b]. Although this gradient is not sufficient to enable fast EDSR in single dots, it is the key mechanism in the present shuttling protocol owing to the much longer distance traveled by the hole. It gives rise to a rotation of the precession axis of the spin as explained in section \ref{sec:singledots}. The coupling between the in-plane and out-of-plane motion of the hole and the cool-down strains shall make negligible contributions to $\Delta g_{zx}$ as long as the dots are sufficiently similar (since the corrections are then about the same in both dots). 

Leaving aside $\Delta g_{xx}$, we can derive a simple analytical expression for $\Psi$ when $b_z=\cos\theta$ is small:
\begin{equation}
\Psi=\frac{|g_{xx}||\Delta g_{zx}|}{g^{2}_{xx}+(\bar{g}_{zx}+g_{zz}b_{z})^2}+{\cal O}\left(\Delta g_{zx}^2\right)\,, 
\label{eq:Psi}
\end{equation}
where $\bar{g}_{zx}=(g_{zx}(x_1)+g_{zx}(x_2))/2$ is the average $g_{zx}$ of the two dots \footnote{Eq.~\eqref{eq:Psi} is exact at $x_c=0$ (where $\Delta g_{xx}=0$ by symmetry). For arbitrary $\Delta g_{xx}$ and $\Delta g_{zx}$,
\begin{equation}
\Psi=\frac{|\bar{g}_{xx}\Delta g_{zx}-(\bar{g}_{zx}+g_{zz}b_{z})\Delta g_{xx}|}{\bar{g}_{xx}^2+(\bar{g}_{zx}+g_{zz}b_{z})^{2}}\,,
\end{equation}
where $\bar{g}_{xx}=(g_{xx}(x_1)+g_{xx}(x_2))/2$.}. This expression shows a single peak at $b_z=-\bar{g}_{zx}/g_{zz}$, with amplitude $\Psi_\mathrm{max}=|\Delta g_{zx}/g_{xx}|$ and FWHM $\delta\theta=2|g_{xx}|/g_{zz}$. As expected, $\Psi_\mathrm{max}$ is proportional to $\Delta g_{zx}$, the driving force for spin rotations, while the FWHM is proportional $|g_{xx}|/g_{zz}$: as explained in section \ref{sec:singledots}, rotations of the principal axes have a significant impact on the precession only when the magnetic field is near the equatorial plane so that any excursion around that plane is amplified by the large $g_{zz}\gg |g_{xx}|$. Increasing $|g_{xx}|$ thus decreases $\Psi_\mathrm{max}$, but broadens the peak, which enables Rabi oscillations over a larger range of magnetic field orientations at roughly constant quality factors (since the magnetic field strength gets smaller for given Larmor frequency). In a mesa, the finite $\Delta g_{zx}$ results from strain relaxation, and is not, therefore, independent from $g_{xx}$.

Finally, we plot in Figure~\ref{fig:shuttle2} the Rabi frequency $\bar{f}_\mathrm{R}$ as a function of the average position of the dots for different etch depths $t$. The magnetic field is oriented along $x$ and its amplitude is chosen (for each $t$) so that the Larmor frequency of the symmetric pair of dots at $x_c=0$ is $f_{\mathrm{L}1}=f_{\mathrm{L}2}=250$\,MHz. When $0.1<t<0.4\,\mu$m, the Rabi frequency is slowly variable around $x_c=0$ where $\varepsilon_{xx}$ and the gradient of $\varepsilon_{xz}$ are almost constant. The Rabi frequency then vanishes and bounces abruptly. This is due to the existence of maxima in $\varepsilon_{xz}$, thus to changes of sign of $\Delta g_{zx}$ (see Fig.~\ref{fig:gzxmesa}). The gradient of $\varepsilon_{xz}$ increases rapidly beyond these maxima, which speeds up the Rabi oscillations, even though a magnetic field along $x$ can not take full advantage of the large $\Delta g_{zx}$ as the peak of $\Psi$ is shifted off the equatorial plane. On the other hand, $\bar{f}_\mathrm{R}(x_c=0)$ decreases with increasing $t>0.2\,\mu$m because $g_{xx}(x_c=0)$ increases faster than $\Delta g_{zx}(x_c=0)$ when the mesa is made deeper. The Rabi frequencies are broadly optimal and uniform for $|x_c|<0.6\,\mu$m for depth $t\approx 0.25\,\mu$m. They are consistently above $4$\,MHz, which is about 60 times larger than in the single dots at $f_\mathrm{L}=250$\,MHz (BT drive with amplitude $V_\mathrm{ac}=1$\,mV at $t=0.25\,\mu$m and $x_0=0$). \footnote{Actually, the Rabi frequency $\bar{f}_\mathrm{R}$ resulting from Eq.~\eqref{eq:Psi} is formally the same as Eq. (3) of Ref.~\cite{martinez2022hole} with the substitution $\Delta g_{zx}\equiv \pi g_{zx}^\prime V_\mathrm{ac}/2$: the shuttling protocol takes advantage (with respect to single dot EDSR) of the much larger displacements of the hole, hence the much larger modulations of $g_{zx}$.}. Larger $\bar{f}_\mathrm{R}$ can even be achieved when the mesa is oriented along $x=[100]$ and $y=[010]$ thanks to a more favorable $\Delta g_{zx}/|g_{xx}|$ ratio (see Appendix \ref{appendix:100}).

\section{Conclusions}

In this work, we have addressed the effects of uniaxial strains on the gyromagnetic factors of holes in Ge/GeSi quantum dots. We have shown that moderate uniaxial strains can considerably enhance the in-plane $g$-factors $g_x^*$ and $g_y^*$ above unity (while $g_z^*$ remains almost constant). This opens the way for strain engineering in Ge/GeSi spin qubit devices. Uniaxial strains may be achieved in various ways, including the deposition of stressors on the heterostructure or the etching of an elongated mesa in a strained buffer. The buffer will indeed drive the relaxation of the whole heterostructure upon etching, which results in inhomogeneous uniaxial strains and $g$-factors across the mesa. The position-dependent enhancement of the in-plane $g$-factors has strong implications for electrical spin manipulation. This not only enables selective, Larmor-frequency dependent addressing of the spins; it also broadens Rabi hot spots and dephasing sweet lines on the unit sphere describing magnetic field orientation. The width of these features is indeed proportional to $|g_{x,y}^*/g_z^*|$ and can be as small as a few degrees in biaxial strains, which calls for a careful alignment of the magnetic field. It can be brought in the $20^\circ$ range in uniaxially strained mesas, easing the constraints on magnetic field alignment and the engineering of sweet spots. The systematic dot-to-dot variations of the $g$-factors induced by uniaxial strains can also be leveraged to drive spins while shuttling holes. Local uniaxial strains resulting from etching operations or ohmic contacts formation may actually be at work in many existing devices and partly explain the wide spread of measured in-plane $g$-factors.

\section*{Acknowledgements}

We thank Romain Maurand for fruitful discussions. This work is supported by the French National Research Agency under the programme ``France 2030'' (PEPR PRESQUILE - ANR-22-PETQ-0002) and by the European Union's Horizon 2020 research and innovation program (grant agreement 951852 QLSI).

\appendix

\section{Uniaxial strains along $[100]$ and $[010]$}
\label{appendix:100}

\begin{figure}[t]
\centering
\includegraphics[width=1.0\linewidth]{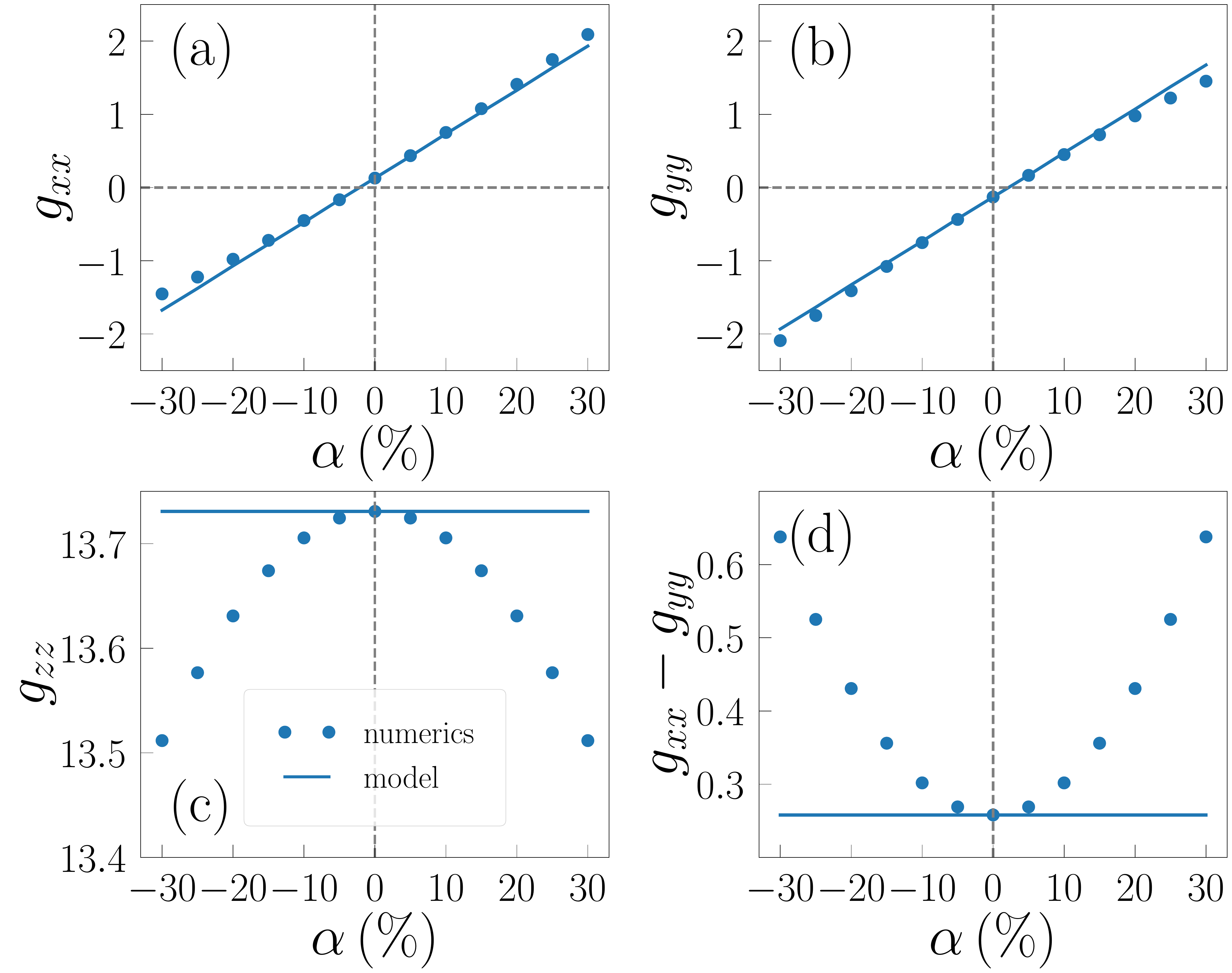}
\caption{The diagonal $g$-matrix elements $g_{ii}(\alpha)$ as a function of the strength $\alpha$ of uniaxial strains applied along $x=[100]$ and $y=[010]$, at $V_\mathrm{C}=-40$\,mV. The results from the full numerical calculations (dots) are compared to the analytical model (solid lines), Eq.~\eqref{eq:dgs100} \cite{martinez2022hole,Abadillo2023}.}
\label{fig:gdelta100}
\end{figure}

We plot in Figure \ref{fig:gdelta100} the diagonal $g$-matrix elements $g_{ii}(\alpha)$ as a function of the strength $\alpha$ of uniaxial strains applied along $x=[100]$ and $y=[010]$ (instead of $x=[110]$ and $y=[\bar{1}10]$ in the main text). The trends are similar to Fig.~\ref{fig:gdelta}, but the $g$-matrix corrections are smaller as $|d_v|>\sqrt{3}|b_v|$. 

\begin{figure}[t]
\centering
\includegraphics[width=0.8\linewidth]{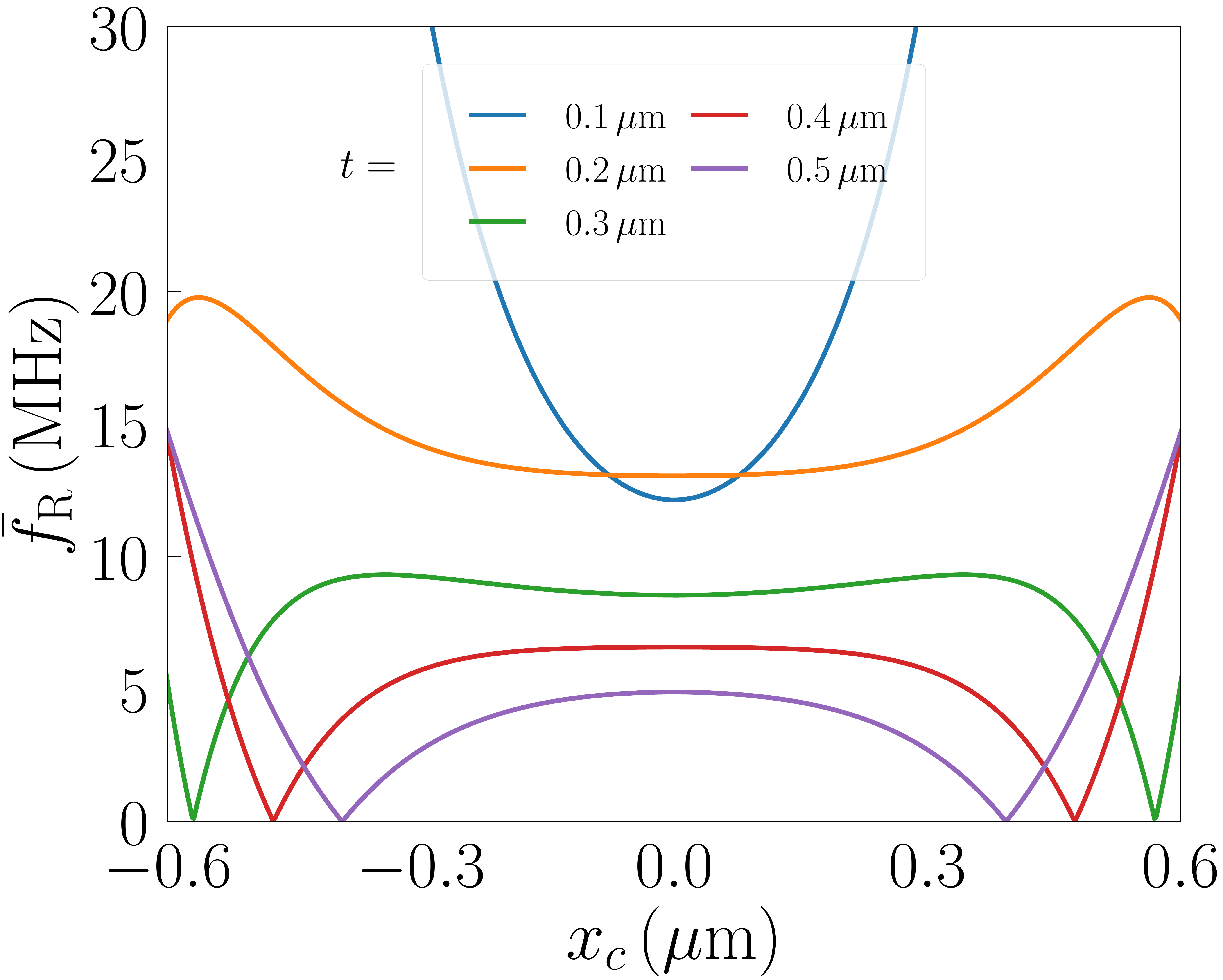}
\caption{Shuttling Rabi frequency $\bar{f}_\mathrm{R}$ as a function of the average position of the dots $x_c=(x_1+x_2)/2$ for mesas with width $W=2\,\mu$m and different depths $t$. The uniaxial strains are applied along $x=[100]$ and $y=[010]$. The magnetic field is oriented along $x$ and the Larmor frequency of the dots at $x_c=0$ is $f_\mathrm{L}=250$\,MHz.}
\label{fig:shuttle2app}
\end{figure}

The shuttling Rabi frequency ${\bar f}_\mathrm{R}$ computed in mesas oriented along $[100]$ and $[010]$ is plotted as a function of the average position of the dots for different etch depths $t$ in Fig.~\ref{fig:shuttle2app} ($W=2\,\mu$m). The magnetic field is oriented along $x$ and its amplitude is chosen (for each $t$) so that the Larmor frequency of the symmetric pair of dots at $x_c=0$ is $f_{\mathrm{L}1}=f_{\mathrm{L}2}=250$\,MHz. The data are, again, qualitatively similar to Fig.~\ref{fig:shuttle2}, but the Rabi frequencies are larger due to a move favorable $\Delta g_{zx}/|g_{xx}|$ ratio (smaller $|g_{xx}|$).



\section{Dependence of the $g$-factors on the mesa width $W$ and depth $t$}
\label{appendix:tW}

\begin{figure}[t]
\centering
\includegraphics[width=0.75\linewidth]{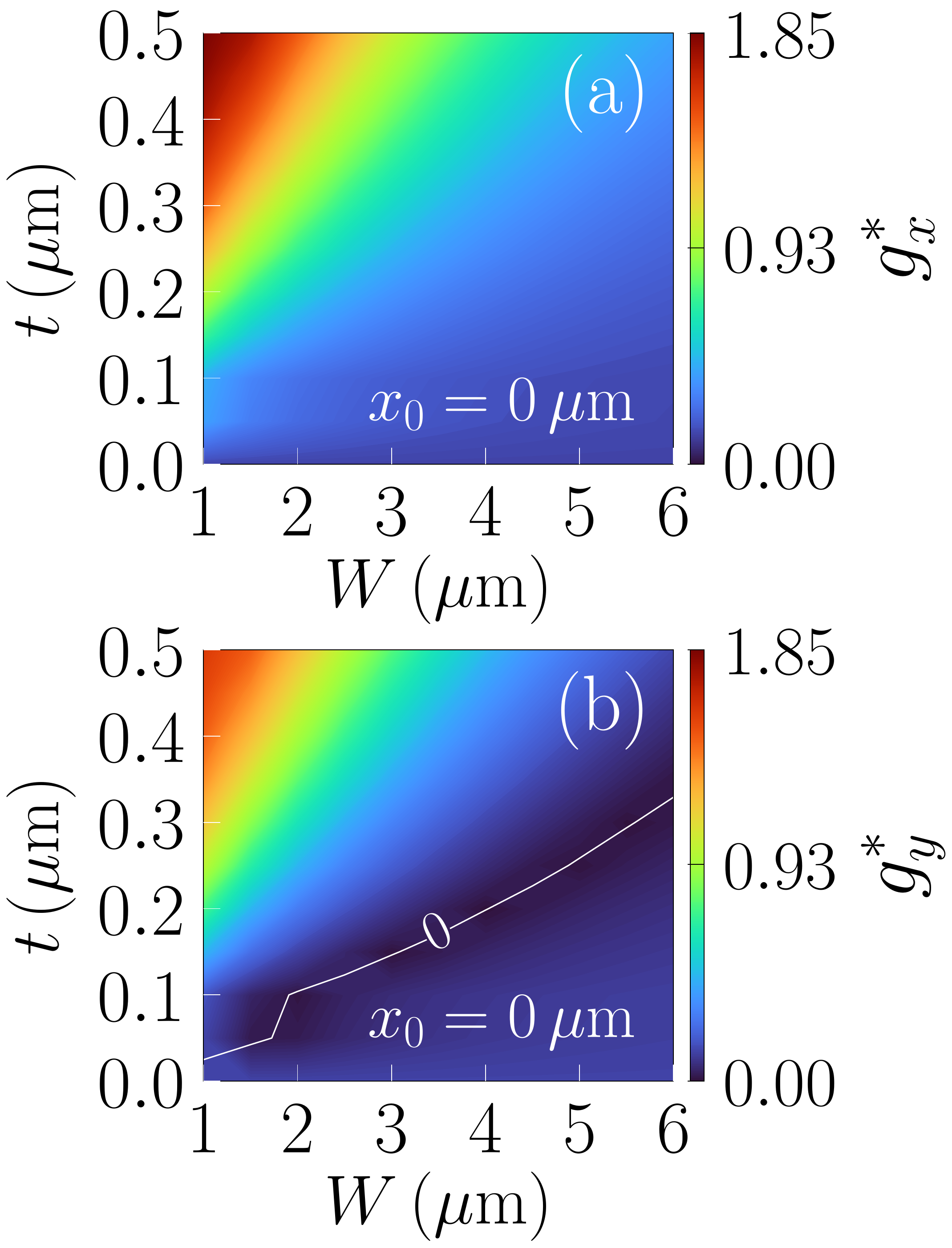}
\caption{The effective in-plane $g$-factors $g_x^*(x_0=0)$ and $g_y^*(x_0=0)$ as a function of the width $W$ and depth $t$ of the mesa, computed from Eqs.~\eqref{eq:gHH}, \eqref{eq:dgs} and \eqref{eq:deltalh}.}
\label{fig:gstar_w}
\end{figure}

In this appendix, we discuss the dependence of the $g$-factors on the mesa width $W$ and depth $t$ in order to provide guidelines for the design of the devices.

\begin{figure}[t]
\centering
\includegraphics[width=0.8\linewidth]{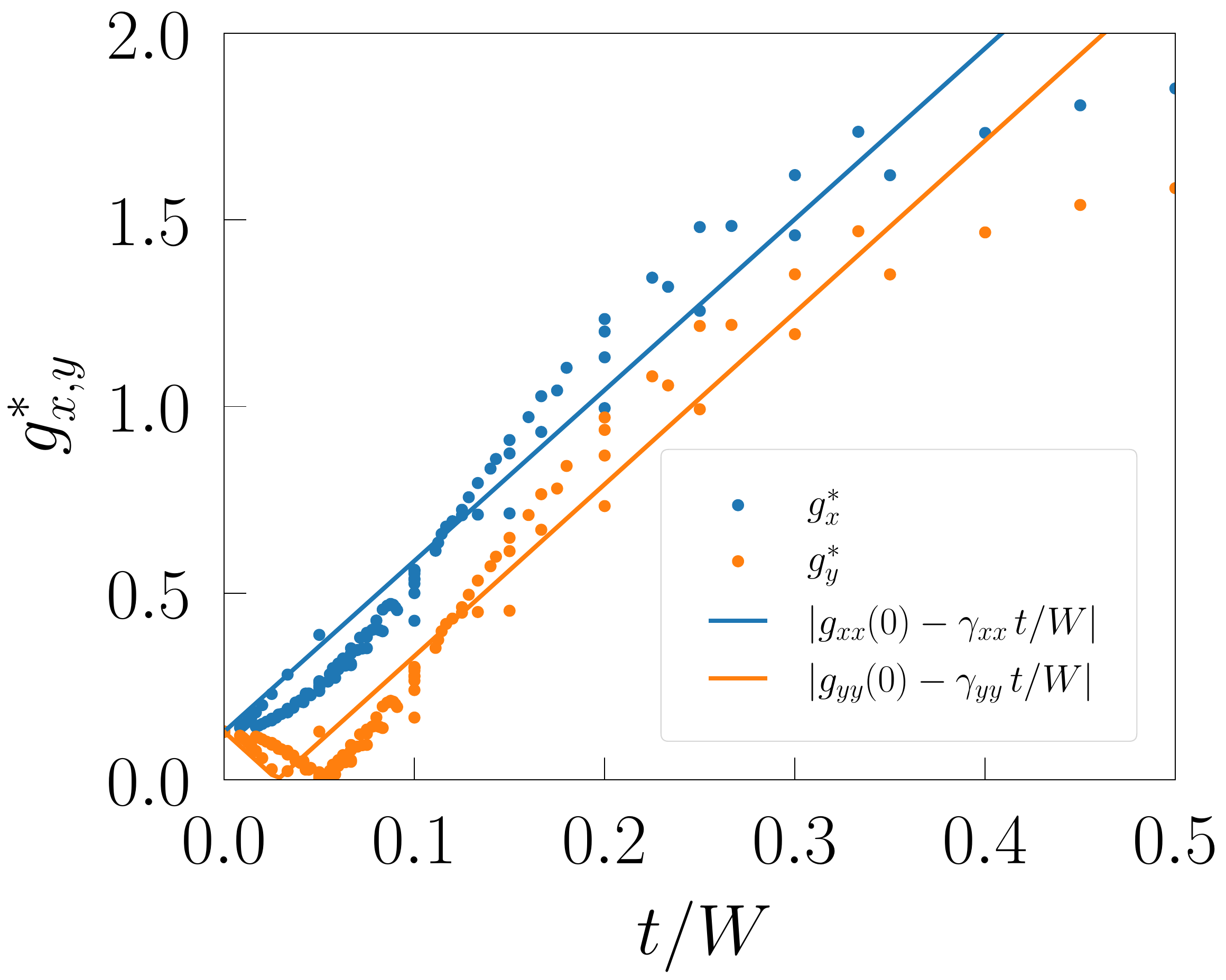}
\caption{The effective in-plane $g$-factors $g_x^*(x_0=0)$ and $g_y^*(x_0=0)$ as a function of the ratio $t/W$ of the mesas (oriented along $x=[110]$ and $y=[\bar{1}10]$). They are fitted to $g_i^*=|g_{ii}(0)-\gamma_{ii}t/W|$ where $\gamma_{ii}=4.6$ for both $g_x^*$ and $g_y^*$.}
\label{fig:gstar_tw}
\end{figure}

We plot in Fig.~\ref{fig:gstar_w} the effective in-plane $g$-factors $g_x^*(x_0=0)$ and $g_y^*(x_0=0)$ as a function of the width $W$ and depth $t$ of the mesa. They are computed at the center of the Ge well using Eqs.~\eqref{eq:gHH}, \eqref{eq:dgs} and \eqref{eq:deltalh}. We assume diagonal $g$-matrix elements $g_{xx}=-3q+\delta g_{xx}^{(c)}=-0.13$ and $g_{yy}=3q+\delta g_{xx}^{(c)}=0.13$ in biaxial strains. On the one hand, the effective $g$-factor $g_x^*=|g_{xx}|$ increases monotonically when $W$ decreases or $t$ increases as the mesa relaxes further. On the other hand, $g_y^*=|g_{yy}|$ shows a line of zeroes in the $(W,t)$ plane where the uniaxial correction $\delta g_{yy}^{(\varepsilon)}$ compensates the biaxial reference. Actually, the effective $g$-factors at the center of the well are primarily functions of $t/W$ over a wide range of mesa widths and depths, as shown in Fig.~\ref{fig:gstar_tw}. They can be approximately fitted to
\begin{equation}
g_i^*\left(\frac{t}{W}\right)\approx\left|g_{ii}(0)-\gamma_{ii}\frac{t}{W}\right|\,,
\end{equation}
where $g_{xx}(0)=-0.13$, $g_{yy}(0)=0.13$ and $\gamma_{xx}\approx\gamma_{yy}=4.6$. 

\section{Impact of finite slopes on mesa relaxation}
\label{appendix:FiniteSlope}

\begin{figure}[t]
\centering
\includegraphics[width=1\linewidth]{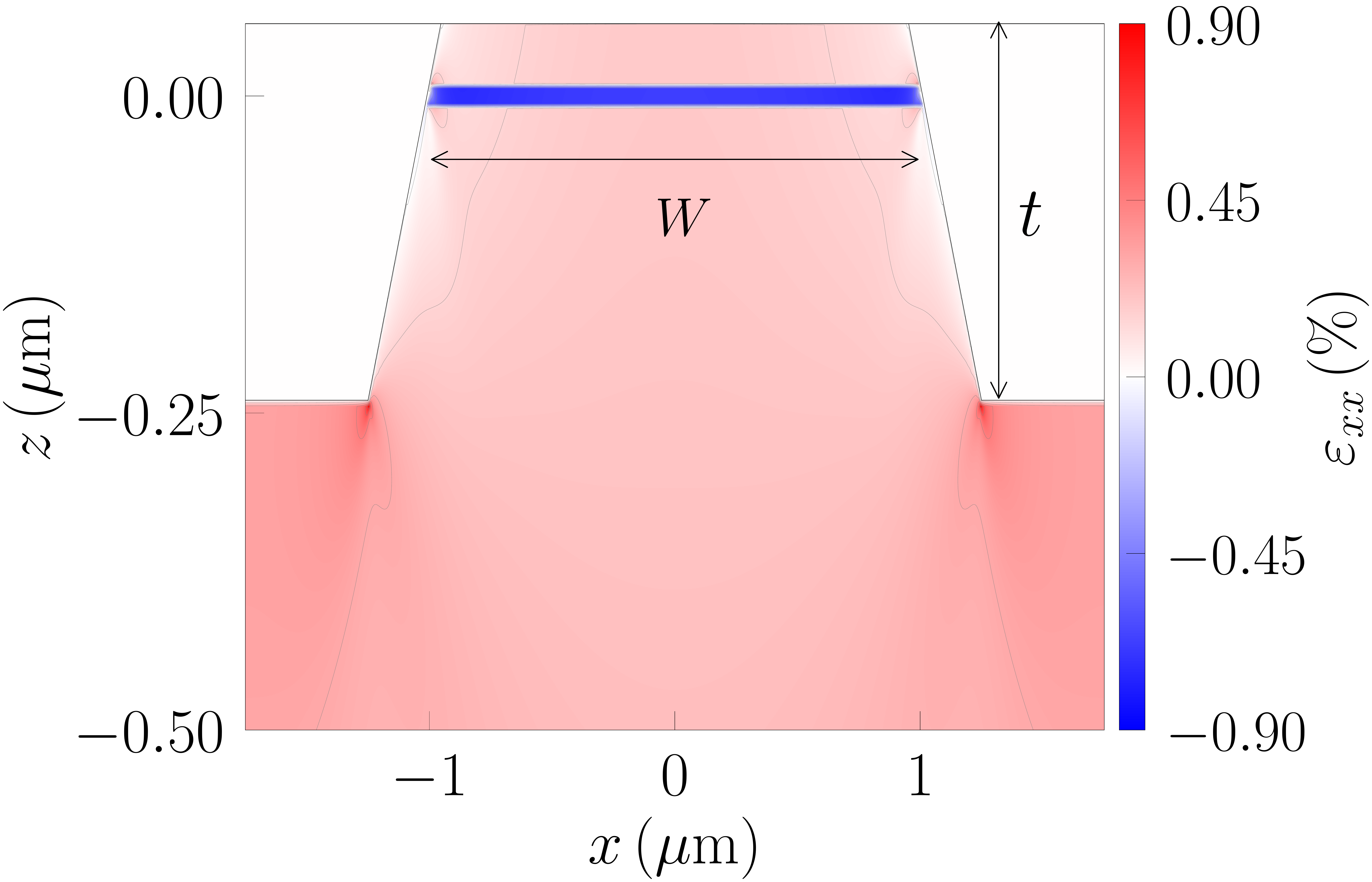}
\caption{Map of the strain $\varepsilon_{xx}$ in the cross section of a tapered mesa with width $W=2\,\mu$m (measured in the Ge well) and depth $t=0.3\,\mu$m. The angle between the side walls and the vertical is $\beta=45^\circ$. The residual strain in the GeSi buffer (before etching) is $\varepsilon_{xx}=\varepsilon_{yy}=\varepsilon_\mathrm{buf}=0.26\%$.}
\label{fig:mesaFiniteSlope}
\end{figure}

\addcomm{We consider in Fig.~\ref{fig:mesaFiniteSlope} a tapered mesa whose side walls make an angle $\beta=45^\circ$ angle with the vertical. This is more representative of experimental structures where metal (bias) lines may need to climb these side walls. Although we did not consider these metal lines in our calculations, we emphasize that they shall, in principle, have little impact on the relaxation of the mesa given their small volume. Ref. \cite{Abadillo2023} actually shows that the strains induced by the metal gates in the heterostructure are small; although these strains can significantly enhance the Rabi frequencies, they have hardly any effect on the $g$-factors.}

\begin{figure}[t]
\centering
\includegraphics[width=0.8\linewidth]{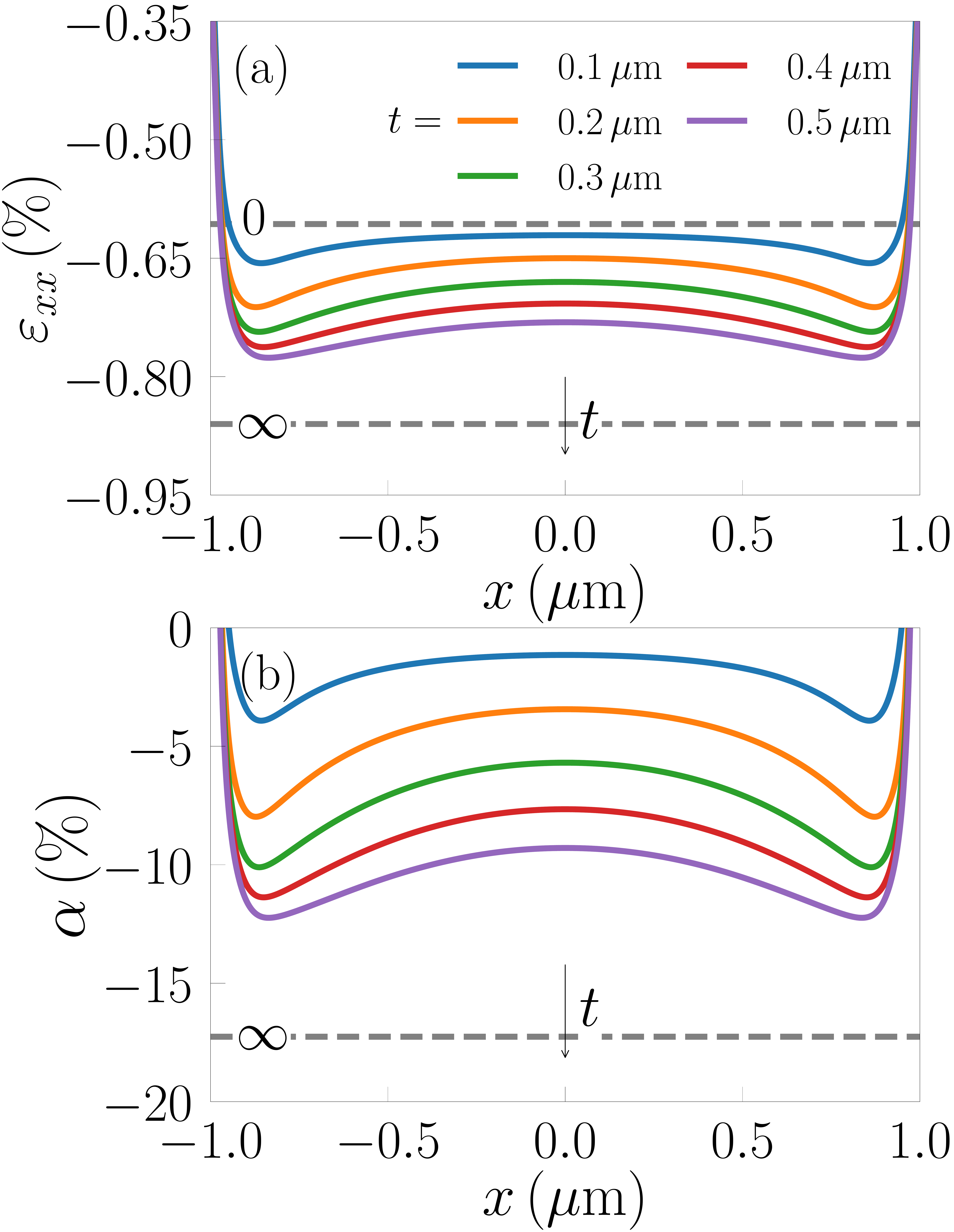}
\caption{(a) Strain $\varepsilon_{xx}(x)$ as a function of the lateral position $x$ along the middle of the Ge well $(z=0)$ for tapered mesas with width $W=2\,\mu$m and depth $t$ varying from $0.1$ to $0.5\,\mu$m (side walls angle $\beta=45^\circ$). (b) The corresponding uniaxial strengths $\alpha(x)$. In both panels, the horizontal dashed line are the same $t\to 0$ and $t\to\infty$ limits as in Fig.~\ref{fig:depth}.}
\label{fig:depthFiniteSlope}
\end{figure}

\begin{figure}[t]
\centering
\includegraphics[width=1.0\linewidth]{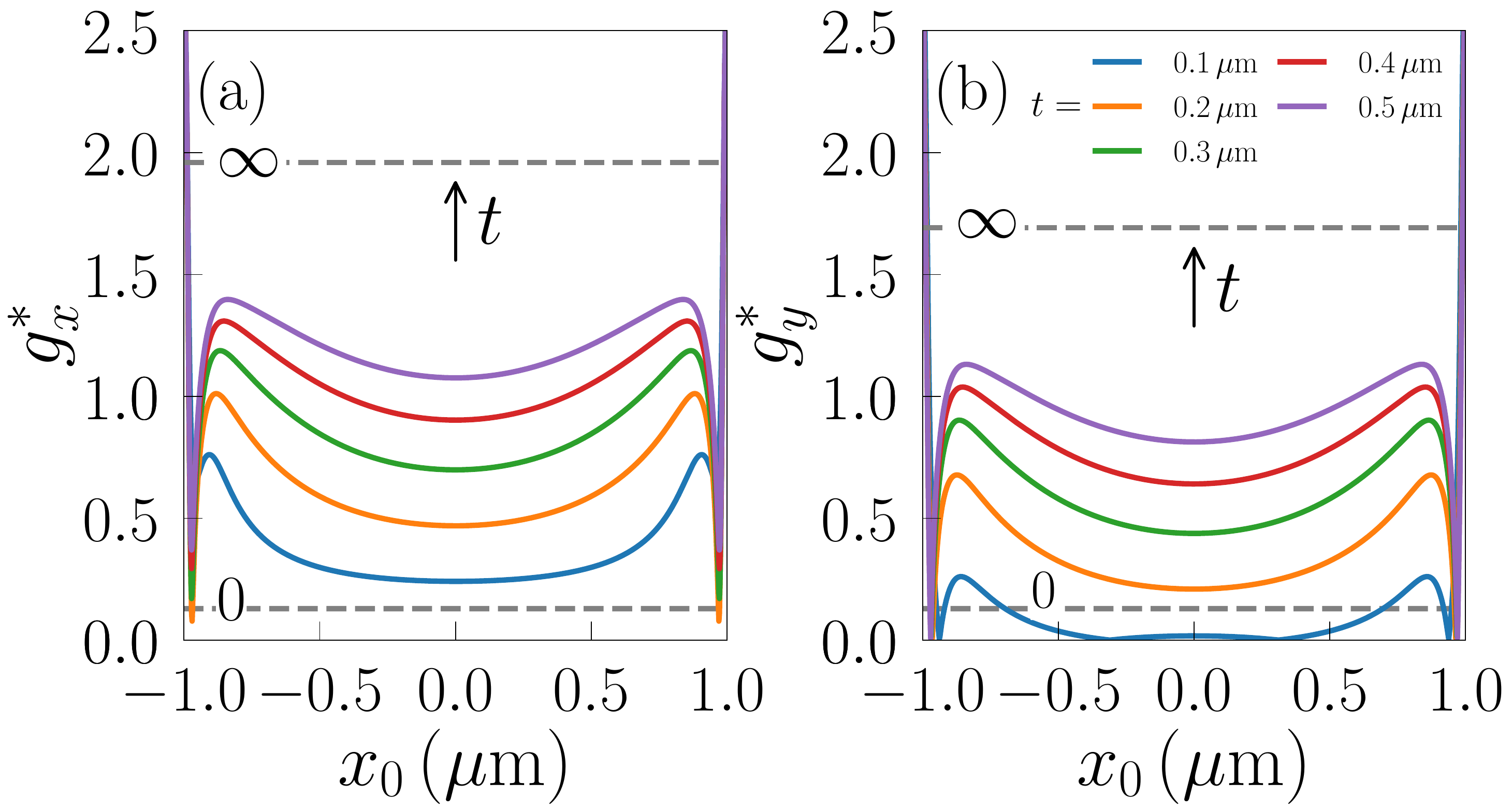}
\caption{The effective in-plane $g$-factors $g_x^*=\sqrt{g_{xx}^2+g_{zx}^2}$ and $g_y^*=|g_{yy}|$ as a function of the lateral position $x_0$ of the dot, for mesas with width $W=2\,\mu$m and depth $t$ from $0.1$ to $0.5\,\mu$m (side walls angle $\beta=45^\circ$). They are computed from Eqs.~\eqref{eq:gHH} and~\eqref{eq:deltalh}. In both panels, the horizontal dashed line are the same $t\to 0$ and $t\to\infty$ limits as in Fig.~\ref{fig:gmesa}}
\label{fig:gmesaFiniteSlope}
\end{figure}

\addcomm{The Ge well of Fig.~\ref{fig:mesaFiniteSlope} has the same width $W=2\,\mu$m as in the main text. The strain $\varepsilon_{xx}(x)$ and the uniaxial strength $\alpha(x)$ in the well are plotted in Fig.~\ref{fig:depthFiniteSlope}, and the effective in-plane $g$-factors $g_x^*$ and $g_y^*$ are plotted in Fig.~\ref{fig:gmesaFiniteSlope}. These figures can be compared to Fig.~\ref{fig:depth} and Fig.~\ref{fig:gmesa} calculated for vertical side walls. The relaxation, uniaxial strength and maximal $g$-factors are smaller for tapered mesas, because the average width of the buffer (which drives the relaxation) is larger. Nonetheless, $g_x^*$ can still be greater than 1 for depths $t\gtrsim 0.4\mu$m. Significant enhancements of the $g$-factors are, therefore, also expected for tapered mesas.}

\FloatBarrier

\bibliography{biblio}

\begin{thebibliography}{65}%
\makeatletter
\providecommand \@ifxundefined [1]{%
 \@ifx{#1\undefined}
}%
\providecommand \@ifnum [1]{%
 \ifnum #1\expandafter \@firstoftwo
 \else \expandafter \@secondoftwo
 \fi
}%
\providecommand \@ifx [1]{%
 \ifx #1\expandafter \@firstoftwo
 \else \expandafter \@secondoftwo
 \fi
}%
\providecommand \natexlab [1]{#1}%
\providecommand \enquote  [1]{``#1''}%
\providecommand \bibnamefont  [1]{#1}%
\providecommand \bibfnamefont [1]{#1}%
\providecommand \citenamefont [1]{#1}%
\providecommand \href@noop [0]{\@secondoftwo}%
\providecommand \href [0]{\begingroup \@sanitize@url \@href}%
\providecommand \@href[1]{\@@startlink{#1}\@@href}%
\providecommand \@@href[1]{\endgroup#1\@@endlink}%
\providecommand \@sanitize@url [0]{\catcode `\\12\catcode `\$12\catcode `\&12\catcode `\#12\catcode `\^12\catcode `\_12\catcode `\%12\relax}%
\providecommand \@@startlink[1]{}%
\providecommand \@@endlink[0]{}%
\providecommand \url  [0]{\begingroup\@sanitize@url \@url }%
\providecommand \@url [1]{\endgroup\@href {#1}{\urlprefix }}%
\providecommand \urlprefix  [0]{URL }%
\providecommand \Eprint [0]{\href }%
\providecommand \doibase [0]{https://doi.org/}%
\providecommand \selectlanguage [0]{\@gobble}%
\providecommand \bibinfo  [0]{\@secondoftwo}%
\providecommand \bibfield  [0]{\@secondoftwo}%
\providecommand \translation [1]{[#1]}%
\providecommand \BibitemOpen [0]{}%
\providecommand \bibitemStop [0]{}%
\providecommand \bibitemNoStop [0]{.\EOS\space}%
\providecommand \EOS [0]{\spacefactor3000\relax}%
\providecommand \BibitemShut  [1]{\csname bibitem#1\endcsname}%
\let\auto@bib@innerbib\@empty
\bibitem [{\citenamefont {Loss}\ and\ \citenamefont {DiVincenzo}(1998)}]{Loss98}%
  \BibitemOpen
  \bibfield  {author} {\bibinfo {author} {\bibfnamefont {D.}~\bibnamefont {Loss}}\ and\ \bibinfo {author} {\bibfnamefont {D.~P.}\ \bibnamefont {DiVincenzo}},\ }\bibfield  {title} {\bibinfo {title} {{Quantum computation with quantum dots}},\ }\href {https://doi.org/10.1103/PhysRevA.57.120} {\bibfield  {journal} {\bibinfo  {journal} {Physical Review A}\ }\textbf {\bibinfo {volume} {57}},\ \bibinfo {pages} {120} (\bibinfo {year} {1998})}\BibitemShut {NoStop}%
\bibitem [{\citenamefont {Burkard}\ \emph {et~al.}(2023)\citenamefont {Burkard}, \citenamefont {Ladd}, \citenamefont {Pan}, \citenamefont {Nichol},\ and\ \citenamefont {Petta}}]{Burkard2023Review}%
  \BibitemOpen
  \bibfield  {author} {\bibinfo {author} {\bibfnamefont {G.}~\bibnamefont {Burkard}}, \bibinfo {author} {\bibfnamefont {T.~D.}\ \bibnamefont {Ladd}}, \bibinfo {author} {\bibfnamefont {A.}~\bibnamefont {Pan}}, \bibinfo {author} {\bibfnamefont {J.~M.}\ \bibnamefont {Nichol}},\ and\ \bibinfo {author} {\bibfnamefont {J.~R.}\ \bibnamefont {Petta}},\ }\bibfield  {title} {\bibinfo {title} {Semiconductor spin qubits},\ }\href {https://doi.org/10.1103/RevModPhys.95.025003} {\bibfield  {journal} {\bibinfo  {journal} {Review of Modern Physics}\ }\textbf {\bibinfo {volume} {95}},\ \bibinfo {pages} {025003} (\bibinfo {year} {2023})}\BibitemShut {NoStop}%
\bibitem [{\citenamefont {Fang}\ \emph {et~al.}(2023)\citenamefont {Fang}, \citenamefont {Philippopoulos}, \citenamefont {Culcer}, \citenamefont {Coish},\ and\ \citenamefont {Chesi}}]{Fang2023Review}%
  \BibitemOpen
  \bibfield  {author} {\bibinfo {author} {\bibfnamefont {Y.}~\bibnamefont {Fang}}, \bibinfo {author} {\bibfnamefont {P.}~\bibnamefont {Philippopoulos}}, \bibinfo {author} {\bibfnamefont {D.}~\bibnamefont {Culcer}}, \bibinfo {author} {\bibfnamefont {W.~A.}\ \bibnamefont {Coish}},\ and\ \bibinfo {author} {\bibfnamefont {S.}~\bibnamefont {Chesi}},\ }\bibfield  {title} {\bibinfo {title} {Recent advances in hole-spin qubits},\ }\href {https://doi.org/10.1088/2633-4356/acb87e} {\bibfield  {journal} {\bibinfo  {journal} {Materials for Quantum Technology}\ }\textbf {\bibinfo {volume} {3}},\ \bibinfo {pages} {012003} (\bibinfo {year} {2023})}\BibitemShut {NoStop}%
\bibitem [{\citenamefont {Sammak}\ \emph {et~al.}(2019)\citenamefont {Sammak}, \citenamefont {Sabbagh}, \citenamefont {Hendrickx}, \citenamefont {Lodari}, \citenamefont {Paquelet~Wuetz}, \citenamefont {Tosato}, \citenamefont {Yeoh}, \citenamefont {Bollani}, \citenamefont {Virgilio}, \citenamefont {Schubert}, \citenamefont {Zaumseil}, \citenamefont {Capellini}, \citenamefont {Veldhorst},\ and\ \citenamefont {Scappucci}}]{Sammak19}%
  \BibitemOpen
  \bibfield  {author} {\bibinfo {author} {\bibfnamefont {A.}~\bibnamefont {Sammak}}, \bibinfo {author} {\bibfnamefont {D.}~\bibnamefont {Sabbagh}}, \bibinfo {author} {\bibfnamefont {N.~W.}\ \bibnamefont {Hendrickx}}, \bibinfo {author} {\bibfnamefont {M.}~\bibnamefont {Lodari}}, \bibinfo {author} {\bibfnamefont {B.}~\bibnamefont {Paquelet~Wuetz}}, \bibinfo {author} {\bibfnamefont {A.}~\bibnamefont {Tosato}}, \bibinfo {author} {\bibfnamefont {L.}~\bibnamefont {Yeoh}}, \bibinfo {author} {\bibfnamefont {M.}~\bibnamefont {Bollani}}, \bibinfo {author} {\bibfnamefont {M.}~\bibnamefont {Virgilio}}, \bibinfo {author} {\bibfnamefont {M.~A.}\ \bibnamefont {Schubert}}, \bibinfo {author} {\bibfnamefont {P.}~\bibnamefont {Zaumseil}}, \bibinfo {author} {\bibfnamefont {G.}~\bibnamefont {Capellini}}, \bibinfo {author} {\bibfnamefont {M.}~\bibnamefont {Veldhorst}},\ and\ \bibinfo {author} {\bibfnamefont {G.}~\bibnamefont {Scappucci}},\ }\bibfield  {title} {\bibinfo {title} {Shallow and undoped germanium quantum wells: A
  playground for spin and hybrid quantum technology},\ }\href {https://doi.org/10.1002/adfm.201807613} {\bibfield  {journal} {\bibinfo  {journal} {Advanced Functional Materials}\ }\textbf {\bibinfo {volume} {29}},\ \bibinfo {pages} {1807613} (\bibinfo {year} {2019})}\BibitemShut {NoStop}%
\bibitem [{\citenamefont {Scappucci}\ \emph {et~al.}(2021)\citenamefont {Scappucci}, \citenamefont {Kloeffel}, \citenamefont {Zwanenburg}, \citenamefont {Loss}, \citenamefont {Myronov}, \citenamefont {Zhang}, \citenamefont {De~Franceschi}, \citenamefont {Katsaros},\ and\ \citenamefont {Veldhorst}}]{Scappucci20}%
  \BibitemOpen
  \bibfield  {author} {\bibinfo {author} {\bibfnamefont {G.}~\bibnamefont {Scappucci}}, \bibinfo {author} {\bibfnamefont {C.}~\bibnamefont {Kloeffel}}, \bibinfo {author} {\bibfnamefont {F.~A.}\ \bibnamefont {Zwanenburg}}, \bibinfo {author} {\bibfnamefont {D.}~\bibnamefont {Loss}}, \bibinfo {author} {\bibfnamefont {M.}~\bibnamefont {Myronov}}, \bibinfo {author} {\bibfnamefont {J.-J.}\ \bibnamefont {Zhang}}, \bibinfo {author} {\bibfnamefont {S.}~\bibnamefont {De~Franceschi}}, \bibinfo {author} {\bibfnamefont {G.}~\bibnamefont {Katsaros}},\ and\ \bibinfo {author} {\bibfnamefont {M.}~\bibnamefont {Veldhorst}},\ }\bibfield  {title} {\bibinfo {title} {The germanium quantum information route},\ }\href {https://doi.org/10.1038/s41578-020-00262-z} {\bibfield  {journal} {\bibinfo  {journal} {Nature Reviews Materials}\ ,\ \bibinfo {pages} {926}} (\bibinfo {year} {2021})}\BibitemShut {NoStop}%
\bibitem [{\citenamefont {Ares}\ \emph {et~al.}(2013)\citenamefont {Ares}, \citenamefont {Golovach}, \citenamefont {Katsaros}, \citenamefont {Stoffel}, \citenamefont {Fournel}, \citenamefont {Glazman}, \citenamefont {Schmidt},\ and\ \citenamefont {De~Franceschi}}]{Ares13}%
  \BibitemOpen
  \bibfield  {author} {\bibinfo {author} {\bibfnamefont {N.}~\bibnamefont {Ares}}, \bibinfo {author} {\bibfnamefont {V.~N.}\ \bibnamefont {Golovach}}, \bibinfo {author} {\bibfnamefont {G.}~\bibnamefont {Katsaros}}, \bibinfo {author} {\bibfnamefont {M.}~\bibnamefont {Stoffel}}, \bibinfo {author} {\bibfnamefont {F.}~\bibnamefont {Fournel}}, \bibinfo {author} {\bibfnamefont {L.~I.}\ \bibnamefont {Glazman}}, \bibinfo {author} {\bibfnamefont {O.~G.}\ \bibnamefont {Schmidt}},\ and\ \bibinfo {author} {\bibfnamefont {S.}~\bibnamefont {De~Franceschi}},\ }\bibfield  {title} {\bibinfo {title} {Nature of tunable hole $g$ factors in quantum dots},\ }\href {https://doi.org/10.1103/PhysRevLett.110.046602} {\bibfield  {journal} {\bibinfo  {journal} {Physical Review Letters}\ }\textbf {\bibinfo {volume} {110}},\ \bibinfo {pages} {046602} (\bibinfo {year} {2013})}\BibitemShut {NoStop}%
\bibitem [{\citenamefont {Watzinger}\ \emph {et~al.}(2018)\citenamefont {Watzinger}, \citenamefont {Kuku\v{c}ka}, \citenamefont {Vuku\v{s}i\'c}, \citenamefont {Gao}, \citenamefont {Wang}, \citenamefont {Sch\"affler}, \citenamefont {Zhang},\ and\ \citenamefont {Katsaros}}]{Watzinger18}%
  \BibitemOpen
  \bibfield  {author} {\bibinfo {author} {\bibfnamefont {H.}~\bibnamefont {Watzinger}}, \bibinfo {author} {\bibfnamefont {J.}~\bibnamefont {Kuku\v{c}ka}}, \bibinfo {author} {\bibfnamefont {L.}~\bibnamefont {Vuku\v{s}i\'c}}, \bibinfo {author} {\bibfnamefont {F.}~\bibnamefont {Gao}}, \bibinfo {author} {\bibfnamefont {T.}~\bibnamefont {Wang}}, \bibinfo {author} {\bibfnamefont {F.}~\bibnamefont {Sch\"affler}}, \bibinfo {author} {\bibfnamefont {J.-J.}\ \bibnamefont {Zhang}},\ and\ \bibinfo {author} {\bibfnamefont {G.}~\bibnamefont {Katsaros}},\ }\bibfield  {title} {\bibinfo {title} {A germanium hole spin qubit},\ }\href {https://doi.org/10.1038/s41467-018-06418-4} {\bibfield  {journal} {\bibinfo  {journal} {Nature Communications}\ }\textbf {\bibinfo {volume} {9}},\ \bibinfo {pages} {3902} (\bibinfo {year} {2018})}\BibitemShut {NoStop}%
\bibitem [{\citenamefont {Hendrickx}\ \emph {et~al.}(2020{\natexlab{a}})\citenamefont {Hendrickx}, \citenamefont {Lawrie}, \citenamefont {Petit}, \citenamefont {Sammak}, \citenamefont {Scappucci},\ and\ \citenamefont {Veldhorst}}]{Hendrickx20b}%
  \BibitemOpen
  \bibfield  {author} {\bibinfo {author} {\bibfnamefont {N.~W.}\ \bibnamefont {Hendrickx}}, \bibinfo {author} {\bibfnamefont {W.~I.~L.}\ \bibnamefont {Lawrie}}, \bibinfo {author} {\bibfnamefont {L.}~\bibnamefont {Petit}}, \bibinfo {author} {\bibfnamefont {A.}~\bibnamefont {Sammak}}, \bibinfo {author} {\bibfnamefont {G.}~\bibnamefont {Scappucci}},\ and\ \bibinfo {author} {\bibfnamefont {M.}~\bibnamefont {Veldhorst}},\ }\bibfield  {title} {\bibinfo {title} {A single-hole spin qubit},\ }\href {https://doi.org/10.1038/s41467-020-17211-7} {\bibfield  {journal} {\bibinfo  {journal} {Nature Communications}\ }\textbf {\bibinfo {volume} {11}},\ \bibinfo {pages} {3478} (\bibinfo {year} {2020}{\natexlab{a}})}\BibitemShut {NoStop}%
\bibitem [{\citenamefont {Hendrickx}\ \emph {et~al.}(2020{\natexlab{b}})\citenamefont {Hendrickx}, \citenamefont {Franke}, \citenamefont {Sammak}, \citenamefont {Scappucci},\ and\ \citenamefont {Veldhorst}}]{Hendrickx20}%
  \BibitemOpen
  \bibfield  {author} {\bibinfo {author} {\bibfnamefont {N.~W.}\ \bibnamefont {Hendrickx}}, \bibinfo {author} {\bibfnamefont {D.~P.}\ \bibnamefont {Franke}}, \bibinfo {author} {\bibfnamefont {A.}~\bibnamefont {Sammak}}, \bibinfo {author} {\bibfnamefont {G.}~\bibnamefont {Scappucci}},\ and\ \bibinfo {author} {\bibfnamefont {M.}~\bibnamefont {Veldhorst}},\ }\bibfield  {title} {\bibinfo {title} {{Fast two-qubit logic with holes in germanium}},\ }\href {https://doi.org/10.1038/s41586-019-1919-3} {\bibfield  {journal} {\bibinfo  {journal} {Nature}\ }\textbf {\bibinfo {volume} {577}},\ \bibinfo {pages} {487} (\bibinfo {year} {2020}{\natexlab{b}})}\BibitemShut {NoStop}%
\bibitem [{\citenamefont {Froning}\ \emph {et~al.}(2021)\citenamefont {Froning}, \citenamefont {Camenzind}, \citenamefont {van~der Molen}, \citenamefont {Li}, \citenamefont {Bakkers}, \citenamefont {Zumbühl},\ and\ \citenamefont {Braakman}}]{Froning21}%
  \BibitemOpen
  \bibfield  {author} {\bibinfo {author} {\bibfnamefont {F.~N.~M.}\ \bibnamefont {Froning}}, \bibinfo {author} {\bibfnamefont {L.~C.}\ \bibnamefont {Camenzind}}, \bibinfo {author} {\bibfnamefont {O.~A.~H.}\ \bibnamefont {van~der Molen}}, \bibinfo {author} {\bibfnamefont {A.}~\bibnamefont {Li}}, \bibinfo {author} {\bibfnamefont {E.~P. A.~M.}\ \bibnamefont {Bakkers}}, \bibinfo {author} {\bibfnamefont {D.~M.}\ \bibnamefont {Zumbühl}},\ and\ \bibinfo {author} {\bibfnamefont {F.~R.}\ \bibnamefont {Braakman}},\ }\bibfield  {title} {\bibinfo {title} {Ultrafast hole spin qubit with gate-tunable spin–orbit switch functionality},\ }\href {https://doi.org/10.1038/s41565-020-00828-6} {\bibfield  {journal} {\bibinfo  {journal} {Nature Nanotechnology}\ }\textbf {\bibinfo {volume} {16}},\ \bibinfo {pages} {308} (\bibinfo {year} {2021})}\BibitemShut {NoStop}%
\bibitem [{\citenamefont {Hendrickx}\ \emph {et~al.}(2021)\citenamefont {Hendrickx}, \citenamefont {Lawrie~William}, \citenamefont {Russ}, \citenamefont {van Riggelen}, \citenamefont {de~Snoo}, \citenamefont {Schouten}, \citenamefont {Sammak}, \citenamefont {Scappucci},\ and\ \citenamefont {Veldhorst}}]{Hendrickx21}%
  \BibitemOpen
  \bibfield  {author} {\bibinfo {author} {\bibfnamefont {N.~W.}\ \bibnamefont {Hendrickx}}, \bibinfo {author} {\bibfnamefont {I.~L.}\ \bibnamefont {Lawrie~William}}, \bibinfo {author} {\bibfnamefont {M.}~\bibnamefont {Russ}}, \bibinfo {author} {\bibfnamefont {F.}~\bibnamefont {van Riggelen}}, \bibinfo {author} {\bibfnamefont {S.~L.}\ \bibnamefont {de~Snoo}}, \bibinfo {author} {\bibfnamefont {R.~N.}\ \bibnamefont {Schouten}}, \bibinfo {author} {\bibfnamefont {A.}~\bibnamefont {Sammak}}, \bibinfo {author} {\bibfnamefont {G.}~\bibnamefont {Scappucci}},\ and\ \bibinfo {author} {\bibfnamefont {M.}~\bibnamefont {Veldhorst}},\ }\bibfield  {title} {\bibinfo {title} {A four-qubit germanium quantum processor},\ }\href {https://doi.org/10.1038/s41586-021-03332-6} {\bibfield  {journal} {\bibinfo  {journal} {Nature}\ }\textbf {\bibinfo {volume} {591}},\ \bibinfo {pages} {580} (\bibinfo {year} {2021})}\BibitemShut {NoStop}%
\bibitem [{\citenamefont {Wang}\ \emph {et~al.}(2022)\citenamefont {Wang}, \citenamefont {Xu}, \citenamefont {Gao}, \citenamefont {Liu}, \citenamefont {Ma}, \citenamefont {Zhang}, \citenamefont {Wang}, \citenamefont {Cao}, \citenamefont {Wang}, \citenamefont {Zhang}, \citenamefont {Culcer}, \citenamefont {Hu}, \citenamefont {Jiang}, \citenamefont {Li}, \citenamefont {Guo},\ and\ \citenamefont {Guo}}]{wang2022ultrafast}%
  \BibitemOpen
  \bibfield  {author} {\bibinfo {author} {\bibfnamefont {K.}~\bibnamefont {Wang}}, \bibinfo {author} {\bibfnamefont {G.}~\bibnamefont {Xu}}, \bibinfo {author} {\bibfnamefont {F.}~\bibnamefont {Gao}}, \bibinfo {author} {\bibfnamefont {H.}~\bibnamefont {Liu}}, \bibinfo {author} {\bibfnamefont {R.-L.}\ \bibnamefont {Ma}}, \bibinfo {author} {\bibfnamefont {X.}~\bibnamefont {Zhang}}, \bibinfo {author} {\bibfnamefont {Z.}~\bibnamefont {Wang}}, \bibinfo {author} {\bibfnamefont {G.}~\bibnamefont {Cao}}, \bibinfo {author} {\bibfnamefont {T.}~\bibnamefont {Wang}}, \bibinfo {author} {\bibfnamefont {J.-J.}\ \bibnamefont {Zhang}}, \bibinfo {author} {\bibfnamefont {D.}~\bibnamefont {Culcer}}, \bibinfo {author} {\bibfnamefont {X.}~\bibnamefont {Hu}}, \bibinfo {author} {\bibfnamefont {H.-W.}\ \bibnamefont {Jiang}}, \bibinfo {author} {\bibfnamefont {H.-O.}\ \bibnamefont {Li}}, \bibinfo {author} {\bibfnamefont {G.-C.}\ \bibnamefont {Guo}},\ and\ \bibinfo {author} {\bibfnamefont {G.-P.}\ \bibnamefont {Guo}},\ }\bibfield
  {title} {\bibinfo {title} {Ultrafast coherent control of a hole spin qubit in a germanium quantum dot},\ }\href {https://doi.org/10.1038/s41467-021-27880-7} {\bibfield  {journal} {\bibinfo  {journal} {Nature Communications}\ }\textbf {\bibinfo {volume} {13}},\ \bibinfo {pages} {206} (\bibinfo {year} {2022})}\BibitemShut {NoStop}%
\bibitem [{\citenamefont {Borsoi}\ \emph {et~al.}(2023)\citenamefont {Borsoi}, \citenamefont {Hendrickx}, \citenamefont {John}, \citenamefont {Meyer}, \citenamefont {Motz}, \citenamefont {van Riggelen}, \citenamefont {Sammak}, \citenamefont {de~Snoo}, \citenamefont {Scappucci},\ and\ \citenamefont {Veldhorst}}]{Borsoi22}%
  \BibitemOpen
  \bibfield  {author} {\bibinfo {author} {\bibfnamefont {F.}~\bibnamefont {Borsoi}}, \bibinfo {author} {\bibfnamefont {N.~W.}\ \bibnamefont {Hendrickx}}, \bibinfo {author} {\bibfnamefont {V.}~\bibnamefont {John}}, \bibinfo {author} {\bibfnamefont {M.}~\bibnamefont {Meyer}}, \bibinfo {author} {\bibfnamefont {S.}~\bibnamefont {Motz}}, \bibinfo {author} {\bibfnamefont {F.}~\bibnamefont {van Riggelen}}, \bibinfo {author} {\bibfnamefont {A.}~\bibnamefont {Sammak}}, \bibinfo {author} {\bibfnamefont {S.~L.}\ \bibnamefont {de~Snoo}}, \bibinfo {author} {\bibfnamefont {G.}~\bibnamefont {Scappucci}},\ and\ \bibinfo {author} {\bibfnamefont {M.}~\bibnamefont {Veldhorst}},\ }\bibfield  {title} {\bibinfo {title} {Shared control of a 16 semiconductor quantum dot crossbar array},\ }\bibfield  {journal} {\bibinfo  {journal} {Nature Nanotechnology}\ }\href {https://doi.org/10.1038/s41565-023-01491-3} {10.1038/s41565-023-01491-3} (\bibinfo {year} {2023})\BibitemShut {NoStop}%
\bibitem [{\citenamefont {Wang}\ \emph {et~al.}(2023)\citenamefont {Wang}, \citenamefont {Déprez}, \citenamefont {Tidjani}, \citenamefont {Lawrie}, \citenamefont {Hendrickx}, \citenamefont {Sammak}, \citenamefont {Scappucci},\ and\ \citenamefont {Veldhorst}}]{Wang2023}%
  \BibitemOpen
  \bibfield  {author} {\bibinfo {author} {\bibfnamefont {C.-A.}\ \bibnamefont {Wang}}, \bibinfo {author} {\bibfnamefont {C.}~\bibnamefont {Déprez}}, \bibinfo {author} {\bibfnamefont {H.}~\bibnamefont {Tidjani}}, \bibinfo {author} {\bibfnamefont {W.~I.~L.}\ \bibnamefont {Lawrie}}, \bibinfo {author} {\bibfnamefont {N.~W.}\ \bibnamefont {Hendrickx}}, \bibinfo {author} {\bibfnamefont {A.}~\bibnamefont {Sammak}}, \bibinfo {author} {\bibfnamefont {G.}~\bibnamefont {Scappucci}},\ and\ \bibinfo {author} {\bibfnamefont {M.}~\bibnamefont {Veldhorst}},\ }\bibfield  {title} {\bibinfo {title} {Probing resonating valence bonds on a programmable germanium quantum simulator},\ }\href {https://doi.org/10.1038/s41534-023-00727-3} {\bibfield  {journal} {\bibinfo  {journal} {npj Quantum Information}\ }\textbf {\bibinfo {volume} {9}},\ \bibinfo {pages} {58} (\bibinfo {year} {2023})}\BibitemShut {NoStop}%
\bibitem [{\citenamefont {Jirovec}\ \emph {et~al.}(2022)\citenamefont {Jirovec}, \citenamefont {Mutter}, \citenamefont {Hofmann}, \citenamefont {Crippa}, \citenamefont {Rychetsky}, \citenamefont {Craig}, \citenamefont {Kukucka}, \citenamefont {Martins}, \citenamefont {Ballabio}, \citenamefont {Ares}, \citenamefont {Chrastina}, \citenamefont {Isella}, \citenamefont {Burkard},\ and\ \citenamefont {Katsaros}}]{Jirovec23}%
  \BibitemOpen
  \bibfield  {author} {\bibinfo {author} {\bibfnamefont {D.}~\bibnamefont {Jirovec}}, \bibinfo {author} {\bibfnamefont {P.~M.}\ \bibnamefont {Mutter}}, \bibinfo {author} {\bibfnamefont {A.}~\bibnamefont {Hofmann}}, \bibinfo {author} {\bibfnamefont {A.}~\bibnamefont {Crippa}}, \bibinfo {author} {\bibfnamefont {M.}~\bibnamefont {Rychetsky}}, \bibinfo {author} {\bibfnamefont {D.~L.}\ \bibnamefont {Craig}}, \bibinfo {author} {\bibfnamefont {J.}~\bibnamefont {Kukucka}}, \bibinfo {author} {\bibfnamefont {F.}~\bibnamefont {Martins}}, \bibinfo {author} {\bibfnamefont {A.}~\bibnamefont {Ballabio}}, \bibinfo {author} {\bibfnamefont {N.}~\bibnamefont {Ares}}, \bibinfo {author} {\bibfnamefont {D.}~\bibnamefont {Chrastina}}, \bibinfo {author} {\bibfnamefont {G.}~\bibnamefont {Isella}}, \bibinfo {author} {\bibfnamefont {G.}~\bibnamefont {Burkard}},\ and\ \bibinfo {author} {\bibfnamefont {G.}~\bibnamefont {Katsaros}},\ }\bibfield  {title} {\bibinfo {title} {Dynamics of hole singlet-triplet qubits with large $g$-factor
  differences},\ }\href {https://doi.org/10.1103/PhysRevLett.128.126803} {\bibfield  {journal} {\bibinfo  {journal} {Physical Review Letters}\ }\textbf {\bibinfo {volume} {128}},\ \bibinfo {pages} {126803} (\bibinfo {year} {2022})}\BibitemShut {NoStop}%
\bibitem [{\citenamefont {Zhang}\ \emph {et~al.}(2024)\citenamefont {Zhang}, \citenamefont {Morozova}, \citenamefont {Rimbach-Russ}, \citenamefont {Jirovec}, \citenamefont {Hsiao}, \citenamefont {Fariña}, \citenamefont {Wang}, \citenamefont {Oosterhout}, \citenamefont {Sammak}, \citenamefont {Scappucci}, \citenamefont {Veldhorst},\ and\ \citenamefont {Vandersypen}}]{Zhang23}%
  \BibitemOpen
  \bibfield  {author} {\bibinfo {author} {\bibfnamefont {X.}~\bibnamefont {Zhang}}, \bibinfo {author} {\bibfnamefont {E.}~\bibnamefont {Morozova}}, \bibinfo {author} {\bibfnamefont {M.}~\bibnamefont {Rimbach-Russ}}, \bibinfo {author} {\bibfnamefont {D.}~\bibnamefont {Jirovec}}, \bibinfo {author} {\bibfnamefont {T.-K.}\ \bibnamefont {Hsiao}}, \bibinfo {author} {\bibfnamefont {P.~C.}\ \bibnamefont {Fariña}}, \bibinfo {author} {\bibfnamefont {C.-A.}\ \bibnamefont {Wang}}, \bibinfo {author} {\bibfnamefont {S.~D.}\ \bibnamefont {Oosterhout}}, \bibinfo {author} {\bibfnamefont {A.}~\bibnamefont {Sammak}}, \bibinfo {author} {\bibfnamefont {G.}~\bibnamefont {Scappucci}}, \bibinfo {author} {\bibfnamefont {M.}~\bibnamefont {Veldhorst}},\ and\ \bibinfo {author} {\bibfnamefont {L.~M.~K.}\ \bibnamefont {Vandersypen}},\ }\bibfield  {title} {\bibinfo {title} {Universal control of four singlet-triplet qubits},\ }\bibfield  {journal} {\bibinfo  {journal} {Nature Nanotechnology}\ }\href
  {https://doi.org/10.1038/s41565-024-01817-9} {10.1038/s41565-024-01817-9} (\bibinfo {year} {2024})\BibitemShut {NoStop}%
\bibitem [{\citenamefont {Hendrickx}\ \emph {et~al.}(2024)\citenamefont {Hendrickx}, \citenamefont {Massai}, \citenamefont {Mergenthaler}, \citenamefont {Schupp}, \citenamefont {Paredes}, \citenamefont {Bedell}, \citenamefont {Salis},\ and\ \citenamefont {Fuhrer}}]{Hendrickx2024}%
  \BibitemOpen
  \bibfield  {author} {\bibinfo {author} {\bibfnamefont {N.~W.}\ \bibnamefont {Hendrickx}}, \bibinfo {author} {\bibfnamefont {L.}~\bibnamefont {Massai}}, \bibinfo {author} {\bibfnamefont {M.}~\bibnamefont {Mergenthaler}}, \bibinfo {author} {\bibfnamefont {F.}~\bibnamefont {Schupp}}, \bibinfo {author} {\bibfnamefont {S.}~\bibnamefont {Paredes}}, \bibinfo {author} {\bibfnamefont {S.~W.}\ \bibnamefont {Bedell}}, \bibinfo {author} {\bibfnamefont {G.}~\bibnamefont {Salis}},\ and\ \bibinfo {author} {\bibfnamefont {A.}~\bibnamefont {Fuhrer}},\ }\bibfield  {title} {\bibinfo {title} {Sweet-spot operation of a germanium hole spin qubit with highly anisotropic noise sensitivity},\ }\bibfield  {journal} {\bibinfo  {journal} {Nature Materials}\ }\href {https://doi.org/10.1038/s41563-024-01857-5} {10.1038/s41563-024-01857-5} (\bibinfo {year} {2024})\BibitemShut {NoStop}%
\bibitem [{\citenamefont {Wang}\ \emph {et~al.}(2024)\citenamefont {Wang}, \citenamefont {John}, \citenamefont {Tidjani}, \citenamefont {Yu}, \citenamefont {Ivlev}, \citenamefont {Déprez}, \citenamefont {van Riggelen-Doelman}, \citenamefont {Woods}, \citenamefont {Hendrickx}, \citenamefont {Lawrie}, \citenamefont {Stehouwer}, \citenamefont {Oosterhout}, \citenamefont {Sammak}, \citenamefont {Friesen}, \citenamefont {Scappucci}, \citenamefont {de~Snoo}, \citenamefont {Rimbach-Russ}, \citenamefont {Borsoi},\ and\ \citenamefont {Veldhorst}}]{Wang2024}%
  \BibitemOpen
  \bibfield  {author} {\bibinfo {author} {\bibfnamefont {C.-A.}\ \bibnamefont {Wang}}, \bibinfo {author} {\bibfnamefont {V.}~\bibnamefont {John}}, \bibinfo {author} {\bibfnamefont {H.}~\bibnamefont {Tidjani}}, \bibinfo {author} {\bibfnamefont {C.~X.}\ \bibnamefont {Yu}}, \bibinfo {author} {\bibfnamefont {A.~S.}\ \bibnamefont {Ivlev}}, \bibinfo {author} {\bibfnamefont {C.}~\bibnamefont {Déprez}}, \bibinfo {author} {\bibfnamefont {F.}~\bibnamefont {van Riggelen-Doelman}}, \bibinfo {author} {\bibfnamefont {B.~D.}\ \bibnamefont {Woods}}, \bibinfo {author} {\bibfnamefont {N.~W.}\ \bibnamefont {Hendrickx}}, \bibinfo {author} {\bibfnamefont {W.~I.~L.}\ \bibnamefont {Lawrie}}, \bibinfo {author} {\bibfnamefont {L.~E.~A.}\ \bibnamefont {Stehouwer}}, \bibinfo {author} {\bibfnamefont {S.~D.}\ \bibnamefont {Oosterhout}}, \bibinfo {author} {\bibfnamefont {A.}~\bibnamefont {Sammak}}, \bibinfo {author} {\bibfnamefont {M.}~\bibnamefont {Friesen}}, \bibinfo {author} {\bibfnamefont {G.}~\bibnamefont {Scappucci}}, \bibinfo
  {author} {\bibfnamefont {S.~L.}\ \bibnamefont {de~Snoo}}, \bibinfo {author} {\bibfnamefont {M.}~\bibnamefont {Rimbach-Russ}}, \bibinfo {author} {\bibfnamefont {F.}~\bibnamefont {Borsoi}},\ and\ \bibinfo {author} {\bibfnamefont {M.}~\bibnamefont {Veldhorst}},\ }\bibfield  {title} {\bibinfo {title} {Operating semiconductor quantum processors with hopping spins},\ }\href {https://doi.org/10.1126/science.ado5915} {\bibfield  {journal} {\bibinfo  {journal} {Science}\ }\textbf {\bibinfo {volume} {385}},\ \bibinfo {pages} {447} (\bibinfo {year} {2024})}\BibitemShut {NoStop}%
\bibitem [{\citenamefont {Martinez}\ and\ \citenamefont {Niquet}(2022)}]{Martinez2022}%
  \BibitemOpen
  \bibfield  {author} {\bibinfo {author} {\bibfnamefont {B.}~\bibnamefont {Martinez}}\ and\ \bibinfo {author} {\bibfnamefont {Y.-M.}\ \bibnamefont {Niquet}},\ }\bibfield  {title} {\bibinfo {title} {Variability of electron and hole spin qubits due to interface roughness and charge traps},\ }\href {https://doi.org/10.1103/PhysRevApplied.17.024022} {\bibfield  {journal} {\bibinfo  {journal} {Physical Review Applied}\ }\textbf {\bibinfo {volume} {17}},\ \bibinfo {pages} {024022} (\bibinfo {year} {2022})}\BibitemShut {NoStop}%
\bibitem [{\citenamefont {Varley}\ \emph {et~al.}(2023)\citenamefont {Varley}, \citenamefont {Ray},\ and\ \citenamefont {Lordi}}]{Varley23}%
  \BibitemOpen
  \bibfield  {author} {\bibinfo {author} {\bibfnamefont {J.~B.}\ \bibnamefont {Varley}}, \bibinfo {author} {\bibfnamefont {K.~G.}\ \bibnamefont {Ray}},\ and\ \bibinfo {author} {\bibfnamefont {V.}~\bibnamefont {Lordi}},\ }\bibfield  {title} {\bibinfo {title} {Dangling bonds as possible contributors to charge noise in silicon and silicon--germanium quantum dot qubits},\ }\href {https://doi.org/10.1021/acsami.3c06725} {\bibfield  {journal} {\bibinfo  {journal} {ACS Applied Materials {\&} Interfaces}\ }\textbf {\bibinfo {volume} {15}},\ \bibinfo {pages} {43111} (\bibinfo {year} {2023})}\BibitemShut {NoStop}%
\bibitem [{\citenamefont {Massai}\ \emph {et~al.}(2023)\citenamefont {Massai}, \citenamefont {Hetényi}, \citenamefont {Mergenthaler}, \citenamefont {Schupp}, \citenamefont {Sommer}, \citenamefont {Paredes}, \citenamefont {Bedell}, \citenamefont {Harvey-Collard}, \citenamefont {Salis}, \citenamefont {Fuhrer},\ and\ \citenamefont {Hendrickx}}]{Massai23}%
  \BibitemOpen
  \bibfield  {author} {\bibinfo {author} {\bibfnamefont {L.}~\bibnamefont {Massai}}, \bibinfo {author} {\bibfnamefont {B.}~\bibnamefont {Hetényi}}, \bibinfo {author} {\bibfnamefont {M.}~\bibnamefont {Mergenthaler}}, \bibinfo {author} {\bibfnamefont {F.~J.}\ \bibnamefont {Schupp}}, \bibinfo {author} {\bibfnamefont {L.}~\bibnamefont {Sommer}}, \bibinfo {author} {\bibfnamefont {S.}~\bibnamefont {Paredes}}, \bibinfo {author} {\bibfnamefont {S.~W.}\ \bibnamefont {Bedell}}, \bibinfo {author} {\bibfnamefont {P.}~\bibnamefont {Harvey-Collard}}, \bibinfo {author} {\bibfnamefont {G.}~\bibnamefont {Salis}}, \bibinfo {author} {\bibfnamefont {A.}~\bibnamefont {Fuhrer}},\ and\ \bibinfo {author} {\bibfnamefont {N.~W.}\ \bibnamefont {Hendrickx}},\ }\bibfield  {title} {\bibinfo {title} {Impact of interface traps on charge noise, mobility and percolation density in {Ge/SiGe} heterostructures},\ }\href {https://arxiv.org/abs/2310.05902} {\bibfield  {journal} {\bibinfo  {journal} {arXiv:2310.05902}\ } (\bibinfo {year}
  {2023})},\ \Eprint {https://arxiv.org/abs/2310.05902} {arXiv:2310.05902} \BibitemShut {NoStop}%
\bibitem [{\citenamefont {Martinez}\ \emph {et~al.}(2024)\citenamefont {Martinez}, \citenamefont {de~Franceschi},\ and\ \citenamefont {Niquet}}]{Martinez24}%
  \BibitemOpen
  \bibfield  {author} {\bibinfo {author} {\bibfnamefont {B.}~\bibnamefont {Martinez}}, \bibinfo {author} {\bibfnamefont {S.}~\bibnamefont {de~Franceschi}},\ and\ \bibinfo {author} {\bibfnamefont {Y.-M.}\ \bibnamefont {Niquet}},\ }\bibfield  {title} {\bibinfo {title} {Mitigating variability in epitaxial-heterostructure-based spin-qubit devices by optimizing gate layout},\ }\href {https://doi.org/10.1103/PhysRevApplied.22.024030} {\bibfield  {journal} {\bibinfo  {journal} {Phys. Rev. Appl.}\ }\textbf {\bibinfo {volume} {22}},\ \bibinfo {pages} {024030} (\bibinfo {year} {2024})}\BibitemShut {NoStop}%
\bibitem [{\citenamefont {Winkler}(2003)}]{Winkler03}%
  \BibitemOpen
  \bibfield  {author} {\bibinfo {author} {\bibfnamefont {R.}~\bibnamefont {Winkler}},\ }\href {https://doi.org/10.1007/b13586} {\emph {\bibinfo {title} {Spin-orbit coupling in two-dimensional electron and hole systems}}}\ (\bibinfo  {publisher} {Springer},\ \bibinfo {address} {Berlin},\ \bibinfo {year} {2003})\BibitemShut {NoStop}%
\bibitem [{\citenamefont {Marcellina}\ \emph {et~al.}(2017)\citenamefont {Marcellina}, \citenamefont {Hamilton}, \citenamefont {Winkler},\ and\ \citenamefont {Culcer}}]{Marcellina17}%
  \BibitemOpen
  \bibfield  {author} {\bibinfo {author} {\bibfnamefont {E.}~\bibnamefont {Marcellina}}, \bibinfo {author} {\bibfnamefont {A.~R.}\ \bibnamefont {Hamilton}}, \bibinfo {author} {\bibfnamefont {R.}~\bibnamefont {Winkler}},\ and\ \bibinfo {author} {\bibfnamefont {D.}~\bibnamefont {Culcer}},\ }\bibfield  {title} {\bibinfo {title} {Spin-orbit interactions in inversion-asymmetric two-dimensional hole systems: A variational analysis},\ }\href {https://doi.org/10.1103/PhysRevB.95.075305} {\bibfield  {journal} {\bibinfo  {journal} {Physical Review B}\ }\textbf {\bibinfo {volume} {95}},\ \bibinfo {pages} {075305} (\bibinfo {year} {2017})}\BibitemShut {NoStop}%
\bibitem [{\citenamefont {Terrazos}\ \emph {et~al.}(2021)\citenamefont {Terrazos}, \citenamefont {Marcellina}, \citenamefont {Wang}, \citenamefont {Coppersmith}, \citenamefont {Friesen}, \citenamefont {Hamilton}, \citenamefont {Hu}, \citenamefont {Koiller}, \citenamefont {Saraiva}, \citenamefont {Culcer},\ and\ \citenamefont {Capaz}}]{Terrazos21}%
  \BibitemOpen
  \bibfield  {author} {\bibinfo {author} {\bibfnamefont {L.~A.}\ \bibnamefont {Terrazos}}, \bibinfo {author} {\bibfnamefont {E.}~\bibnamefont {Marcellina}}, \bibinfo {author} {\bibfnamefont {Z.}~\bibnamefont {Wang}}, \bibinfo {author} {\bibfnamefont {S.~N.}\ \bibnamefont {Coppersmith}}, \bibinfo {author} {\bibfnamefont {M.}~\bibnamefont {Friesen}}, \bibinfo {author} {\bibfnamefont {A.~R.}\ \bibnamefont {Hamilton}}, \bibinfo {author} {\bibfnamefont {X.}~\bibnamefont {Hu}}, \bibinfo {author} {\bibfnamefont {B.}~\bibnamefont {Koiller}}, \bibinfo {author} {\bibfnamefont {A.~L.}\ \bibnamefont {Saraiva}}, \bibinfo {author} {\bibfnamefont {D.}~\bibnamefont {Culcer}},\ and\ \bibinfo {author} {\bibfnamefont {R.~B.}\ \bibnamefont {Capaz}},\ }\bibfield  {title} {\bibinfo {title} {Theory of hole-spin qubits in strained germanium quantum dots},\ }\href {https://doi.org/10.1103/PhysRevB.103.125201} {\bibfield  {journal} {\bibinfo  {journal} {Physical Review B}\ }\textbf {\bibinfo {volume} {103}},\ \bibinfo {pages} {125201}
  (\bibinfo {year} {2021})}\BibitemShut {NoStop}%
\bibitem [{\citenamefont {Wang}\ \emph {et~al.}(2021)\citenamefont {Wang}, \citenamefont {Marcellina}, \citenamefont {Hamilton}, \citenamefont {Cullen}, \citenamefont {Rogge}, \citenamefont {Salfi},\ and\ \citenamefont {Culcer}}]{Wang21}%
  \BibitemOpen
  \bibfield  {author} {\bibinfo {author} {\bibfnamefont {Z.}~\bibnamefont {Wang}}, \bibinfo {author} {\bibfnamefont {E.}~\bibnamefont {Marcellina}}, \bibinfo {author} {\bibfnamefont {A.~R.}\ \bibnamefont {Hamilton}}, \bibinfo {author} {\bibfnamefont {J.~H.}\ \bibnamefont {Cullen}}, \bibinfo {author} {\bibfnamefont {S.}~\bibnamefont {Rogge}}, \bibinfo {author} {\bibfnamefont {J.}~\bibnamefont {Salfi}},\ and\ \bibinfo {author} {\bibfnamefont {D.}~\bibnamefont {Culcer}},\ }\bibfield  {title} {\bibinfo {title} {Optimal operation points for ultrafast, highly coherent {Ge} hole spin-orbit qubits},\ }\href {https://doi.org/10.1038/s41534-021-00386-2} {\bibfield  {journal} {\bibinfo  {journal} {npj Quantum Information}\ }\textbf {\bibinfo {volume} {7}},\ \bibinfo {pages} {54} (\bibinfo {year} {2021})}\BibitemShut {NoStop}%
\bibitem [{\citenamefont {Bosco}\ \emph {et~al.}(2021)\citenamefont {Bosco}, \citenamefont {Benito}, \citenamefont {Adelsberger},\ and\ \citenamefont {Loss}}]{Bosco21b}%
  \BibitemOpen
  \bibfield  {author} {\bibinfo {author} {\bibfnamefont {S.}~\bibnamefont {Bosco}}, \bibinfo {author} {\bibfnamefont {M.}~\bibnamefont {Benito}}, \bibinfo {author} {\bibfnamefont {C.}~\bibnamefont {Adelsberger}},\ and\ \bibinfo {author} {\bibfnamefont {D.}~\bibnamefont {Loss}},\ }\bibfield  {title} {\bibinfo {title} {Squeezed hole spin qubits in {Ge} quantum dots with ultrafast gates at low power},\ }\href {https://doi.org/10.1103/PhysRevB.104.115425} {\bibfield  {journal} {\bibinfo  {journal} {Physical Review B}\ }\textbf {\bibinfo {volume} {104}},\ \bibinfo {pages} {115425} (\bibinfo {year} {2021})}\BibitemShut {NoStop}%
\bibitem [{\citenamefont {Martinez}\ \emph {et~al.}(2022)\citenamefont {Martinez}, \citenamefont {Abadillo-Uriel}, \citenamefont {Rodr\'{\i}guez-Mena},\ and\ \citenamefont {Niquet}}]{martinez2022hole}%
  \BibitemOpen
  \bibfield  {author} {\bibinfo {author} {\bibfnamefont {B.}~\bibnamefont {Martinez}}, \bibinfo {author} {\bibfnamefont {J.~C.}\ \bibnamefont {Abadillo-Uriel}}, \bibinfo {author} {\bibfnamefont {E.~A.}\ \bibnamefont {Rodr\'{\i}guez-Mena}},\ and\ \bibinfo {author} {\bibfnamefont {Y.-M.}\ \bibnamefont {Niquet}},\ }\bibfield  {title} {\bibinfo {title} {Hole spin manipulation in inhomogeneous and nonseparable electric fields},\ }\href {https://doi.org/10.1103/PhysRevB.106.235426} {\bibfield  {journal} {\bibinfo  {journal} {Physical Review B}\ }\textbf {\bibinfo {volume} {106}},\ \bibinfo {pages} {235426} (\bibinfo {year} {2022})}\BibitemShut {NoStop}%
\bibitem [{\citenamefont {Sarkar}\ \emph {et~al.}(2023)\citenamefont {Sarkar}, \citenamefont {Wang}, \citenamefont {Rendell}, \citenamefont {Hendrickx}, \citenamefont {Veldhorst}, \citenamefont {Scappucci}, \citenamefont {Khalifa}, \citenamefont {Salfi}, \citenamefont {Saraiva}, \citenamefont {Dzurak}, \citenamefont {Hamilton},\ and\ \citenamefont {Culcer}}]{Sarkar23}%
  \BibitemOpen
  \bibfield  {author} {\bibinfo {author} {\bibfnamefont {A.}~\bibnamefont {Sarkar}}, \bibinfo {author} {\bibfnamefont {Z.}~\bibnamefont {Wang}}, \bibinfo {author} {\bibfnamefont {M.}~\bibnamefont {Rendell}}, \bibinfo {author} {\bibfnamefont {N.~W.}\ \bibnamefont {Hendrickx}}, \bibinfo {author} {\bibfnamefont {M.}~\bibnamefont {Veldhorst}}, \bibinfo {author} {\bibfnamefont {G.}~\bibnamefont {Scappucci}}, \bibinfo {author} {\bibfnamefont {M.}~\bibnamefont {Khalifa}}, \bibinfo {author} {\bibfnamefont {J.}~\bibnamefont {Salfi}}, \bibinfo {author} {\bibfnamefont {A.}~\bibnamefont {Saraiva}}, \bibinfo {author} {\bibfnamefont {A.~S.}\ \bibnamefont {Dzurak}}, \bibinfo {author} {\bibfnamefont {A.~R.}\ \bibnamefont {Hamilton}},\ and\ \bibinfo {author} {\bibfnamefont {D.}~\bibnamefont {Culcer}},\ }\bibfield  {title} {\bibinfo {title} {Electrical operation of planar {Ge} hole spin qubits in an in-plane magnetic field},\ }\href {https://doi.org/10.1103/PhysRevB.108.245301} {\bibfield  {journal} {\bibinfo  {journal}
  {Physical Review B}\ }\textbf {\bibinfo {volume} {108}},\ \bibinfo {pages} {245301} (\bibinfo {year} {2023})}\BibitemShut {NoStop}%
\bibitem [{\citenamefont {Abadillo-Uriel}\ \emph {et~al.}(2023)\citenamefont {Abadillo-Uriel}, \citenamefont {Rodr\'{\i}guez-Mena}, \citenamefont {Martinez},\ and\ \citenamefont {Niquet}}]{Abadillo2023}%
  \BibitemOpen
  \bibfield  {author} {\bibinfo {author} {\bibfnamefont {J.~C.}\ \bibnamefont {Abadillo-Uriel}}, \bibinfo {author} {\bibfnamefont {E.~A.}\ \bibnamefont {Rodr\'{\i}guez-Mena}}, \bibinfo {author} {\bibfnamefont {B.}~\bibnamefont {Martinez}},\ and\ \bibinfo {author} {\bibfnamefont {Y.-M.}\ \bibnamefont {Niquet}},\ }\bibfield  {title} {\bibinfo {title} {Hole-spin driving by strain-induced spin-orbit interactions},\ }\href {https://doi.org/10.1103/PhysRevLett.131.097002} {\bibfield  {journal} {\bibinfo  {journal} {Physical Review Letters}\ }\textbf {\bibinfo {volume} {131}},\ \bibinfo {pages} {097002} (\bibinfo {year} {2023})}\BibitemShut {NoStop}%
\bibitem [{\citenamefont {Rodr\'{\i}guez-Mena}\ \emph {et~al.}(2023)\citenamefont {Rodr\'{\i}guez-Mena}, \citenamefont {Abadillo-Uriel}, \citenamefont {Veste}, \citenamefont {Martinez}, \citenamefont {Li}, \citenamefont {Skl\'enard},\ and\ \citenamefont {Niquet}}]{Rodriguez2023}%
  \BibitemOpen
  \bibfield  {author} {\bibinfo {author} {\bibfnamefont {E.~A.}\ \bibnamefont {Rodr\'{\i}guez-Mena}}, \bibinfo {author} {\bibfnamefont {J.~C.}\ \bibnamefont {Abadillo-Uriel}}, \bibinfo {author} {\bibfnamefont {G.}~\bibnamefont {Veste}}, \bibinfo {author} {\bibfnamefont {B.}~\bibnamefont {Martinez}}, \bibinfo {author} {\bibfnamefont {J.}~\bibnamefont {Li}}, \bibinfo {author} {\bibfnamefont {B.}~\bibnamefont {Skl\'enard}},\ and\ \bibinfo {author} {\bibfnamefont {Y.-M.}\ \bibnamefont {Niquet}},\ }\bibfield  {title} {\bibinfo {title} {Linear-in-momentum spin orbit interactions in planar {Ge/GeSi} heterostructures and spin qubits},\ }\href {https://doi.org/10.1103/PhysRevB.108.205416} {\bibfield  {journal} {\bibinfo  {journal} {Physical Review B}\ }\textbf {\bibinfo {volume} {108}},\ \bibinfo {pages} {205416} (\bibinfo {year} {2023})}\BibitemShut {NoStop}%
\bibitem [{\citenamefont {Mauro}\ \emph {et~al.}(2024)\citenamefont {Mauro}, \citenamefont {Rodr\'{\i}guez-Mena}, \citenamefont {Bassi}, \citenamefont {Schmitt},\ and\ \citenamefont {Niquet}}]{Mauro24}%
  \BibitemOpen
  \bibfield  {author} {\bibinfo {author} {\bibfnamefont {L.}~\bibnamefont {Mauro}}, \bibinfo {author} {\bibfnamefont {E.~A.}\ \bibnamefont {Rodr\'{\i}guez-Mena}}, \bibinfo {author} {\bibfnamefont {M.}~\bibnamefont {Bassi}}, \bibinfo {author} {\bibfnamefont {V.}~\bibnamefont {Schmitt}},\ and\ \bibinfo {author} {\bibfnamefont {Y.-M.}\ \bibnamefont {Niquet}},\ }\bibfield  {title} {\bibinfo {title} {Geometry of the dephasing sweet spots of spin-orbit qubits},\ }\href {https://doi.org/10.1103/PhysRevB.109.155406} {\bibfield  {journal} {\bibinfo  {journal} {Physical Review B}\ }\textbf {\bibinfo {volume} {109}},\ \bibinfo {pages} {155406} (\bibinfo {year} {2024})}\BibitemShut {NoStop}%
\bibitem [{\citenamefont {Venitucci}\ \emph {et~al.}(2018)\citenamefont {Venitucci}, \citenamefont {Bourdet}, \citenamefont {Pouzada},\ and\ \citenamefont {Niquet}}]{Venitucci18}%
  \BibitemOpen
  \bibfield  {author} {\bibinfo {author} {\bibfnamefont {B.}~\bibnamefont {Venitucci}}, \bibinfo {author} {\bibfnamefont {L.}~\bibnamefont {Bourdet}}, \bibinfo {author} {\bibfnamefont {D.}~\bibnamefont {Pouzada}},\ and\ \bibinfo {author} {\bibfnamefont {Y.-M.}\ \bibnamefont {Niquet}},\ }\bibfield  {title} {\bibinfo {title} {Electrical manipulation of semiconductor spin qubits within the $g$-matrix formalism},\ }\href {https://doi.org/10.1103/PhysRevB.98.155319} {\bibfield  {journal} {\bibinfo  {journal} {Physical Review B}\ }\textbf {\bibinfo {volume} {98}},\ \bibinfo {pages} {155319} (\bibinfo {year} {2018})}\BibitemShut {NoStop}%
\bibitem [{\citenamefont {Mansir}\ \emph {et~al.}(2018)\citenamefont {Mansir}, \citenamefont {Conti}, \citenamefont {Zeng}, \citenamefont {Pla}, \citenamefont {Bertet}, \citenamefont {Swift}, \citenamefont {Van~de Walle}, \citenamefont {Thewalt}, \citenamefont {Sklenard}, \citenamefont {Niquet},\ and\ \citenamefont {Morton}}]{Morton18}%
  \BibitemOpen
  \bibfield  {author} {\bibinfo {author} {\bibfnamefont {J.}~\bibnamefont {Mansir}}, \bibinfo {author} {\bibfnamefont {P.}~\bibnamefont {Conti}}, \bibinfo {author} {\bibfnamefont {Z.}~\bibnamefont {Zeng}}, \bibinfo {author} {\bibfnamefont {J.~J.}\ \bibnamefont {Pla}}, \bibinfo {author} {\bibfnamefont {P.}~\bibnamefont {Bertet}}, \bibinfo {author} {\bibfnamefont {M.~W.}\ \bibnamefont {Swift}}, \bibinfo {author} {\bibfnamefont {C.~G.}\ \bibnamefont {Van~de Walle}}, \bibinfo {author} {\bibfnamefont {M.~L.~W.}\ \bibnamefont {Thewalt}}, \bibinfo {author} {\bibfnamefont {B.}~\bibnamefont {Sklenard}}, \bibinfo {author} {\bibfnamefont {Y.~M.}\ \bibnamefont {Niquet}},\ and\ \bibinfo {author} {\bibfnamefont {J.~J.~L.}\ \bibnamefont {Morton}},\ }\bibfield  {title} {\bibinfo {title} {Linear hyperfine tuning of donor spins in silicon using hydrostatic strain},\ }\href {https://doi.org/10.1103/PhysRevLett.120.167701} {\bibfield  {journal} {\bibinfo  {journal} {Physical Review Letters}\ }\textbf {\bibinfo {volume} {120}},\
  \bibinfo {pages} {167701} (\bibinfo {year} {2018})}\BibitemShut {NoStop}%
\bibitem [{\citenamefont {Pla}\ \emph {et~al.}(2018)\citenamefont {Pla}, \citenamefont {Bienfait}, \citenamefont {Pica}, \citenamefont {Mansir}, \citenamefont {Mohiyaddin}, \citenamefont {Zeng}, \citenamefont {Niquet}, \citenamefont {Morello}, \citenamefont {Schenkel}, \citenamefont {Morton},\ and\ \citenamefont {Bertet}}]{Pla18}%
  \BibitemOpen
  \bibfield  {author} {\bibinfo {author} {\bibfnamefont {J.~J.}\ \bibnamefont {Pla}}, \bibinfo {author} {\bibfnamefont {A.}~\bibnamefont {Bienfait}}, \bibinfo {author} {\bibfnamefont {G.}~\bibnamefont {Pica}}, \bibinfo {author} {\bibfnamefont {J.}~\bibnamefont {Mansir}}, \bibinfo {author} {\bibfnamefont {F.~A.}\ \bibnamefont {Mohiyaddin}}, \bibinfo {author} {\bibfnamefont {Z.}~\bibnamefont {Zeng}}, \bibinfo {author} {\bibfnamefont {Y.-M.}\ \bibnamefont {Niquet}}, \bibinfo {author} {\bibfnamefont {A.}~\bibnamefont {Morello}}, \bibinfo {author} {\bibfnamefont {T.}~\bibnamefont {Schenkel}}, \bibinfo {author} {\bibfnamefont {J.~J.~L.}\ \bibnamefont {Morton}},\ and\ \bibinfo {author} {\bibfnamefont {P.}~\bibnamefont {Bertet}},\ }\bibfield  {title} {\bibinfo {title} {Strain-induced spin-resonance shifts in silicon devices},\ }\href {https://doi.org/10.1103/PhysRevApplied.9.044014} {\bibfield  {journal} {\bibinfo  {journal} {Physical Review Applied}\ }\textbf {\bibinfo {volume} {9}},\ \bibinfo {pages} {044014}
  (\bibinfo {year} {2018})}\BibitemShut {NoStop}%
\bibitem [{\citenamefont {Liles}\ \emph {et~al.}(2021)\citenamefont {Liles}, \citenamefont {Martins}, \citenamefont {Miserev}, \citenamefont {Kiselev}, \citenamefont {Thorvaldson}, \citenamefont {Rendell}, \citenamefont {Jin}, \citenamefont {Hudson}, \citenamefont {Veldhorst}, \citenamefont {Itoh}, \citenamefont {Sushkov}, \citenamefont {Ladd}, \citenamefont {Dzurak},\ and\ \citenamefont {Hamilton}}]{Liles20}%
  \BibitemOpen
  \bibfield  {author} {\bibinfo {author} {\bibfnamefont {S.~D.}\ \bibnamefont {Liles}}, \bibinfo {author} {\bibfnamefont {F.}~\bibnamefont {Martins}}, \bibinfo {author} {\bibfnamefont {D.~S.}\ \bibnamefont {Miserev}}, \bibinfo {author} {\bibfnamefont {A.~A.}\ \bibnamefont {Kiselev}}, \bibinfo {author} {\bibfnamefont {I.~D.}\ \bibnamefont {Thorvaldson}}, \bibinfo {author} {\bibfnamefont {M.~J.}\ \bibnamefont {Rendell}}, \bibinfo {author} {\bibfnamefont {I.~K.}\ \bibnamefont {Jin}}, \bibinfo {author} {\bibfnamefont {F.~E.}\ \bibnamefont {Hudson}}, \bibinfo {author} {\bibfnamefont {M.}~\bibnamefont {Veldhorst}}, \bibinfo {author} {\bibfnamefont {K.~M.}\ \bibnamefont {Itoh}}, \bibinfo {author} {\bibfnamefont {O.~P.}\ \bibnamefont {Sushkov}}, \bibinfo {author} {\bibfnamefont {T.~D.}\ \bibnamefont {Ladd}}, \bibinfo {author} {\bibfnamefont {A.~S.}\ \bibnamefont {Dzurak}},\ and\ \bibinfo {author} {\bibfnamefont {A.~R.}\ \bibnamefont {Hamilton}},\ }\bibfield  {title} {\bibinfo {title} {Electrical control of the $g$
  tensor of the first hole in a silicon {MOS} quantum dot},\ }\href {https://doi.org/10.1103/PhysRevB.104.235303} {\bibfield  {journal} {\bibinfo  {journal} {Physical Review B}\ }\textbf {\bibinfo {volume} {104}},\ \bibinfo {pages} {235303} (\bibinfo {year} {2021})}\BibitemShut {NoStop}%
\bibitem [{\citenamefont {Adelsberger}\ \emph {et~al.}(2024)\citenamefont {Adelsberger}, \citenamefont {Bosco}, \citenamefont {Klinovaja},\ and\ \citenamefont {Loss}}]{Adelsberger23}%
  \BibitemOpen
  \bibfield  {author} {\bibinfo {author} {\bibfnamefont {C.}~\bibnamefont {Adelsberger}}, \bibinfo {author} {\bibfnamefont {S.}~\bibnamefont {Bosco}}, \bibinfo {author} {\bibfnamefont {J.}~\bibnamefont {Klinovaja}},\ and\ \bibinfo {author} {\bibfnamefont {D.}~\bibnamefont {Loss}},\ }\bibfield  {title} {\bibinfo {title} {Valley-free silicon fins caused by shear strain},\ }\href {https://doi.org/10.1103/PhysRevLett.133.037001} {\bibfield  {journal} {\bibinfo  {journal} {Physical Review Letters}\ }\textbf {\bibinfo {volume} {133}},\ \bibinfo {pages} {037001} (\bibinfo {year} {2024})}\BibitemShut {NoStop}%
\bibitem [{\citenamefont {Woods}\ \emph {et~al.}(2024)\citenamefont {Woods}, \citenamefont {Soomro}, \citenamefont {Joseph}, \citenamefont {Frink}, \citenamefont {Joynt}, \citenamefont {Eriksson},\ and\ \citenamefont {Friesen}}]{Woods2024}%
  \BibitemOpen
  \bibfield  {author} {\bibinfo {author} {\bibfnamefont {B.~D.}\ \bibnamefont {Woods}}, \bibinfo {author} {\bibfnamefont {H.}~\bibnamefont {Soomro}}, \bibinfo {author} {\bibfnamefont {E.~S.}\ \bibnamefont {Joseph}}, \bibinfo {author} {\bibfnamefont {C.~C.~D.}\ \bibnamefont {Frink}}, \bibinfo {author} {\bibfnamefont {R.}~\bibnamefont {Joynt}}, \bibinfo {author} {\bibfnamefont {M.~A.}\ \bibnamefont {Eriksson}},\ and\ \bibinfo {author} {\bibfnamefont {M.}~\bibnamefont {Friesen}},\ }\bibfield  {title} {\bibinfo {title} {Coupling conduction-band valleys in sige heterostructures via shear strain and ge concentration oscillations},\ }\href {https://doi.org/10.1038/s41534-024-00853-6} {\bibfield  {journal} {\bibinfo  {journal} {npj Quantum Information}\ }\textbf {\bibinfo {volume} {10}},\ \bibinfo {pages} {54} (\bibinfo {year} {2024})}\BibitemShut {NoStop}%
\bibitem [{\citenamefont {Corley-Wiciak}\ \emph {et~al.}(2023)\citenamefont {Corley-Wiciak}, \citenamefont {Richter}, \citenamefont {Zoellner}, \citenamefont {Zaitsev}, \citenamefont {Manganelli}, \citenamefont {Zatterin}, \citenamefont {Sch{\"u}lli}, \citenamefont {Corley-Wiciak}, \citenamefont {Katzer}, \citenamefont {Reichmann}, \citenamefont {Klesse}, \citenamefont {Hendrickx}, \citenamefont {Sammak}, \citenamefont {Veldhorst}, \citenamefont {Scappucci}, \citenamefont {Virgilio},\ and\ \citenamefont {Capellini}}]{Corley2023}%
  \BibitemOpen
  \bibfield  {author} {\bibinfo {author} {\bibfnamefont {C.}~\bibnamefont {Corley-Wiciak}}, \bibinfo {author} {\bibfnamefont {C.}~\bibnamefont {Richter}}, \bibinfo {author} {\bibfnamefont {M.~H.}\ \bibnamefont {Zoellner}}, \bibinfo {author} {\bibfnamefont {I.}~\bibnamefont {Zaitsev}}, \bibinfo {author} {\bibfnamefont {C.~L.}\ \bibnamefont {Manganelli}}, \bibinfo {author} {\bibfnamefont {E.}~\bibnamefont {Zatterin}}, \bibinfo {author} {\bibfnamefont {T.~U.}\ \bibnamefont {Sch{\"u}lli}}, \bibinfo {author} {\bibfnamefont {A.~A.}\ \bibnamefont {Corley-Wiciak}}, \bibinfo {author} {\bibfnamefont {J.}~\bibnamefont {Katzer}}, \bibinfo {author} {\bibfnamefont {F.}~\bibnamefont {Reichmann}}, \bibinfo {author} {\bibfnamefont {W.~M.}\ \bibnamefont {Klesse}}, \bibinfo {author} {\bibfnamefont {N.~W.}\ \bibnamefont {Hendrickx}}, \bibinfo {author} {\bibfnamefont {A.}~\bibnamefont {Sammak}}, \bibinfo {author} {\bibfnamefont {M.}~\bibnamefont {Veldhorst}}, \bibinfo {author} {\bibfnamefont {G.}~\bibnamefont {Scappucci}},
  \bibinfo {author} {\bibfnamefont {M.}~\bibnamefont {Virgilio}},\ and\ \bibinfo {author} {\bibfnamefont {G.}~\bibnamefont {Capellini}},\ }\bibfield  {title} {\bibinfo {title} {Nanoscale mapping of the {3D} strain tensor in a germanium quantum well hosting a functional spin qubit device},\ }\href {https://doi.org/10.1021/acsami.2c17395} {\bibfield  {journal} {\bibinfo  {journal} {ACS Applied Materials {\&} Interfaces}\ }\textbf {\bibinfo {volume} {15}},\ \bibinfo {pages} {3119} (\bibinfo {year} {2023})}\BibitemShut {NoStop}%
\bibitem [{\citenamefont {Abragam}\ and\ \citenamefont {Bleaney}(1971)}]{Abragam1970}%
  \BibitemOpen
  \bibfield  {author} {\bibinfo {author} {\bibfnamefont {A.}~\bibnamefont {Abragam}}\ and\ \bibinfo {author} {\bibfnamefont {B.}~\bibnamefont {Bleaney}},\ }\href@noop {} {\emph {\bibinfo {title} {Electron Paramagnetic Resonance of Transition Ions}}},\ Resonance Paramagnetique Electronique Des Ions de Transition\ (\bibinfo  {publisher} {{Presses Universitaires de France}},\ \bibinfo {address} {{France}},\ \bibinfo {year} {1971})\BibitemShut {NoStop}%
\bibitem [{\citenamefont {Crippa}\ \emph {et~al.}(2018)\citenamefont {Crippa}, \citenamefont {Maurand}, \citenamefont {Bourdet}, \citenamefont {Kotekar-Patil}, \citenamefont {Amisse}, \citenamefont {Jehl}, \citenamefont {Sanquer}, \citenamefont {Laviéville}, \citenamefont {Bohuslavskyi}, \citenamefont {Hutin}, \citenamefont {Barraud}, \citenamefont {Vinet}, \citenamefont {Niquet},\ and\ \citenamefont {{De Franceschi}}}]{Crippa18}%
  \BibitemOpen
  \bibfield  {author} {\bibinfo {author} {\bibfnamefont {A.}~\bibnamefont {Crippa}}, \bibinfo {author} {\bibfnamefont {R.}~\bibnamefont {Maurand}}, \bibinfo {author} {\bibfnamefont {L.}~\bibnamefont {Bourdet}}, \bibinfo {author} {\bibfnamefont {D.}~\bibnamefont {Kotekar-Patil}}, \bibinfo {author} {\bibfnamefont {A.}~\bibnamefont {Amisse}}, \bibinfo {author} {\bibfnamefont {X.}~\bibnamefont {Jehl}}, \bibinfo {author} {\bibfnamefont {M.}~\bibnamefont {Sanquer}}, \bibinfo {author} {\bibfnamefont {R.}~\bibnamefont {Laviéville}}, \bibinfo {author} {\bibfnamefont {H.}~\bibnamefont {Bohuslavskyi}}, \bibinfo {author} {\bibfnamefont {L.}~\bibnamefont {Hutin}}, \bibinfo {author} {\bibfnamefont {S.}~\bibnamefont {Barraud}}, \bibinfo {author} {\bibfnamefont {M.}~\bibnamefont {Vinet}}, \bibinfo {author} {\bibfnamefont {Y.-M.}\ \bibnamefont {Niquet}},\ and\ \bibinfo {author} {\bibfnamefont {S.}~\bibnamefont {{De Franceschi}}},\ }\bibfield  {title} {\bibinfo {title} {Electrical spin driving by $g$-matrix modulation in
  spin-orbit qubits},\ }\href {https://doi.org/10.1103/PhysRevLett.120.137702} {\bibfield  {journal} {\bibinfo  {journal} {Physical Review Letters}\ }\textbf {\bibinfo {volume} {120}},\ \bibinfo {pages} {137702} (\bibinfo {year} {2018})}\BibitemShut {NoStop}%
\bibitem [{\citenamefont {Michal}\ \emph {et~al.}(2021)\citenamefont {Michal}, \citenamefont {Venitucci},\ and\ \citenamefont {Niquet}}]{Michal21}%
  \BibitemOpen
  \bibfield  {author} {\bibinfo {author} {\bibfnamefont {V.~P.}\ \bibnamefont {Michal}}, \bibinfo {author} {\bibfnamefont {B.}~\bibnamefont {Venitucci}},\ and\ \bibinfo {author} {\bibfnamefont {Y.-M.}\ \bibnamefont {Niquet}},\ }\bibfield  {title} {\bibinfo {title} {Longitudinal and transverse electric field manipulation of hole spin-orbit qubits in one-dimensional channels},\ }\href {https://doi.org/10.1103/PhysRevB.103.045305} {\bibfield  {journal} {\bibinfo  {journal} {Physical Review B}\ }\textbf {\bibinfo {volume} {103}},\ \bibinfo {pages} {045305} (\bibinfo {year} {2021})}\BibitemShut {NoStop}%
\bibitem [{Note1()}]{Note1}%
  \BibitemOpen
  \bibinfo {note} {We assume here that $\langle p_xp_y\rangle =0$; if not the case there are additional $\delta g_{xy}^{(c)}$ and $\delta g_{yx}^{(c)}$ corrections given in Ref.~\cite {Michal21}. We have also discarded small $\propto \langle p_x^2\rangle /\Delta _\protect \mathrm {LH}$ and $\propto \langle p_y^2\rangle /\Delta _\protect \mathrm {LH}$ corrections to $g_{zz}$.}\BibitemShut {Stop}%
\bibitem [{\citenamefont {Piot}\ \emph {et~al.}(2022)\citenamefont {Piot}, \citenamefont {Brun}, \citenamefont {Schmitt}, \citenamefont {Zihlmann}, \citenamefont {Michal}, \citenamefont {Apra}, \citenamefont {Abadillo-Uriel}, \citenamefont {Jehl}, \citenamefont {Bertrand}, \citenamefont {Niebojewski}, \citenamefont {Hutin}, \citenamefont {Vinet}, \citenamefont {Urdampilleta}, \citenamefont {Meunier}, \citenamefont {Niquet}, \citenamefont {Maurand},\ and\ \citenamefont {De~Franceschi}}]{Piot22}%
  \BibitemOpen
  \bibfield  {author} {\bibinfo {author} {\bibfnamefont {N.}~\bibnamefont {Piot}}, \bibinfo {author} {\bibfnamefont {B.}~\bibnamefont {Brun}}, \bibinfo {author} {\bibfnamefont {V.}~\bibnamefont {Schmitt}}, \bibinfo {author} {\bibfnamefont {S.}~\bibnamefont {Zihlmann}}, \bibinfo {author} {\bibfnamefont {V.~P.}\ \bibnamefont {Michal}}, \bibinfo {author} {\bibfnamefont {A.}~\bibnamefont {Apra}}, \bibinfo {author} {\bibfnamefont {J.~C.}\ \bibnamefont {Abadillo-Uriel}}, \bibinfo {author} {\bibfnamefont {X.}~\bibnamefont {Jehl}}, \bibinfo {author} {\bibfnamefont {B.}~\bibnamefont {Bertrand}}, \bibinfo {author} {\bibfnamefont {H.}~\bibnamefont {Niebojewski}}, \bibinfo {author} {\bibfnamefont {L.}~\bibnamefont {Hutin}}, \bibinfo {author} {\bibfnamefont {M.}~\bibnamefont {Vinet}}, \bibinfo {author} {\bibfnamefont {M.}~\bibnamefont {Urdampilleta}}, \bibinfo {author} {\bibfnamefont {T.}~\bibnamefont {Meunier}}, \bibinfo {author} {\bibfnamefont {Y.-M.}\ \bibnamefont {Niquet}}, \bibinfo {author} {\bibfnamefont
  {R.}~\bibnamefont {Maurand}},\ and\ \bibinfo {author} {\bibfnamefont {S.}~\bibnamefont {De~Franceschi}},\ }\bibfield  {title} {\bibinfo {title} {A single hole spin with enhanced coherence in natural silicon},\ }\href {https://doi.org/10.1038/s41565-022-01196-z} {\bibfield  {journal} {\bibinfo  {journal} {Nature Nanotechnology}\ }\textbf {\bibinfo {volume} {17}},\ \bibinfo {pages} {1072} (\bibinfo {year} {2022})}\BibitemShut {NoStop}%
\bibitem [{\citenamefont {Fischetti}\ and\ \citenamefont {Laux}(1996)}]{Fischetti96}%
  \BibitemOpen
  \bibfield  {author} {\bibinfo {author} {\bibfnamefont {M.~V.}\ \bibnamefont {Fischetti}}\ and\ \bibinfo {author} {\bibfnamefont {S.~E.}\ \bibnamefont {Laux}},\ }\bibfield  {title} {\bibinfo {title} {Band structure, deformation potentials, and carrier mobility in strained {Si}, {Ge}, and {SiGe} alloys},\ }\href {https://doi.org/10.1063/1.363052} {\bibfield  {journal} {\bibinfo  {journal} {Journal of Applied Physics}\ }\textbf {\bibinfo {volume} {80}},\ \bibinfo {pages} {2234} (\bibinfo {year} {1996})}\BibitemShut {NoStop}%
\bibitem [{Note2()}]{Note2}%
  \BibitemOpen
  \bibinfo {note} {For completeness, \begin {equation} \delta g_{zy}^{(\varepsilon )}=-\protect \frac {4\protect \sqrt {3}\kappa d_v}{\Delta _\protect \mathrm {LH}}\langle \varepsilon _{yz}\rangle \protect \,, \end {equation} and: \begin {equation} \delta g_{xy}^{(\varepsilon )}=-\delta g_{yx}^{(\varepsilon )}=\protect \frac {12b_v\kappa }{\Delta _\protect \mathrm {LH}}\langle \varepsilon _{xy}\rangle \protect \,. \end {equation}}\BibitemShut {NoStop}%
\bibitem [{Note3()}]{Note3}%
  \BibitemOpen
  \bibinfo {note} {{In the geometry of Fig.~\ref {fig:device}, the dot can be squeezed along $y$ by applying a positive voltage on the B and T gates. However, the squeezed hole tends to be pulled out of the well by the B and T gates, and ultimately localizes at the top GeSi/Al$_2$O$_3$ interface where the action of the side gates is screened by the C gate. In the present device, we can at best reach $\langle p_y^2\rangle \approx 9\langle p_x^2\rangle $ and $|g_{xx}|\approx 0.3$, $|g_{yy}|\approx 0$ before losing control of the squeezed dot \cite {martinez2022hole,Mauro24}.}}\BibitemShut {Stop}%
\bibitem [{\citenamefont {Luttinger}(1956)}]{Luttinger56}%
  \BibitemOpen
  \bibfield  {author} {\bibinfo {author} {\bibfnamefont {J.~M.}\ \bibnamefont {Luttinger}},\ }\bibfield  {title} {\bibinfo {title} {Quantum theory of cyclotron resonance in semiconductors: General theory},\ }\href {https://doi.org/10.1103/PhysRev.102.1030} {\bibfield  {journal} {\bibinfo  {journal} {Physical Review}\ }\textbf {\bibinfo {volume} {102}},\ \bibinfo {pages} {1030} (\bibinfo {year} {1956})}\BibitemShut {NoStop}%
\bibitem [{\citenamefont {Lew Yan~Voon}\ and\ \citenamefont {Willatzen}(2009)}]{KP09}%
  \BibitemOpen
  \bibfield  {author} {\bibinfo {author} {\bibfnamefont {L.~C.}\ \bibnamefont {Lew Yan~Voon}}\ and\ \bibinfo {author} {\bibfnamefont {M.}~\bibnamefont {Willatzen}},\ }\href {https://doi.org/10.1007/978-3-540-92872-0} {\emph {\bibinfo {title} {The k p Method}}}\ (\bibinfo  {publisher} {Springer},\ \bibinfo {address} {Berlin},\ \bibinfo {year} {2009})\BibitemShut {NoStop}%
\bibitem [{Note4()}]{Note4}%
  \BibitemOpen
  \bibinfo {note} {Indeed, $\varepsilon _{zz}=-(c_{12}/c_{11})(\varepsilon _{xx}+\varepsilon _{yy})$ with $c_{11}$ and $c_{12}$ the elastic constants of the material, so that $\Delta _\protect \mathrm {LH}$ as given by Eq.~\protect \eqref {eq:deltalh} is independent on $\alpha $ as long as $\varepsilon _{xx}+\varepsilon _{yy}$ is constant.}\BibitemShut {Stop}%
\bibitem [{Note5()}]{Note5}%
  \BibitemOpen
  \bibinfo {note} {The fitted $\Delta _\protect \mathrm {LH}$ is actually larger than the $\approx 70$\protect \,meV splitting between the uncoupled, ground HH and LH states expected from Eq.~\protect \eqref {eq:delta} because Eqs.~\protect \eqref {eq:dgc} and \protect \eqref {eq:dgs} are approximations to sums over LH states with different excitation energies.}\BibitemShut {Stop}%
\bibitem [{\citenamefont {Thompson}\ \emph {et~al.}(2004)\citenamefont {Thompson}, \citenamefont {Armstrong}, \citenamefont {Auth}, \citenamefont {Cea}, \citenamefont {Chau}, \citenamefont {Glass}, \citenamefont {Hoffman}, \citenamefont {Klaus}, \citenamefont {Ma}, \citenamefont {Mcintyre}, \citenamefont {Murthy}, \citenamefont {Obradovic}, \citenamefont {Shifren}, \citenamefont {Sivakumar}, \citenamefont {Tyagi}, \citenamefont {Ghani}, \citenamefont {Mistry}, \citenamefont {Bohr},\ and\ \citenamefont {El-Mansy}}]{Thompson04}%
  \BibitemOpen
  \bibfield  {author} {\bibinfo {author} {\bibfnamefont {S.}~\bibnamefont {Thompson}}, \bibinfo {author} {\bibfnamefont {M.}~\bibnamefont {Armstrong}}, \bibinfo {author} {\bibfnamefont {C.}~\bibnamefont {Auth}}, \bibinfo {author} {\bibfnamefont {S.}~\bibnamefont {Cea}}, \bibinfo {author} {\bibfnamefont {R.}~\bibnamefont {Chau}}, \bibinfo {author} {\bibfnamefont {G.}~\bibnamefont {Glass}}, \bibinfo {author} {\bibfnamefont {T.}~\bibnamefont {Hoffman}}, \bibinfo {author} {\bibfnamefont {J.}~\bibnamefont {Klaus}}, \bibinfo {author} {\bibfnamefont {Z.}~\bibnamefont {Ma}}, \bibinfo {author} {\bibfnamefont {B.}~\bibnamefont {Mcintyre}}, \bibinfo {author} {\bibfnamefont {A.}~\bibnamefont {Murthy}}, \bibinfo {author} {\bibfnamefont {B.}~\bibnamefont {Obradovic}}, \bibinfo {author} {\bibfnamefont {L.}~\bibnamefont {Shifren}}, \bibinfo {author} {\bibfnamefont {S.}~\bibnamefont {Sivakumar}}, \bibinfo {author} {\bibfnamefont {S.}~\bibnamefont {Tyagi}}, \bibinfo {author} {\bibfnamefont {T.}~\bibnamefont {Ghani}}, \bibinfo
  {author} {\bibfnamefont {K.}~\bibnamefont {Mistry}}, \bibinfo {author} {\bibfnamefont {M.}~\bibnamefont {Bohr}},\ and\ \bibinfo {author} {\bibfnamefont {Y.}~\bibnamefont {El-Mansy}},\ }\bibfield  {title} {\bibinfo {title} {A logic nanotechnology featuring strained-silicon},\ }\href {https://doi.org/10.1109/LED.2004.825195} {\bibfield  {journal} {\bibinfo  {journal} {IEEE Electron Device Letters}\ }\textbf {\bibinfo {volume} {25}},\ \bibinfo {pages} {191} (\bibinfo {year} {2004})}\BibitemShut {NoStop}%
\bibitem [{\citenamefont {Tsutsui}\ \emph {et~al.}(2019)\citenamefont {Tsutsui}, \citenamefont {Mochizuki}, \citenamefont {Loubet}, \citenamefont {Bedell},\ and\ \citenamefont {Sadana}}]{Tsutsui19}%
  \BibitemOpen
  \bibfield  {author} {\bibinfo {author} {\bibfnamefont {G.}~\bibnamefont {Tsutsui}}, \bibinfo {author} {\bibfnamefont {S.}~\bibnamefont {Mochizuki}}, \bibinfo {author} {\bibfnamefont {N.}~\bibnamefont {Loubet}}, \bibinfo {author} {\bibfnamefont {S.~W.}\ \bibnamefont {Bedell}},\ and\ \bibinfo {author} {\bibfnamefont {D.~K.}\ \bibnamefont {Sadana}},\ }\bibfield  {title} {\bibinfo {title} {{Strain engineering in functional materials}},\ }\href {https://doi.org/10.1063/1.5075637} {\bibfield  {journal} {\bibinfo  {journal} {AIP Advances}\ }\textbf {\bibinfo {volume} {9}},\ \bibinfo {pages} {030701} (\bibinfo {year} {2019})}\BibitemShut {NoStop}%
\bibitem [{\citenamefont {Ito}\ \emph {et~al.}(2000)\citenamefont {Ito}, \citenamefont {Namba}, \citenamefont {Yamaguchi}, \citenamefont {Hirata}, \citenamefont {Ando}, \citenamefont {Koyama}, \citenamefont {Kuroki}, \citenamefont {Ikezawa}, \citenamefont {Suzuki}, \citenamefont {Saitoh},\ and\ \citenamefont {Horiuchi}}]{Ito00}%
  \BibitemOpen
  \bibfield  {author} {\bibinfo {author} {\bibfnamefont {S.}~\bibnamefont {Ito}}, \bibinfo {author} {\bibfnamefont {H.}~\bibnamefont {Namba}}, \bibinfo {author} {\bibfnamefont {K.}~\bibnamefont {Yamaguchi}}, \bibinfo {author} {\bibfnamefont {T.}~\bibnamefont {Hirata}}, \bibinfo {author} {\bibfnamefont {K.}~\bibnamefont {Ando}}, \bibinfo {author} {\bibfnamefont {S.}~\bibnamefont {Koyama}}, \bibinfo {author} {\bibfnamefont {S.}~\bibnamefont {Kuroki}}, \bibinfo {author} {\bibfnamefont {N.}~\bibnamefont {Ikezawa}}, \bibinfo {author} {\bibfnamefont {T.}~\bibnamefont {Suzuki}}, \bibinfo {author} {\bibfnamefont {T.}~\bibnamefont {Saitoh}},\ and\ \bibinfo {author} {\bibfnamefont {T.}~\bibnamefont {Horiuchi}},\ }\bibfield  {title} {\bibinfo {title} {Mechanical stress effect of etch-stop nitride and its impact on deep submicron transistor design},\ }in\ \href {https://doi.org/10.1109/IEDM.2000.904303} {\emph {\bibinfo {booktitle} {International Electron Devices Meeting}}}\ (\bibinfo {year} {2000})\ p.\ \bibinfo {pages}
  {247}\BibitemShut {NoStop}%
\bibitem [{\citenamefont {Shimizu}\ \emph {et~al.}(2001)\citenamefont {Shimizu}, \citenamefont {Hachimine}, \citenamefont {Ohki}, \citenamefont {Ohta}, \citenamefont {Koguchi}, \citenamefont {Nonaka}, \citenamefont {Sato},\ and\ \citenamefont {Ootsuka}}]{Shimizu01}%
  \BibitemOpen
  \bibfield  {author} {\bibinfo {author} {\bibfnamefont {A.}~\bibnamefont {Shimizu}}, \bibinfo {author} {\bibfnamefont {K.}~\bibnamefont {Hachimine}}, \bibinfo {author} {\bibfnamefont {N.}~\bibnamefont {Ohki}}, \bibinfo {author} {\bibfnamefont {H.}~\bibnamefont {Ohta}}, \bibinfo {author} {\bibfnamefont {M.}~\bibnamefont {Koguchi}}, \bibinfo {author} {\bibfnamefont {Y.}~\bibnamefont {Nonaka}}, \bibinfo {author} {\bibfnamefont {H.}~\bibnamefont {Sato}},\ and\ \bibinfo {author} {\bibfnamefont {F.}~\bibnamefont {Ootsuka}},\ }\bibfield  {title} {\bibinfo {title} {Local mechanical-stress control ({LMC}): a new technique for {CMOS}-performance enhancement},\ }in\ \href {https://doi.org/10.1109/IEDM.2001.979529} {\emph {\bibinfo {booktitle} {International Electron Devices Meeting}}}\ (\bibinfo {year} {2001})\ p.\ \bibinfo {pages} {19.4.1}\BibitemShut {NoStop}%
\bibitem [{\citenamefont {Armand~Pilon}\ \emph {et~al.}(2019)\citenamefont {Armand~Pilon}, \citenamefont {Lyasota}, \citenamefont {Niquet}, \citenamefont {Reboud}, \citenamefont {Calvo}, \citenamefont {Pauc}, \citenamefont {Widiez}, \citenamefont {Bonzon}, \citenamefont {Hartmann}, \citenamefont {Chelnokov}, \citenamefont {Faist},\ and\ \citenamefont {Sigg}}]{Pilon2019}%
  \BibitemOpen
  \bibfield  {author} {\bibinfo {author} {\bibfnamefont {F.~T.}\ \bibnamefont {Armand~Pilon}}, \bibinfo {author} {\bibfnamefont {A.}~\bibnamefont {Lyasota}}, \bibinfo {author} {\bibfnamefont {Y.-M.}\ \bibnamefont {Niquet}}, \bibinfo {author} {\bibfnamefont {V.}~\bibnamefont {Reboud}}, \bibinfo {author} {\bibfnamefont {V.}~\bibnamefont {Calvo}}, \bibinfo {author} {\bibfnamefont {N.}~\bibnamefont {Pauc}}, \bibinfo {author} {\bibfnamefont {J.}~\bibnamefont {Widiez}}, \bibinfo {author} {\bibfnamefont {C.}~\bibnamefont {Bonzon}}, \bibinfo {author} {\bibfnamefont {J.~M.}\ \bibnamefont {Hartmann}}, \bibinfo {author} {\bibfnamefont {A.}~\bibnamefont {Chelnokov}}, \bibinfo {author} {\bibfnamefont {J.}~\bibnamefont {Faist}},\ and\ \bibinfo {author} {\bibfnamefont {H.}~\bibnamefont {Sigg}},\ }\bibfield  {title} {\bibinfo {title} {Lasing in strained germanium microbridges},\ }\href {https://doi.org/10.1038/s41467-019-10655-6} {\bibfield  {journal} {\bibinfo  {journal} {Nature Communuications}\ }\textbf {\bibinfo {volume}
  {10}},\ \bibinfo {pages} {54} (\bibinfo {year} {2019})}\BibitemShut {NoStop}%
\bibitem [{\citenamefont {Richter}\ \emph {et~al.}(2008)\citenamefont {Richter}, \citenamefont {Arnoldus}, \citenamefont {Hansen},\ and\ \citenamefont {Thomsen}}]{Richter08}%
  \BibitemOpen
  \bibfield  {author} {\bibinfo {author} {\bibfnamefont {J.}~\bibnamefont {Richter}}, \bibinfo {author} {\bibfnamefont {M.~B.}\ \bibnamefont {Arnoldus}}, \bibinfo {author} {\bibfnamefont {O.}~\bibnamefont {Hansen}},\ and\ \bibinfo {author} {\bibfnamefont {E.~V.}\ \bibnamefont {Thomsen}},\ }\bibfield  {title} {\bibinfo {title} {{Four point bending setup for characterization of semiconductor piezoresistance}},\ }\href {https://doi.org/10.1063/1.2908428} {\bibfield  {journal} {\bibinfo  {journal} {Review of Scientific Instruments}\ }\textbf {\bibinfo {volume} {79}},\ \bibinfo {pages} {044703} (\bibinfo {year} {2008})}\BibitemShut {NoStop}%
\bibitem [{\citenamefont {Chen}\ \emph {et~al.}(2011)\citenamefont {Chen}, \citenamefont {Euaruksakul}, \citenamefont {Liu}, \citenamefont {Himpsel}, \citenamefont {Liu},\ and\ \citenamefont {Lagally}}]{Chen11}%
  \BibitemOpen
  \bibfield  {author} {\bibinfo {author} {\bibfnamefont {F.}~\bibnamefont {Chen}}, \bibinfo {author} {\bibfnamefont {C.}~\bibnamefont {Euaruksakul}}, \bibinfo {author} {\bibfnamefont {Z.}~\bibnamefont {Liu}}, \bibinfo {author} {\bibfnamefont {F.~J.}\ \bibnamefont {Himpsel}}, \bibinfo {author} {\bibfnamefont {F.}~\bibnamefont {Liu}},\ and\ \bibinfo {author} {\bibfnamefont {M.~G.}\ \bibnamefont {Lagally}},\ }\bibfield  {title} {\bibinfo {title} {Conduction band structure and electron mobility in uniaxially strained si via externally applied strain in nanomembranes},\ }\href {https://doi.org/10.1088/0022-3727/44/32/325107} {\bibfield  {journal} {\bibinfo  {journal} {Journal of Physics D: Applied Physics}\ }\textbf {\bibinfo {volume} {44}},\ \bibinfo {pages} {325107} (\bibinfo {year} {2011})}\BibitemShut {NoStop}%
\bibitem [{Note6()}]{Note6}%
  \BibitemOpen
  \bibinfo {note} {This definition is introduced for convenience. When $t\to \infty $, the strain $\varepsilon _{xx}$ in the buffer is not expected to tend to zero (but to a negative value, as the buffer remains strained along $y$) and remains inhomogeneous along $x$ owing, notably, to the competition between the relaxation of the buffer and Ge well.}\BibitemShut {Stop}%
\bibitem [{\citenamefont {Rashba}\ and\ \citenamefont {Efros}(2003)}]{Rashba03}%
  \BibitemOpen
  \bibfield  {author} {\bibinfo {author} {\bibfnamefont {E.~I.}\ \bibnamefont {Rashba}}\ and\ \bibinfo {author} {\bibfnamefont {A.~L.}\ \bibnamefont {Efros}},\ }\bibfield  {title} {\bibinfo {title} {Orbital mechanisms of electron-spin manipulation by an electric field},\ }\href {https://doi.org/10.1103/PhysRevLett.91.126405} {\bibfield  {journal} {\bibinfo  {journal} {Physical Review Letters}\ }\textbf {\bibinfo {volume} {91}},\ \bibinfo {pages} {126405} (\bibinfo {year} {2003})}\BibitemShut {NoStop}%
\bibitem [{\citenamefont {Kato}\ \emph {et~al.}(2003)\citenamefont {Kato}, \citenamefont {Myers}, \citenamefont {Driscoll}, \citenamefont {Gossard}, \citenamefont {Levy},\ and\ \citenamefont {Awschalom}}]{Kato03}%
  \BibitemOpen
  \bibfield  {author} {\bibinfo {author} {\bibfnamefont {Y.}~\bibnamefont {Kato}}, \bibinfo {author} {\bibfnamefont {R.~C.}\ \bibnamefont {Myers}}, \bibinfo {author} {\bibfnamefont {D.~C.}\ \bibnamefont {Driscoll}}, \bibinfo {author} {\bibfnamefont {A.~C.}\ \bibnamefont {Gossard}}, \bibinfo {author} {\bibfnamefont {J.}~\bibnamefont {Levy}},\ and\ \bibinfo {author} {\bibfnamefont {D.~D.}\ \bibnamefont {Awschalom}},\ }\bibfield  {title} {\bibinfo {title} {Gigahertz electron spin manipulation using voltage-controlled $g$-tensor modulation},\ }\href {https://doi.org/10.1126/science.1080880} {\bibfield  {journal} {\bibinfo  {journal} {Science}\ }\textbf {\bibinfo {volume} {299}},\ \bibinfo {pages} {1201} (\bibinfo {year} {2003})}\BibitemShut {NoStop}%
\bibitem [{\citenamefont {Golovach}\ \emph {et~al.}(2006)\citenamefont {Golovach}, \citenamefont {Borhani},\ and\ \citenamefont {Loss}}]{Golovach06}%
  \BibitemOpen
  \bibfield  {author} {\bibinfo {author} {\bibfnamefont {V.~N.}\ \bibnamefont {Golovach}}, \bibinfo {author} {\bibfnamefont {M.}~\bibnamefont {Borhani}},\ and\ \bibinfo {author} {\bibfnamefont {D.}~\bibnamefont {Loss}},\ }\bibfield  {title} {\bibinfo {title} {Electric-dipole-induced spin resonance in quantum dots},\ }\href {https://doi.org/10.1103/PhysRevB.74.165319} {\bibfield  {journal} {\bibinfo  {journal} {Physical Review B}\ }\textbf {\bibinfo {volume} {74}},\ \bibinfo {pages} {165319} (\bibinfo {year} {2006})}\BibitemShut {NoStop}%
\bibitem [{\citenamefont {Rashba}(2008)}]{Rashba08}%
  \BibitemOpen
  \bibfield  {author} {\bibinfo {author} {\bibfnamefont {E.~I.}\ \bibnamefont {Rashba}},\ }\bibfield  {title} {\bibinfo {title} {Theory of electric dipole spin resonance in quantum dots: Mean field theory with gaussian fluctuations and beyond},\ }\href {https://doi.org/10.1103/PhysRevB.78.195302} {\bibfield  {journal} {\bibinfo  {journal} {Physical Review B}\ }\textbf {\bibinfo {volume} {78}},\ \bibinfo {pages} {195302} (\bibinfo {year} {2008})}\BibitemShut {NoStop}%
\bibitem [{Note7()}]{Note7}%
  \BibitemOpen
  \bibinfo {note} {Eq.~\protect \eqref {eq:Psi} is exact at $x_c=0$ (where $\Delta g_{xx}=0$ by symmetry). For arbitrary $\Delta g_{xx}$ and $\Delta g_{zx}$, \begin {equation} \Psi =\protect \frac {|\protect \bar {g}_{xx}\Delta g_{zx}-(\protect \bar {g}_{zx}+g_{zz}b_{z})\Delta g_{xx}|}{\protect \bar {g}_{xx}^2+(\protect \bar {g}_{zx}+g_{zz}b_{z})^{2}}\protect \,, \end {equation} where $\protect \bar {g}_{xx}=(g_{xx}(x_1)+g_{xx}(x_2))/2$.}\BibitemShut {Stop}%
\bibitem [{Note8()}]{Note8}%
  \BibitemOpen
  \bibinfo {note} {Actually, the Rabi frequency $\protect \bar {f}_\protect \mathrm {R}$ resulting from Eq.~\protect \eqref {eq:Psi} is formally the same as Eq. (3) of Ref.~\cite {martinez2022hole} with the substitution $\Delta g_{zx}\equiv \pi g_{zx}^\prime V_\protect \mathrm {ac}/2$: the shuttling protocol takes advantage (with respect to single dot EDSR) of the much larger displacements of the hole, hence the much larger modulations of $g_{zx}$.}\BibitemShut {Stop}%
\end{thebibliography}%

\end{document}